\documentclass[pre,groupedaddress,twocolumn,a4paper,showpacs]{revtex4-1}
\usepackage{graphics}
\usepackage{amssymb}
\usepackage{wasysym}
\usepackage{latexsym}
\usepackage{graphicx}
\usepackage{epsfig}
\usepackage{subfigure}
\usepackage{float}
\begin{document}

\title{Local adaptation, phenotypic plasticity, and species coexistence}

  \author{Jos\'e F.  Fontanari}
\affiliation{Instituto de F\'{\i}sica de S\~ao Carlos,
  Universidade de S\~ao Paulo,
  Caixa Postal 369, 13560-970 S\~ao Carlos, S\~ao Paulo, Brazil}
  
 \author{Margarida Matos}
  \affiliation{cE3c - Centre for Ecology, Evolution and Environmental Changes \& CHANGE - Global Change and Sustainability Institute, Lisboa, Portugal}
   \affiliation{Departamento de Biologia Animal, Faculdade de Ci\^encias, Universidade de Lisboa, Lisboa, Portugal}
   
   \author{Mauro Santos}
           \affiliation{Departament de Gen\`etica i de Microbiologia, Grup de Gen\`omica, Bioinform\`atica i Biologia Evolutiva (GBBE), Universitat Aut\`onoma de Barcelona, Spain}
    \affiliation{cE3c - Centre for Ecology, Evolution and Environmental Changes \& CHANGE - Global Change and Sustainability Institute, Lisboa, Portugal}

\begin{abstract}
Understanding the mechanisms of species coexistence has always been a fundamental topic in ecology. Classical theory predicts that interspecific competition may select for traits that stabilize niche differences, although recent work shows that this is not strictly necessary. Here we ask whether adaptive phenotypic plasticity could allow species coexistence (i.e., some stability at an equilibrium point) without ecological differentiation in habitat use. We used individual-based stochastic simulations defining a landscape composed of spatially uncorrelated or autocorrelated environmental patches, where two species with the same competitive strategies, not able to coexist without some form of phenotypic plasticity, expanded their ranges in the absence of a competition-colonization trade-off (a well-studied mechanism for species diversity). Each patch is characterized by a random environmental value that determines the optimal phenotype of its occupants.  In such a scenario, only local adaptation and gene flow (migration) may interact to promote genetic variation and coexistence in the metapopulation.  Results show that a competitively inferior species with adaptive phenotypic plasticity can coexist in a same patch with a  competitively superior, non-plastic species, provided the    migration rates  and  variances of the  patches'  environmental values are sufficiently large. 
\end{abstract}

\maketitle

\section{Introduction}

Spatial variation in the direction and strength of natural selection may often lead to eco-evolutionary dynamics. Local selection for the optimum phenotype could be hindered not only by gene flow but also from competitive interactions with other species; interactions which, in turn, could also be affected by the dual processes of gene flow and divergent selection \citep{Hendry(2017)}. Furthermore, theoretical models have shown that phenotypic plasticity, the ability of organisms to express different phenotypes depending on environmental conditions \citep{Bradshaw(1965),Schlichting(1986)}, readily evolves when selective conditions are variable, whether in time or space \citep{Hendry(2017),Pfennig(2021)}. In particular, spatial environmental variation where selection favors different phenotypes in each environment facilitates the evolution of adaptive phenotypic plasticity given some movement between environments\citep{Via(1985),Scheiner(1998),Gomulkiewicz(1992)}. Since phenotypic plasticity occurs within an ecological context - e.g., a normal or helmet morph in water fleas depending on the absence or presence of predators \citep{Agrawal(1999)}, or quorum sensing in bacteria according to surrounding bacterial cell density \citep{Miller(2001)} - much current interest focuses on how variation in phenotypic plasticity can affect the dynamics of interacting populations or species \citep{Fischer(2014),Turcotte(2016),Perez(2019),Muthukrishnan(2020),Start(2020),Gomez(2021)}.

Classical theory predicts that interspecific competition may select for traits that stabilize niche differences, weakening competitive interactions and therefore promoting species coexistence \citep{Macarthur(1967),Slatkin(1980),Doebeli(1996)}. However, recent work shows that ecological niche differentiation is not a requirement for species coexistence, and ecologically equivalent species can coexist when behaviors associated with reproductive interactions and sexual selection affect species demography in a frequency-dependent way \citep{Gomez(2021)}. On the other hand, the effect of phenotypic plasticity on species coexistence has been mainly framed within the classic context, in the sense that plasticity for ecologically relevant traits can eventually stabilize niche differentiation \citep{Turcotte(2016)}. Our aim here is to tackle the following question: does phenotypic plasticity affect species coexistence to the point that a competitively inferior plastic species can coexist with a competitively superior nonplastic one in the absence of niche differences? Specifically, we consider the following thought experiment: take an ecological model and contrast the community dynamics with or without intraspecific expressed variation for plasticity. When and why does variation change the dynamics? \citep{Bolnick(2011)}.

Here we develop a computational model to investigate the effect of phenotypic plasticity on species’ coexistence. We assume a density-compensating process which controls the size of the population (i.e., density-dependent population growth), coupled with a density- and frequency-independent viability selection for a local optimum that can be attained by adaptive phenotypic plasticity. We ran individual-based stochastic simulations using a two-dimensional landscape composed of spatially uncorrelated or autocorrelated environmental patches. We assumed two species: a competitively superior nonplastic species 1, that will always displace a second, phenotypically plastic species 2, in a single patch as well as in a two-dimensional landscape, with migration between patches but without environmental heterogeneity, which is temporally constant (i.e., when each patch has the same environmental value at each time step that defines the optimum phenotype). We mainly focus on the scenario where the two species are placed in a single random patch of a spatially heterogeneous and empty landscape, and thereafter are allowed to expand their ranges without being subjected to a competition-colonization trade-off (a well-studied mechanism for species diversity maintenance; \citet{Hastings(1980),Calcagno(2006),Muthukrishnan(2020)}.  Individuals of both species migrate to adjacent patches with the same probability per generation and have the same competitive strategies (i.e., the same absolute intra- and interspecific competition coefficients all over the patches) across the spatially varying landscape, but those patches with the highest average fitness contribute the most individuals (hard selection; \citet{Christiansen(1975)}). A brief digression: here we refer to fitness in the evolutionary context of population genetics, and not as the average competitive ability as used in the framework of ``modern coexistence theory'' \citep{Barabas(2018)}. Intuition suggests that expressing phenotypic plasticity will enhance local adaptation \citep{Scheiner(1998),Scheiner(2013)}, which could give some fitness advantage to the ecologically inferior plastic species and facilitate coexistence. Quantitative numerical results as well as qualitative analytical arguments  support this intuition. In particular, we show that both species coexist in most patches provided the variance of the optimum phenotypes across patches and the migration probability are sufficiently large. This conclusion holds true even when plasticity was to a certain extent costly.

\section{Model}\label{sec:model}

Here we describe an eco-evolutionary scenario to investigate the possibility of coexistence between two species when the ecological competition matrix  violates the mutual invasibility condition for any given patch. 

\subsection{Spatial setting}\label{sec:env}

We constructed an individual-based model to simulate a metapopulation of two multi-locus, haploid species that occupy discrete patches located on a 2-dimensional grid of linear length $L$ and toroidal shape (a doughnut) to avoid edge effects. Each patch on the grid is characterized by an environmental value $E_i, ~i=1, \ldots,L^2$, which are random variables distributed by the multivariate normal distribution
\begin{equation}\label{eq:multi}
f \left ( E; \mu, \Sigma \right ) = \frac{\exp \left [ -\frac{1}{2} (E-\mu)^T \Sigma^{-1} (E - \mu) \right ]}{(2\pi)^{L^2/2} |\Sigma|^{1/2}} ,
\end{equation}
where $E^T = (E_1, \ldots, E_{L^2})$ and  $\mu^T = (\mu_1, \ldots, \mu_{L^2})$  is a vector whose elements are the expected values of the environmental values, i.e., $\mathbb{E}(E_i) = \mu_i$. Here $\Sigma$ is the covariance matrix whose elements are $\Sigma_{ij} = \sqrt{\sigma_i^2 \sigma_j^2} \rho_{ij}$  where $\sigma_i^2$  is the variance of the environmental value at patch $i$ and $\rho_{ij}$ is the correlation between the environmental values at patches $i$ and $j$, which we choose to depend on the Euclidian distance $d_{ij}$
between those patches. Explicitly, we set $\rho_{ij} = \rho^{d_{ij}}$   where $\rho \in [0,1]$  is the correlation  between the environmental values of patches for which $d_{ij}=1$. Of course, $d_{ij}=1$ is the smallest distance between any two patches in the grid. We note that the correlation decreases exponentially with the distance between patches, i.e., $\rho_{ij} = \exp (-d_{ij}/\xi)$  where $\xi = 1/|\ln \rho |$ is the correlation length of the environment. 

A word is in order about the calculation of the Euclidean distance $d$ between two points $(i_1,i_2)$ and $(j_1,j_2)$ in a rectangular grid with cyclic boundary conditions (toroid). Let us assume that the open grid is $L_1 \times L_2$, i.e., that there are $L_1$ patches in the horizontal direction and  $L_2$ in the vertical direction ($L_1=L_2=L$ for the grid considered in this paper), so that   $i_1, j_1 = 1, \ldots,  L_1$  and  $i_2, j_2 = 1, \ldots,  L_2$. The horizontal and vertical distances between these points are given by the equations $d_h = \min ( |j_1 - i_1|, L_1 - |j_1 - i_1|)$ and $d_v = \min ( |j_2 - i_2|, L_2 - |j_2 - i_2|)$, from where we can readily calculate the Euclidean distance, viz., $d= \sqrt{d_h^2 + d_v^2}$ .

To avoid a profusion of parameters we assume that the patches are statistically identical, i.e., $\mu_i=\mu$  and $\sigma_i^2 = \sigma_e^2$ for $i=1, \ldots, L^2$.  With this assumption we can set $\mu = 0$ without loss of generality, since a different choice of $\mu$ would amount to a uniform shift on the environmental values and so it would be inconsequential. Although we set $\sigma_e^2=2$ in most of our simulations, we have also analyzed the effect of $\sigma_e^2$ on species’ coexistence. We note that for $\rho=0$ the environmental values $E_i$ are statistically independent normal random variables with mean zero and variance $\sigma_e^2$. Most of our analysis will focus on the uncorrelated environment, but we have also analyzed the possibility of coexistence in environmentally autocorrelated landscapes (i.e., $\rho > 0$).  As a brief technical note, we mention that in the case the matrix $\Sigma$ is symmetric and positive definite we can readily produce samples of the random vector $\mathbf{E}$ by setting $\mathbf{E}= \mu + \Sigma^{1/2} \mathbf{X}$ where $ \mathbf{X}^T = (X_1,\ldots,X_{L^2})$ is a random vector whose components are statistically independent standard normal random variables \citep{Wasserman(2004)}. The main difficulty here is the calculation of the square root of the matrix $\Sigma$, which can be done using its spectral decomposition.

\subsection{Viability selection}\label{sec:sel}

Following \citet{Scheiner(2020)}, the  phenotype $Z_i$  of an individual located at patch $i$ at the time of development was determined by 40 haploid loci as
\begin{equation}\label{Z}
Z_i = \sum_{k=1}^{m_{r}} R_k + E_i b  \sum_{k=1}^{m_p} P_k + \epsilon ,
\end{equation}
where $R_k$ are  the allelic values at the $m_{r} =20$ nonplastic or rigid loci (i.e., loci whose  phenotypic expression does not ontogenetically react to the environmental value), $P_k$ are the allelic values at the $m_p=20$ plastic loci (their phenotypic expression depends on external environmental cues that influence development) and $\epsilon$  is a normally distributed environmental effect with mean $0$ and variance $\sigma_\epsilon = 1/10$.  Here $b$ is the plasticity parameter that takes on the value $b=0$ for  the nonplastic species (species 1) and $b=1$ for the plastic species (species 2). There is no lack of generality in this choice because the allelic values $P_k$ (and $R_k$ as well) are real-valued variables and so any other choice of the plasticity parameter  can be reset to $b=1$  by a proper  rescaling of $P_k$.

The initial allelic values for all loci were also independently drawn from a normal distribution with mean $0$ and variance $1/10$. Hence the sum of allelic effects for each set of loci is a normal random variable of mean $0$ and variance $20/10 = 2$, which matches our typical choice for the variance of the environmental values, viz., $\sigma_e^2 = 2$. For a  given genotype, the phenotype $ Z_i$ at  patch $i$ is a linear function of the environment value $E_i$ and so $\sum_{k=1}^{m_{r}} R_k$ 
 is the intercept  and $ b  \sum_{k=1}^{m_p} P_k$  is the slope \citep{Scheiner(2013),Scheiner(2020)}. In the initial setup, the expected values of these quantities are zero.

Selection is only for viability, and the survival probability of an individual at patch $i$ depends on its phenotype and the cost of plasticity. Here we assume a Gaussian fitness model \citep{Scheiner(2020)}
\begin{equation}\label{W}
W_i = \exp \left [  - \frac{(Z_i-E_i)^2}{2w^2} - \frac{c}{2} \left (  \sum_{k=1}^{m_p} P_k \right )^2 \right ] ,
\end{equation}
where $E_i$ is the optimum phenotype that coincides with the patch’s environmental value
(i.e., stabilizing selection with a moving optimum), $w^2$  is inversely proportional to the strength of stabilizing selection and $c \geq 0$ determines the cost of plasticity.  Here we set $w^2=1$ without loss of generality. In fact, we can easily eliminate the parameter $w^2$ from the model by rescaling the adaptive nonplastic allele values $R_k '= R_k/w$  and the patches environmental values $E_i' = E_i/w$. The adaptive plastic allele values $P_k$ do not change. Since $E_i \sim N(0,\sigma_e^2)$, we have $E_i' \sim N(0,\sigma_e^2/w^2)$. Hence the effect of $w^2$ is simply a rescaling of the variance of the patches environmental values $\sigma_e^2$. In other words, increasing the strength of selection (i.e., decreasing $w^2$) is equivalent to increasing   $\sigma_e^2$ and hence to increasing the  roughness of the landscape. 
Equation (\ref{W})  includes maintenance costs of plasticity \citep{DeWitt(1998)} because there is a proportional reduction in survival when $c> 0$ even if plasticity is not expressed as it is  the case for species 1. However, in this paper we assume that individuals belonging to  species 1 do not carry plastic alleles, so effectively $b=0$ and $c=0$ for them.

We note that there are no intra and interspecific interactions during the viability selection process, whose net effect is to decrease the population of both species.  After passing the viability selection process, the surviving individuals compete among themselves to repopulate their patch, as described next.

\subsection{Ecological competition}\label{sec:comp}

We assume that when the two species occupy the same patch they interact and there is  interference or scramble competition. Given the species abundances $N_{1i}$ and $N_{2i}$ in patch $i=1,\ldots,L^2$   after viability selection, the total number of offspring $N'_{1i}$ and $N'_{2i}$ produced by the individuals of each species is determined by Ricker's equations \citep{Ricker(1954)}
\begin{eqnarray}
        N'_{1i} & =   &   N_{1i} \exp \left [ r \left ( 1 - \frac{a_{11}N_{1i} + a_{12} N_{2i}}{K_{max}} \right ) \right ]  \label{N1}\\
      N'_{2i} & =   & N_{2i} \exp \left [ r \left ( 1 - \frac{a_{21}N_{1i} + a_{22} N_{2i}}{K_{max}} \right ) \right ] \label{N2},
 \end{eqnarray}
where $\mbox{e}^r$  is the maximum growth rate in a low-density population, $K_{max}$ is the carrying capacity of each species when alone (assuming $a_{11}=a_{22}=1$), and $a_{ij}$ is the per capita effect of species $j$ on species $i$ \citep{Godfray(1991)}. For simplicity, here we assume that both species have the same maximum growth rate and equilibrium population size when alone in a patch.

In the absence of migration, so the individuals are confined to their birth patches, we consider the scenario where the nonplastic species (i.e., species 1) outcompetes the plastic species (i.e., species 2). This scenario is achieved by setting $r \in (0,2)$, $ a_{21} > a_{11}$ and $a_{12} <a_{22}$. Furthermore, since we do not want to distinguish between the two species when only one species is present in the metapopulation, we set $a_{11}=a_{22}=1$. To investigate the possibility of species coexistence (or lack thereof) when the ecologically inferior species 2 displays plasticity we set $a_{21} = 3/2 $, $a_{12}=1/2$   and $r =0.6$ in our simulations. Assuming a relatively large $r$ is reasonable in the case of expanding species as, e.g., insects \citep{Frazier(2006)} and plants (Appendix in \citet{Franco(2004)}). We emphasize that our choice for the competition matrix $a$ excludes the possibility of mutual invasion, which is a standard requisite for species coexistence \citep{Pasztor(2006)}. In fact, since $a_{21} =3/2 >   1= a_{11}$ species 1 can invade a resident   population of individuals of species 2 at equilibrium, but since $a_{12} = 1/2 <   1= a_{22}$ species 2 cannot invade  a resident   population of individuals of species 1 at equilibrium. It is instructive to note that $\det(a) =1/4 >0$, so our competition matrix offers a counterexample to the  fallacious statement that $\det(a)  >0$ implies negative frequency dependence (i.e. rare advantage) and hence ensures mutual invasibility (TBox 9.1 in \citet{Pasztor(2006)}).

We note that Ricker equations (\ref{N1}) and (\ref{N2})  yield real values for the number of offspring of each species in  patch $i=1, \ldots,L ^2$ and, in fact, the conditions that guarantee the superiority of species 1 over species 2 presented above for the single-patch situation are valid only when species numbers or abundances  $N_{1i}$ and $N_{2i}$  are real variables. It turns out that transforming those real variables into integer variables that are necessary for our individual-based simulations may produce spurious results, such as permanence of a few individuals of species 2 in patches dominated by individuals of species 1 or the stability of patches dominated by individuals of species 2 against the invasion of a few individuals of species 1. Here we circumvent this difficulty by taking the ceiling function in equation (\ref{N1})  (i.e., the least integer greater than or equal to $N'_{1i}$) and the floor function in equation (\ref{N2}) (i.e., the greatest integer less than or equal to $N'_{2i}$), which biases the competition in favor of species 1. In fact, this procedure impairs considerably the ability of species 2 to colonize a vacant patch $i$ even when there is no competition (i.e., $N_{1i} =0$) because a single founder of species 2 cannot produce more than one offspring for our choice of growth rate ($r = 0.6$). For instance,  if there is a  single adult individual of species  2 in an otherwise empty patch $i$ (i.e., $N_{2i} =1$ and $N_{1i}=0$), then equation (\ref{N2}) yields $N'_{2i} < \mbox{e}^r \approx 1.8$. Since $\mbox{floor} (1.8) = 1$, a single founder of species 2 cannot populate a vacant patch. 
This effect is mitigated when the migration rate is large since  in this case  there is a good chance that several individuals of species 2 migrate together to the same patch. However, this is actually a convenient scenario for our purposes since the more ecologically impaired species 2 is, the more remarkable the finding that plasticity can guarantee its permanence in the metapopulation.

In addition, we set $K_{max}$ as a hard upper bound to the number of offspring $N'_{1i}$  and$N'_{2i}$ in patch $i$. In other words, whenever $N'_{li} > K_{max}$ we set $N'_{li} = K_{max}$  for $l=1, 2$. This procedure is actually inconsequential because the populations at each patch approach the support capacity $K_{max}$ from below because of our choice of the growth rate (viz., $r = 0.6$).  For Ricker's growth equation,  overshooting and  the possibility of limit cycles happens for $r>2$ only  (see, e.g., \citet{Franco(2017)}). In the Supplementary Material, we present  several instances of the time evolution of the species abundances that support our claim that $N_{li} < K_{max}$ for $l=1, 2$.

\subsection{Reproduction}\label{sec:rep}

The ecological competition procedure described above determines the number of offspring of each species $N'_{1i}$ and $N'_{2i}$ in patch $i=1,\ldots,L^2$.  We  re-emphasize that although equations (\ref{N1}) and (\ref{N2}) produce  real values for the species abundances, we take the integer values of those abundances using the floor and ceiling functions with the care to
bias the competition in favor of species 1 (see subsection \ref{sec:comp}). Now we need to specify the phenotypes of the $N'_{1i}$ offspring of species 1 and  of the  $N'_{2i}$ offspring of species 2 in patch $i$. We assume that the  individuals that passed the viability selection sieve  reproduce asexually (see section \ref{sec:S7} of the Supplementary Material for a brief discussion of the effect of recombination)  and that the mother of each offspring is chosen randomly, with replacement,  among the survivors. We recall that the numbers of surviving individuals of species  1 and species 2 in patch $i$ are   $N_{1i}$ and $N_{2i}$, respectively, and that all survivors have the same probability of being chosen as mothers regardless of their fitness. 
Hence,  selection works at the level of survival (viability selection) only and not at the level of the  (genetic) differences of reproduction of survivors. In this way  the reproductive output  reflects the ecological dynamics and not the population composition at each generation.

The differences between mother and offspring are due solely to mutations in the $m_{r}$  nonplastic loci and in the $m_p$  plastic loci, which were implemented as follows. Each allele of the offspring can mutate with probability $u_r$ or $u_p$ depending on whether it is a nonplastic or a plastic allele. (Here we assume $u_r=u_p= 5/1000$, which gives a genome-wide mutation rate $U =0.2$.)  Once a mutation occurs, say at the plastic locus $k$, we add a normal random variable $\xi$  of mean zero and variance $1/100$ to the existing allelic value which then becomes $P_k + \xi$. This is Kimura's continuum-of-alleles model \citep{Kimura(1965)}. As usual, generations were discrete and nonoverlapping. During the development of an individual in a particular patch, we ignored any potential influence of parental phenotypes as, e.g., transgenerational plasticity \citep{Uller(2008)}.

In sum, the offspring generation of species $l=1,2$ in patch $i$  is obtained by selecting with replacement $N'_{li}$ individuals  from the $N_{li}$ survivors of the selection sieve.   The selected survivors are referred to as mothers. The phenotype differences between  offspring and mothers are the mutations in the    nonplastic and plastic loci. 

\subsection{Migration}\label{sec:mig}

Each individual within each patch can migrate to one of the eight surrounding patches (Moore neighborhood) with probability $p_{mig}$, and its destination is equally likely to be any of the eight patches. We were careful in keeping track migrant and non-migrant individuals that remained in their natal patch \citep{Hassell(1995)}. The flow of migrants between patches takes place simultaneously and it may result in some patches becoming empty or exceeding the carrying capacity.  After arriving at their destination patches, the migrants as well as the residents of those patches pass the viability selection sieve as described in subsection \ref{sec:sel}. Again, some patches may become empty at this stage.

\subsection{Metapopulation dynamics}\label{sec:meta}

As originally defined by \citet{Levins(1969)}, metapopulation dynamics consists of the extinction and colonization of local populations. Early models analogous to Levins' showed that two competitors could coexist globally even if coexistence was impossible in a single patch \citep{Levin(1974),Slatkin(1974),Nee(1992)}. However, here we 
do not impose a random extinction probability; only local adaptation and gene flow may interact to promote genetic variation and coexistence in the metapopulation. Actually, real metapopulations may contain local populations that never go extinct \citep{Schoener(1987)}. A cautionary note: \citet[p. 1232]{Kawecki(2004)} rightly pointed out that ``local adaptation is about genetic differentiation'', but warned to minimize non-genetic effects such as plasticity (thus considering it a ``nuisance parameter'') when studying local adaptation. However, at the metapopulation level studied here the total phenotypic variation for plastic species 2 is the result of the variation in the reaction norm intercepts (first term on the right side of equation (\ref{Z}) ) and slopes (second term on the right side of equation (\ref{Z})), both of which have a genetic basis \citep{Scheiner(1993),Sommer(2020)}. Therefore, local adaptation is better understood as how close the mean phenotype matches the patch’s environmental optimum value.

In this paper we consider only  an expanding population scenario. More pointedly, 
the initial population was located on a randomly selected patch of the 2-dimensional grid at carrying capacity $K_{max}$  for the two  species, each at equal frequency (i.e., $K_{max}/2$ individuals from each species), and all  other patches were empty. We recall that the carrying capacity of the metapopulation is $K_{max} L^2$ individuals, so there is plenty of room for expansion from this initial setup.   For each species independently, there was an equilibration period of 2000 generations at the seed patch before the colonization of the empty patches started. We initialize the allelic values $R_k, ~k=1,\ldots,m_r$ and $P_k, ~k=1,\ldots,m_p$ in equation (\ref{Z}) for each individual as described before. In the equilibration period the two species evolve independently in the seed patch, i.e., there is no  interspecific (as well as intraspecific) competition since after viability selection   one guarantees that there will be exactly $K_{max}/ 2$ offspring of each species. Thus, equations (\ref{N1}) and (\ref{N2})  are  not used in the equilibration period. In sum, during equilibration viability selection  decreases the population of each species by eliminating the less fit individuals and  reproduction resets the population of the seed patch to its original size.

After the  equilibration period, the colonization of empty patches starts. The order of events is migration, phenotype determination, viability selection, ecological competition and reproduction. The sequence of these five events comprises one generation.  Note that  this  sequence of events  guarantees that a given individual undergoes the processes of selection, competition and reproduction within the same patch and that only their offspring have the possibility to migrate to neighboring patches.

A word is in order about  the ecology that our model describes. Consider a particular patch, say patch $i$, at a moment just after migration, so  its population  consists  of the offspring that stayed in patch $i$ and those that migrated to patch $i$. We recall that the  model assumes that only the offspring migrate.  To reach the reproductive age, these offspring must pass the selection sieve in patch $i$. Those who passed this sieve become adults: they are the survivors, which amount to $N_{1i}$  individuals of species 1 and  $N_{2i}$  individuals of species 2. The survivors  compete among themselves in patch $i$ to secure the resources to support their potential offspring. This competition is described  in a   coarse-grained manner  by equations (\ref{N1}) and (\ref{N2}), which output the number of offspring that each species can give rise to and sustain in patch $i$, viz., $N_{1i}'$ and $N_{2i}'$. At this point,  we could  argue  that the survivors produce an infinite number of offspring but only  $N_{1i}'+ N_{2i}'$  of them survive because of  resources  limitation. Alternatively, we could argue that the survivors produce exactly the number of offspring determined by equations (\ref{N1}) and (\ref{N2}). This last interpretation is the one adopted in  population dynamics  \citep{Godfray(1991),Pasztor(2006)}, from where we have borrowed those Ricker-like equations.   In any case, assuming one or the other scenario would not affect the outcomes. Next, the  mothers of the $N_{1i}'+ N_{2i}'$  offspring are chosen randomly with replacement among the survivors of each species. Behind the coarse-grained approach  is the  assumption that adults of the same species are indistinguishable with respect to their  competitive and reproductive abilities. Finally, each  offspring  decides if it will stay in patch $i$ or move to one the neighboring patches.

\subsection{Computer simulations}

Individual-based simulations were independently implemented in Fortran and in \citet{MATLAB:2020} algebra environment using tools supplied by the Statistics Toolbox. Simulation results were double-checked by different authors to avoid any potential error. The results presented here are based in the Fortran code because it has speed advantages over MATLAB. The variable parameters were: $L$ (landscape dimensionality), $\sigma_e^2$ (variance of the environmental values $E_i$), $\rho$  (environmental  correlation), $K_{max}$ (patch’s carrying capacity), $c$ (plasticity cost), and $p_{mig}$ (migration probability). For each set of conditions, we run 1000 
independent simulations (a random landscape for each simulation). The metapopulation dynamics was run for at most 2100 generations and 
we used the last 100 generations to average over  the quantities of interest (e.g., the abundance of each species) in the equilibrium regime.  If one of the two species fixed before that upper limit, we halted the dynamics. Otherwise, we considered that coexistence was achieved. However, in the study of  the single-species metapopulation dynamics all runs reached the upper limit of 2100 generations. In the Supplementary Material we present many instances of the time evolution of both species (e.g., figure \ref{fig:S12}), which show that  the  running time of  $2000$ generations is sufficient to guarantee that the metapopulation dynamics reaches the equilibrium regime.

Since the  quantities used to characterize coexistence at equilibrium are averages over patches (typically $L^2=400$), last generations of the colonization stage  (100) and runs  (typically 500 runs result in coexistence), the number of samples used to estimate their mean values is very large, resulting  in error bars  smaller than the sizes of the symbols used in the figures.
However, in order to assess the variability of the equilibrium variables described next,  in  section \ref{sec:S8} of the Supplementary Material  we offer a variety of scatter plots for selected values of the model parameters.

\subsection{Equilibrium variables}

In this paper  we aim at  the characterization of the metapopulation in the  equilibrium  regime,  defined as the regime between generations $t=2000$ and $t=2100$. In  the Supplementary Material  we present results for the  time evolution of both species in a variety of scenarios, but here we consider  the equilibrium regime only. 
We focus on the following  four variables.
\begin{itemize}
\item The   mean relative abundances of each species, which we denote by  $\langle \langle n_l \rangle \rangle$ for $l=1,2$.  These are the natural variables to describe the metapopulation at equilibrium. For $p_{mig} > 0$, $\langle \langle n_l \rangle \rangle $ is measured by averaging  the  number of individuals of species $l$ (just after viability selection)  over all patches during  the last 100 generations of the 2100 generations runs. The result is then divided by the number of patches ($L^2$)  and by the patch's carrying capacity ($K_{max}$).  The same procedure applies for  $p_{mig} = 0$, except that we must omit  the division by the number of patches since the population cannot leave the seed patch in this case. The final result is then averaged over the independent runs. We represent all those averages by a double brackets notation. In the Supplementary Material we introduce a single bracket notation to discuss results for single runs.  We note that all patches are considered in the computation of the mean relative abundances, regardless of whether they are empty, contain a single species or contain both species. 
\item The fraction of runs $\Gamma$ for which there is coexistence at generation $2100$. This quantity essentially measures the fraction of runs for which species 2 is not extinct, since even for rugged environments and large migration probabilities, species 1 is rarely extinct. For a run to result in coexistence it is enough that both species are present in the metapopulation at $t=2100$.  Hence  $\Gamma$ offers no information whatsoever on the nature of the coexistence, i.e., whether the two species coexist within a same patch or inhabit different patches. We stress that there is no averaging procedure involved in the evaluation of $\Gamma$.
\item The  mean fraction of  patches $\langle \langle \Pi \rangle \rangle$  that  carry  both species for the  runs that led to coexistence. For each run, an average  is calculated  over the last 100 generations of the run and then the result is averaged over runs. Hence the double brackets notation.
Clearly,   $\langle \langle \Pi \rangle \rangle$  offers valuable information on the nature of coexistence. Values of  $\langle \langle \Pi \rangle \rangle$  close to 1 indicate that most patches harbor both species, whereas values of  $\langle \langle \Pi \rangle \rangle$   close to 0 indicate that coexistence may take place in only a few patches due perhaps to their extreme environmental values that prevent their colonization by species 1. This latter type  coexistence, which we refer to as accidental coexistence, is not  interesting and  $\langle \langle \Pi \rangle \rangle$ allows us to distinguish it from the relevant case where coexistence happens within patches.
\end{itemize}
To facilitate the interpretation of these variables, in  section \ref{sec:S3} of the Supplementary Material we offer snapshots of the grid where the abundances the two species in each patch is shown in a color scale.

\section{Results}

\subsection{Single-species  metapopulation dynamics}

As the uncertain heterogeneous environment poses an adaptive  challenge to both species through the viability selection sieve, it is instructive to study  the  metapopulation dynamics separately for each species before considering the competition between them.  In addition, for the runs that  do not result in coexistence,  the equilibrium  of the metapopulation is described by the single-species dynamics.  As before, the initial single-species population was located on a randomly selected patch of the 2-dimensional grid at carrying capacity $K_{max}$ and there was an equilibration period of 2000 generations before the individuals  were allowed to migrate to the neighboring patches.

\subsubsection{Nonplastic species}

Let us consider first the dynamics of the nonplastic species 1, which is obtained by setting  $b=0$ in equation (\ref{Z}), $c=0$ in equation (\ref{W}), and $N_{2i}=0$ in equations (\ref{N1}) and (\ref{N2}).     

\begin{figure}
\includegraphics[width=0.48\textwidth]{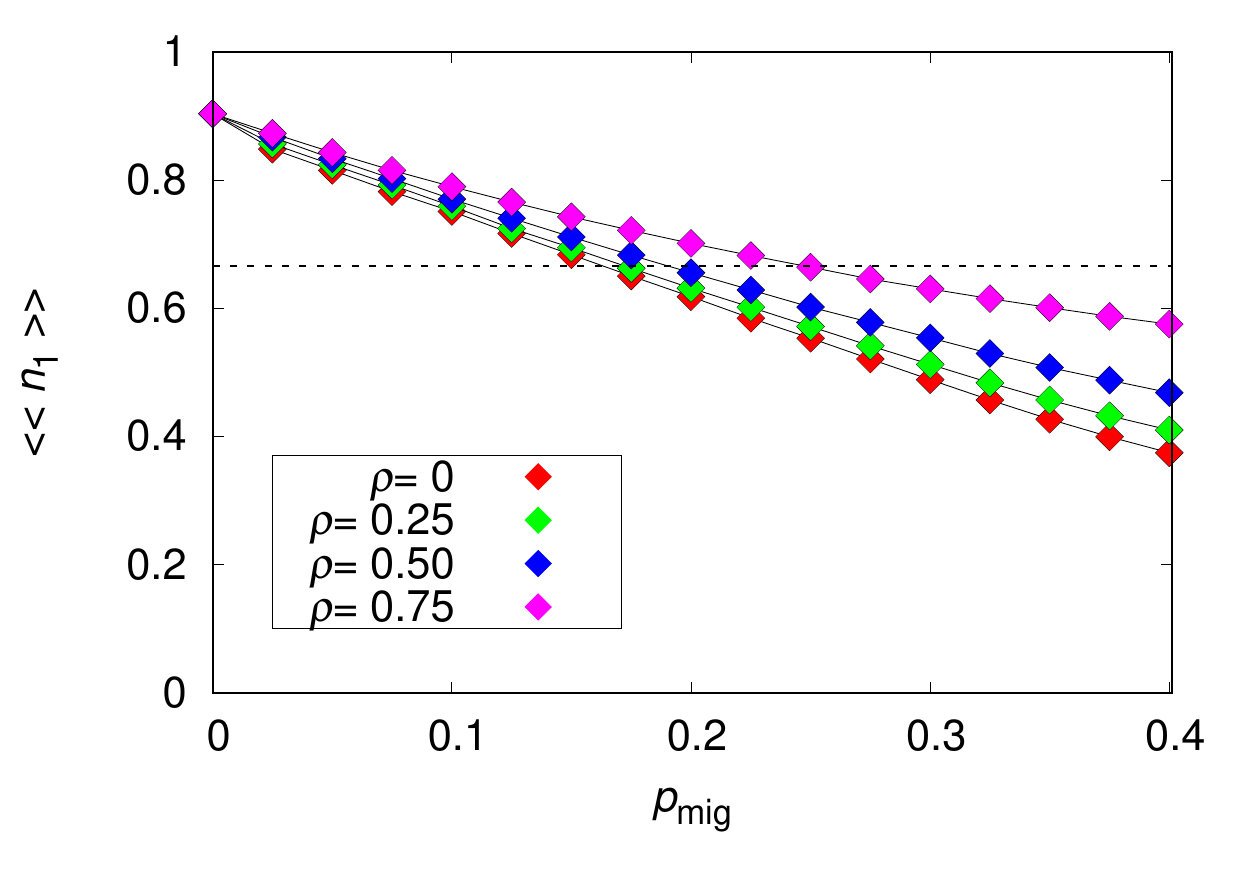}
\caption{Mean patch relative abundance of the nonplastic species  $\langle \langle n_1  \rangle \rangle$ for the single-species metapopulation dynamics as  function of the migration probability $p_{mig}$ for the  environmental correlation $\rho =0, 0.25, 0.5$ and $0.75$,  as indicated.  The other parameters are $L=20$, $K_{max} = 100$ and $\sigma_e^2 =2$. The lines connecting the symbols are guides to the eye. The dashed horizontal line is $\langle \langle n_1 \rangle  \rangle = 1/a_{21} = 2/3$.
}\label{fig:1}
\end{figure}

The effects of the migration probability and environmental correlation on  the mean  relative abundance of species  1 are summarized in figure \ref{fig:1}. There is a steady decrease of $\langle \langle n_1 \rangle  \rangle$ with increasing $p_{mig}$, which is clearly a consequence of the difficulty of the nonplastic species  to adapt to the heterogeneous patches. 
This happens in part because some lineage branches of a migrant individual (ancestor) have not enough time to adapt to their local environment since  the individuals  are forced to migrate to neighboring patches.  However, some lineage branches are likely to stay and to  adapt to their local environment.  But a fraction of the population of these well-adapted lineages are continually transferred  to patches where they are poorly adapted and  the individuals have little chances of surviving and hence of sending offspring back to the patch of their ancestors. In that sense, migration  produces an effective fitness independent culling of individuals of species 1. 
This problem is mitigated when the environment is highly correlated, i.e., the environmental values at neighboring patches are likely to be very similar,  and disappears altogether for a homogeneous environment ($\rho=1$). The finding that the nonplastic species reaches only  a fraction of the  maximal patch occupancy  is key to explaining coexistence in our model: the dashed horizontal line in figure \ref{fig:1} indicates the population density below which the nonplastic species cannot prevent the invasion of the plastic species, as  will be shown in subsection  \ref{sec:mec}.

  In the case the population is confined to the seed patch (i.e., for $p_{mig} =0$)  we find  $\langle \langle n_1 \rangle \rangle \approx 0.9$. The adaptation is not perfect  due to the  noise $\epsilon$ in equation (\ref{Z}) and to the nonzero genome-wide mutation probability $U$. (We note that since the genome of species 1  is determined by  the $m_r=20$ nonplastic alleles $R_k$ only, and since each allele has probability $u_r=5/1000$ of mutating we have $U=0.1$.) It is instructive to quantify the effect of $\epsilon$ on the survival probability of an individual of species 1 carrying the optimal phenotype in the seed patch $i$. In this case,  $Z_i^{opt}= E_i + \epsilon$ and so $W_i^{opt} = \mbox{e}^{-\epsilon^2/2}$. Recalling that $ \epsilon \sim N(0,\sigma_\epsilon)$, the expected survival probability of the optimal phenotype is 
  \begin{eqnarray}
  \mathbb{E}(W_i^{opt})  &  =  & \int_{-\infty}^\infty \frac{d \epsilon}{\sqrt{2 \pi \sigma_\epsilon^2}}  \exp \left [ - \frac{1}{2} ( 1 + \frac{1}{\sigma_\epsilon^2} ) \epsilon^2 \right ]  \nonumber \\
  & = & \frac{1}{\sqrt{1 +\sigma_\epsilon^2}}, 
  \end{eqnarray}
which yields $\mathbb{E}(W_i^{opt}) \approx 0.95$ for $\sigma_\epsilon^2=1/10$. 

The probability of metapopulation extinction was essentially zero for species 1, except for large values of the migration probability (i.e.,
$p_{mig} > 0.35$). For instance, for $p_{mig} = 0.4$ we find that only  $8$ out of  the $1000$  runs resulted in  extinction for $\rho=0$, whereas no extinction was observed for $\rho=0.75$. In section \ref{sec:S1} of the Supplementary Material we discuss the adaptation process of species 1 with emphasis on  the time dependence of the sum of the nonplastic allelic values $\sum_k R_k$ and to the mean fitness of the population.

\subsubsection{Plastic species}

We turn now to the dynamics of the plastic species 2, which is obtained by setting  $b=1$ in equation (\ref{Z}), and $N_{1i}=0$ in equations (\ref{N1}) and (\ref{N2}). The setup is the same as described in the study of the nonplastic species.

\begin{figure}
\includegraphics[width=0.48\textwidth]{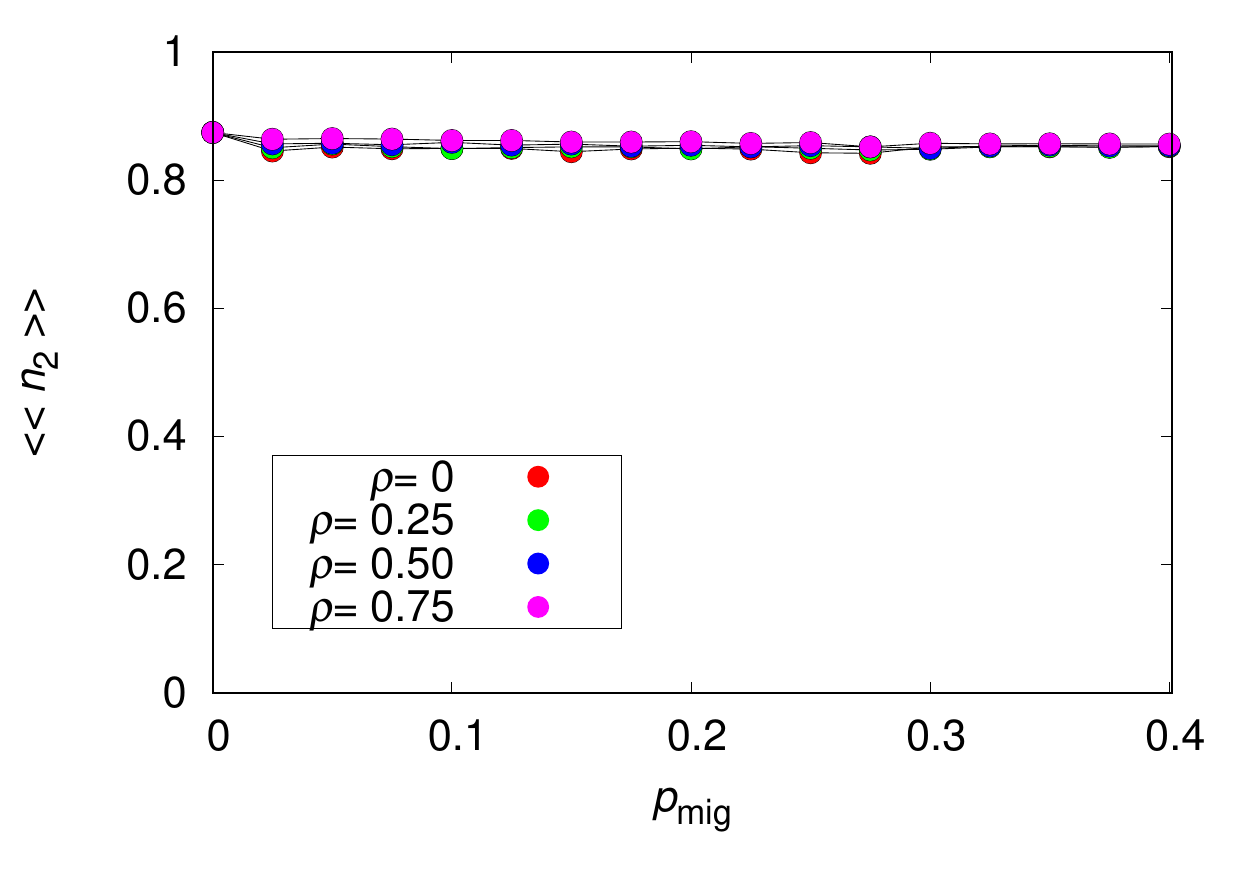}
\caption{Mean patch relative abundance of the plastic species  $\langle \langle n_2 \rangle \rangle$ for the single-species metapopulation dynamics as  function of the migration probability $p_{mig}$ for the  environmental correlation $\rho =0, 0.25, 0.5$ and $0.75$,  as indicated, and plasticity cost $c=0$.  The other parameters are $L=20$, $K_{max} = 100$ and $\sigma_e^2 =2$. The lines connecting the symbols are guides to the eye.
}\label{fig:2}
\end{figure}

Figure \ref{fig:2} shows that   the migration probability and the environmental correlation have no effect on the  relative abundance of the plastic species 2 in the case plasticity is costless ($c=0$). This unexciting finding is actually  important because it validates 
our modeling of the plastic species. In fact, a plastic species should thrive equally well in all patches (hence the unresponsiveness to changes on $p_{mig}$), regardless of the environment (hence the  unresponsiveness to $\rho$), as observed in figure \ref{fig:2}.  In addition,  these results  already illustrate the fitness advantage of the plastic species 2 over the nonplastic species 1, specially for large migration probability. Here we use the  relative abundance of the species after viability selection as a proxy for the fitness of the species.  Of course, adaptation of species 2 mainly happens via the contribution of the plastic components $P_k$  to the mean optimum phenotype and this is achieved by setting the nonplastic components $R_k$ as close to zero as possible.  In section \ref{sec:S2} of the Supplementary Material we offer  a study  of the adaptation process of species 2 with emphasis on  the time dependence of the sum of  both  nonplastic $\sum_k R_k$ and plastic $\sum_k P_k$ allelic values as well as  of the mean fitness of the population.
We note that for $p_{mig} =0$,  we find  $\langle \langle n_2 \rangle  \rangle \approx 0.87$, which indicates that species 2 is slightly less well adapted to the environment of the seed patch than species 1. The probable reason for this is that the genome-wide mutation probability for species 2 is twice that of species 1.

\begin{figure}
\includegraphics[width=0.48\textwidth]{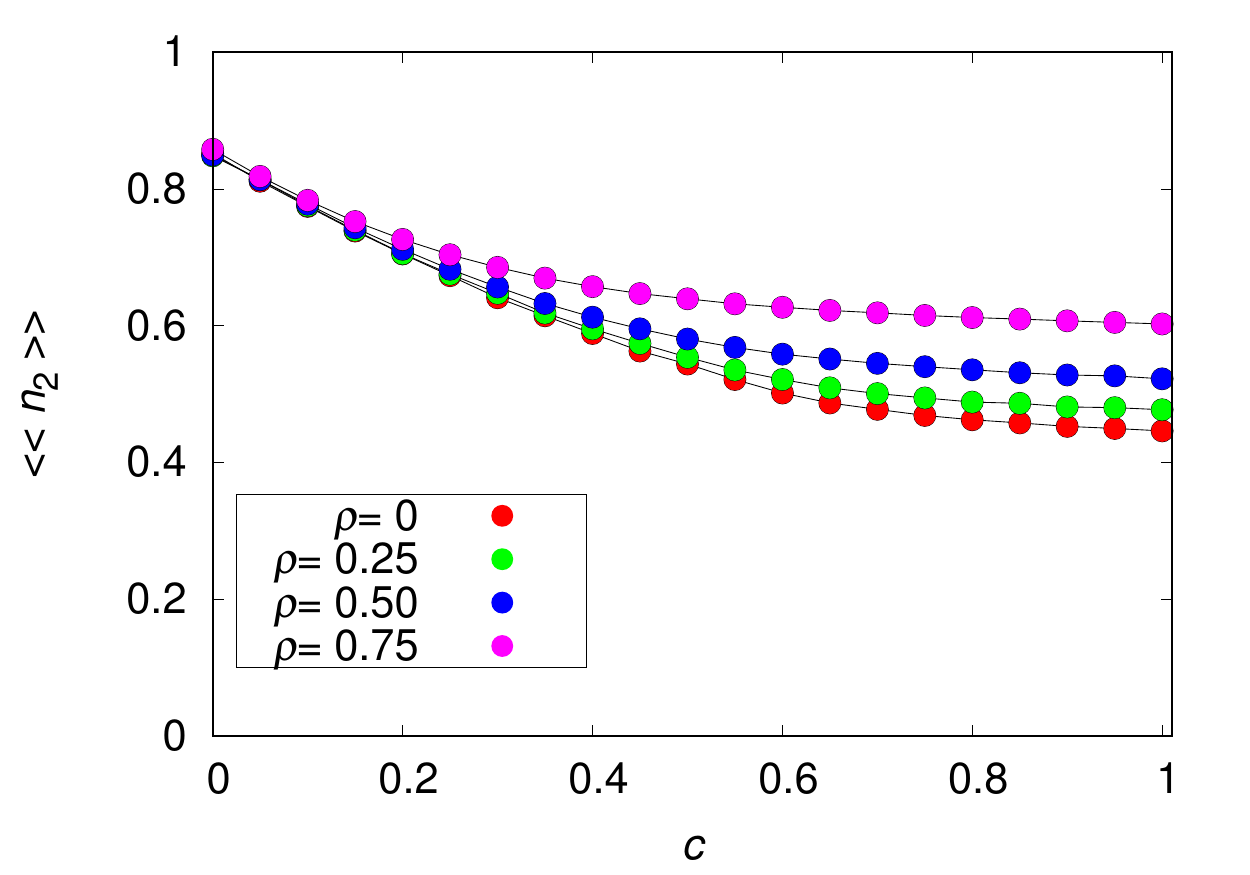}
\caption{Mean patch relative abundance of the plastic species  $\langle \langle n_2 \rangle \rangle$ for the single-species metapopulation dynamics as  function of the plasticity cost $c$  for the  environmental correlation $\rho =0, 0.25, 0.5$ and $0.75$,  as indicated, and migration probability $p_{mig}=0.3$.  The other parameters are $L=20$, $K_{max} = 100$ and $\sigma_e^2 =2$. The lines connecting the symbols are guides to the eye.
}\label{fig:3}
\end{figure}

  The  invariance of $\langle \langle n_2 \rangle \rangle$ to changes in $p_{mig}$ and $\rho$  does not hold when  there is a cost to plasticity (i.e., $c >0$), as shown in figure \ref{fig:3}. This is expected   because  introducing a cost to plasticity makes species 2 less plastic and hence more similar to species 1. In fact,  in order to maximize survival for large $c$, the  allelic values $P_k$ must tend to zero, thus reducing the influence of the penalty term in equation (\ref{W}).  Of course, setting the values of the plastic alleles to zero is equivalent to turning species 2 into a nonplastic species  (see figure \ref{fig:S6} of the Supplementary Material). For $p_{mig}=0$ and $c>0$ the optimal phenotype is $R_k = E_i, \forall k$ and $P_k =0, \forall k $  where $i$ the seed patch. This result can be obtained by the direct maximization of $W_i$, given in equation (\ref{W}), with respect to $R_k$ and $P_k$.  For  $p_{mig}>0$, there is a trade-off between $R_k$ and  $P_k$: for small $c$ it is advantageous to explore plasticity (see figures \ref{fig:1} and \ref{fig:2}), whereas for large $c$ it is advantageous to turn  off the plastic alleles. Although in the latter case species 2 becomes essentially a nonplastic species, we note that $\langle \langle n_2 \rangle \rangle$ is slightly below $\langle \langle n_1 \rangle \rangle$ because of the practical impossibility to keep $P_k$ close to zero due to the persistent perturbations produced by  the mutation process.

 We advance that, somewhat surprisingly, the plasticity cost will be crucial  to  the interpretation of the results of the interspecies competition  in our model.  In fact, as  already mentioned without evidence, if the relative abundance of species 1 in a given patch is less than some threshold value,  the resident species cannot prevent the invasion of (and the consequent coexistence with)  a competitively inferior species.  However,   we  will show next that control of the fitness of species 2  using the parameter $c$ (see figure \ref{fig:3})  indicates  that successful invasion  requires  the invading species  to be very well adapted to the patchy environment. 
 
 In time, we say that a species is competitively inferior if it cannot invade a resident population of the other species in a single-patch scenario (i.e., for $p_{mig}=0$). In that sense, competitive superiority or inferiority is completely determined by the competition matrix $a$ introduced in subsection \ref{sec:comp}. Also, by  fitness of a species we mean the  relative abundance of the species after viability selection, which is given by  averaging   the survival probability, equation  (\ref{W}), over individuals, patches, and generations at equilibrium.

\subsection{Two-species metapopulation dynamics}

We consider now the general setup where the two species  are   first let to reach equilibrium independently of each other in the seed patch and then are allowed to compete and migrate to the neighboring patches.  Of course, the focus here is on the  runs that led to coexistence since  the runs that do not lead to coexistence were already fully  characterized in the previous subsection.

\begin{figure}
 \subfigure{\includegraphics[width=0.48\textwidth]{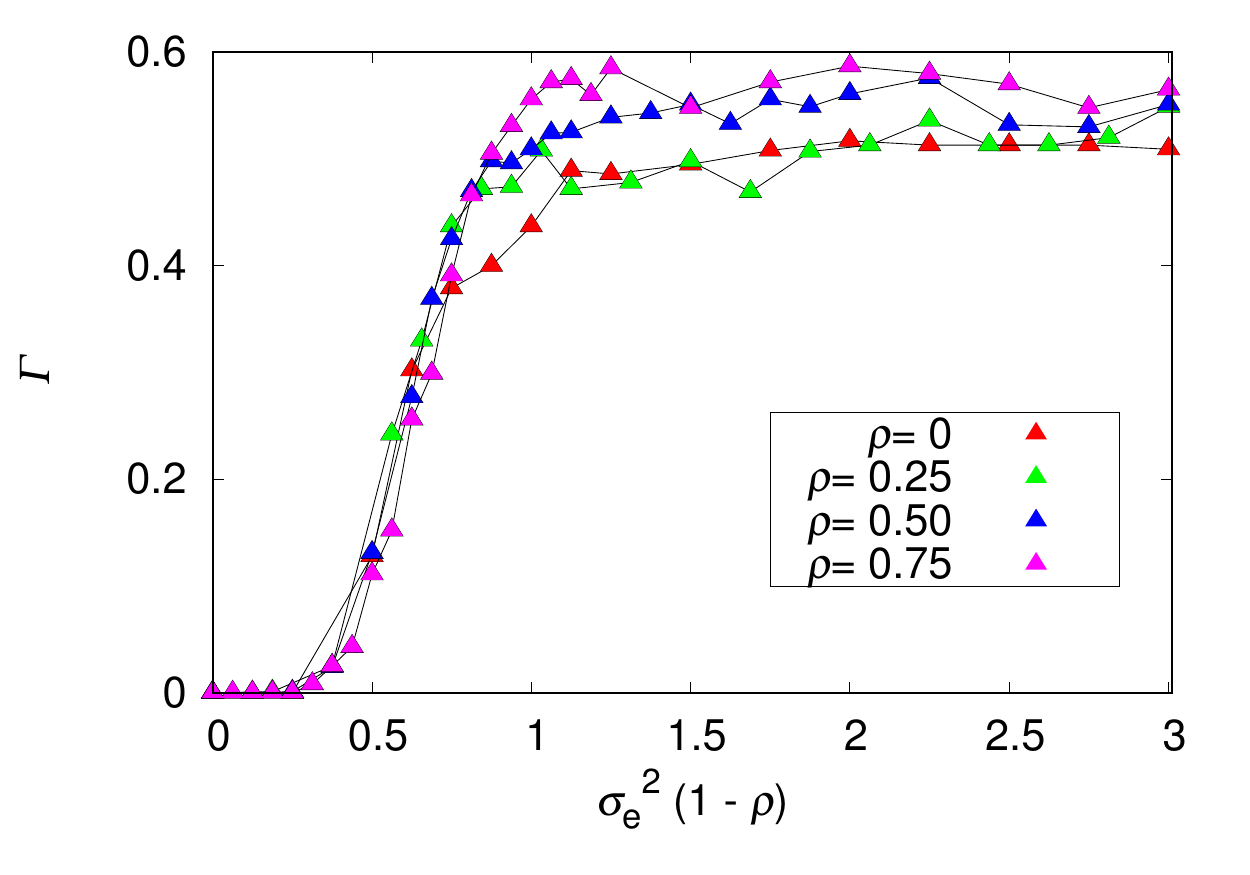}} 
 \subfigure{\includegraphics[width=0.48\textwidth]{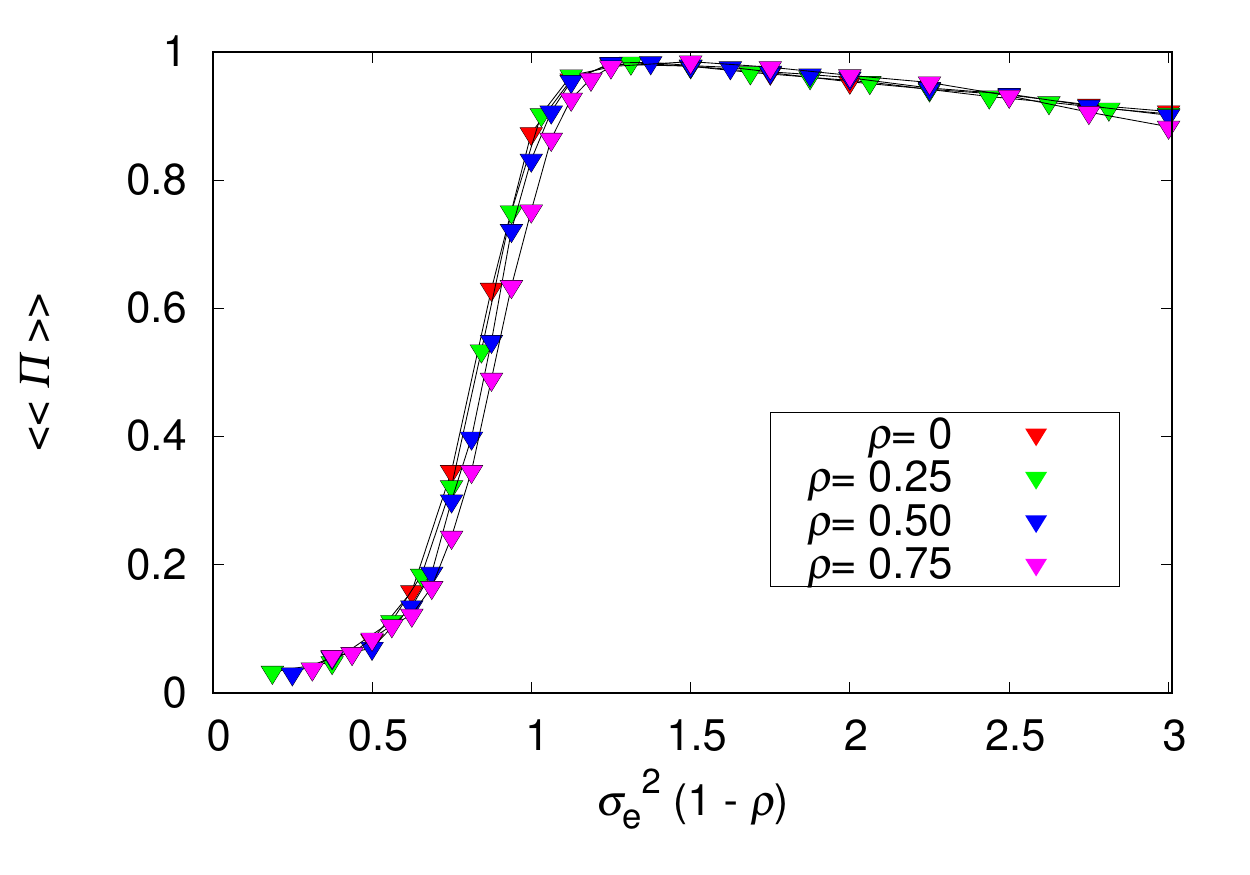}}
\caption{Influence of the variance of environmental values $\sigma_e^2$ and patch’s environmental correlation $\rho$ on species coexistence. \textbf{Upper Panel:} Fraction of runs that led to species coexistence. \textbf{Lower Panel:} Fraction of patches where there is species coexistence. The parameters are $L=20$, $K_{max} = 100$, $c=0$ and $p_{mig} =0.3$. The lines connecting the symbols are guides to the eye.
 }  
\label{fig:4}  
\end{figure}

Figure \ref{fig:4} summarizes the effects of the environment  on the probability that a run results in coexistence, which is measured by $\Gamma$ (upper panel of figure \ref{fig:4}),  and on the fraction of patches that harbor the two species, which is measured by $\langle \langle \Pi \rangle \rangle$ (lower panel of figure \ref{fig:4}). To a good approximation the effect of the environment is  represented by the single  variable $\sigma_e^2 (1-\rho)$, which means that $\rho$  can be absorbed in $\sigma_e^2$  and we can study the  uncorrelated landscape only without loss of generality.  In other words, increasing the correlation between patches is equivalent to decreasing the variance of environmental values in an uncorrelated landscape.
The important message from figure \ref{fig:4} is that the plastic species 2 is extinct in a quasi-homogeneous or smooth environment (i.e., for $\sigma_e^2 (1-\rho) \approx 0$). We note that in  this region  there are no data for $\langle \langle \Pi \rangle \rangle$ because no run resulted in coexistence.  

Interestingly, increase of  the  environment roughness  has only a limited effect on the probability of coexistence $\Gamma$, which quickly levels out and remains unaffected by further changes on $\sigma_e^2$ (upper panel of figure \ref{fig:4}). The probability that a patch exhibits coexistence $\langle \langle \Pi \rangle \rangle$  displays a more interesting behavior (lower panel of figure \ref{fig:4}). For smooth environments,  most patches are occupied by species 1 only, but as the environment roughness increases, those patches  begin to harbor both species. The slow decrease  of $\langle \langle \Pi \rangle \rangle$ we observe for large $\sigma_e^2$  is due to the appearance of patches occupied by species 2 only (see figures \ref{fig:S9}, \ref{fig:S10} and \ref{fig:S11} of the Supplementary Material).

\begin{figure}
 \subfigure{\includegraphics[width=0.48\textwidth]{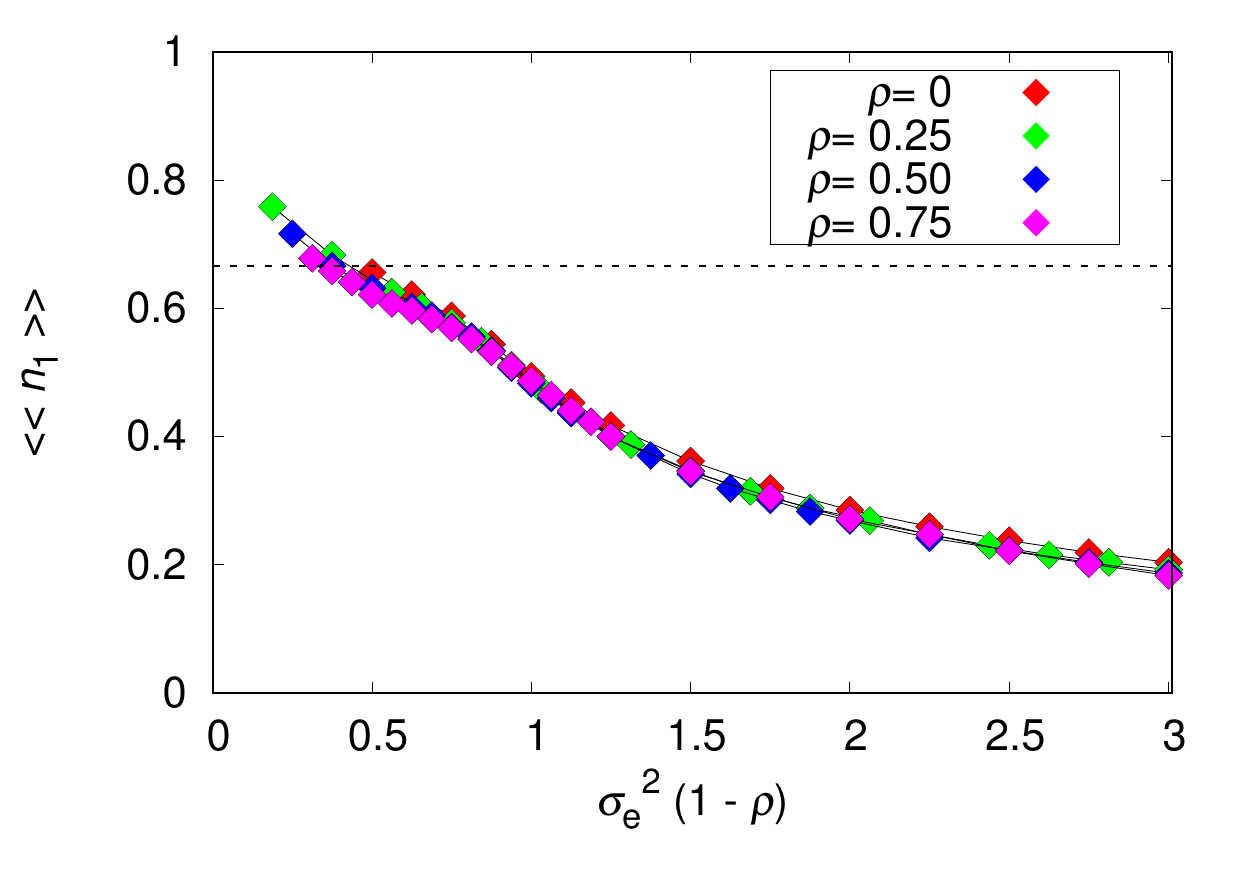}} 
 \subfigure{\includegraphics[width=0.48\textwidth]{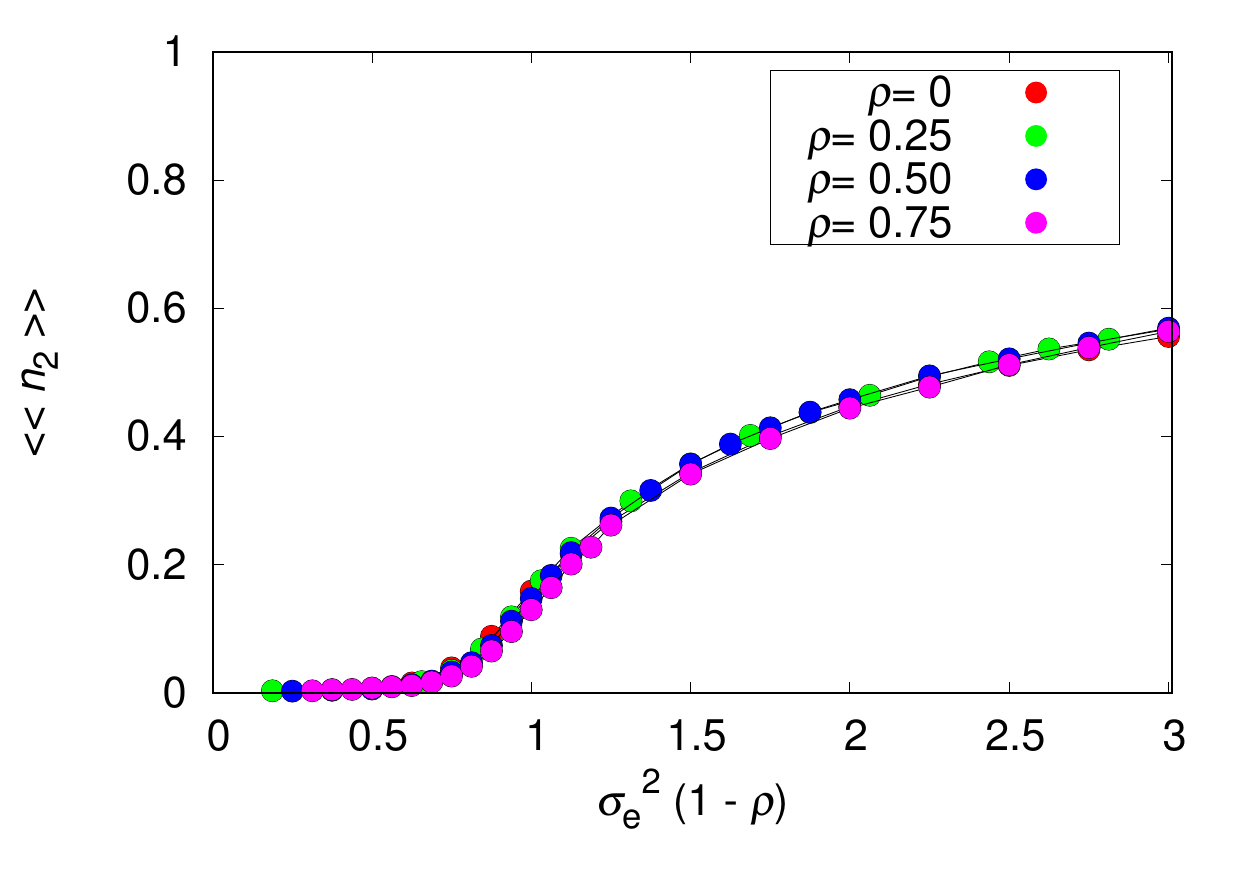}}
\caption{Influence of the variance of environmental values $\sigma_e^2$ and patch’s environmental correlation $\rho$ on the  mean patch relative abundances. \textbf{Upper Panel:} Nonplastic species 1. \textbf{Lower Panel:} Plastic species 2. The parameters are $L=20$, $K_{max} = 100$, $c=0$ and $p_{mig} =0.3$. The lines connecting the symbols are guides to the eye. The dashed horizontal line is $\langle \langle n_1 \rangle  \rangle = 1/a_{21} = 2/3$.
 }  
\label{fig:5}  
\end{figure}

Figure \ref{fig:5} shows the environmental effect on the relative abundances of both species. For  smooth environments,  species 2  is present in a few patches only (lower panel of figure \ref{fig:4}) and so its relative abundance $\langle \langle n_2 \rangle \rangle $ must necessarily be small, even if its density is high in the patches where it is present. In fact, the  relative abundances are informative only when $\langle \langle \Pi \rangle \rangle \approx 1$, in which case they represent the proportions of each species within a patch. The low density of species 1 for rugged environments is an indication that there may be patches occupied by species 2 only, which supports our explanation for the decreasing of  $\langle \langle \Pi \rangle \rangle$ for increasing $\sigma_e^2$. We recall that robust  species coexistence can happen only if  the density of species 1   is below the threshold $1/a_{21} = 2/3$, which is indicated by the dashed horizontal  line in the upper panel of figure {\ref{fig:5}. Otherwise, the observed coexistence is accidental, in the sense that species 2 occupies patches characterized by extreme environment values that are not suitable to species 1.

\begin{figure} 
 \subfigure{\includegraphics[width=0.48\textwidth]{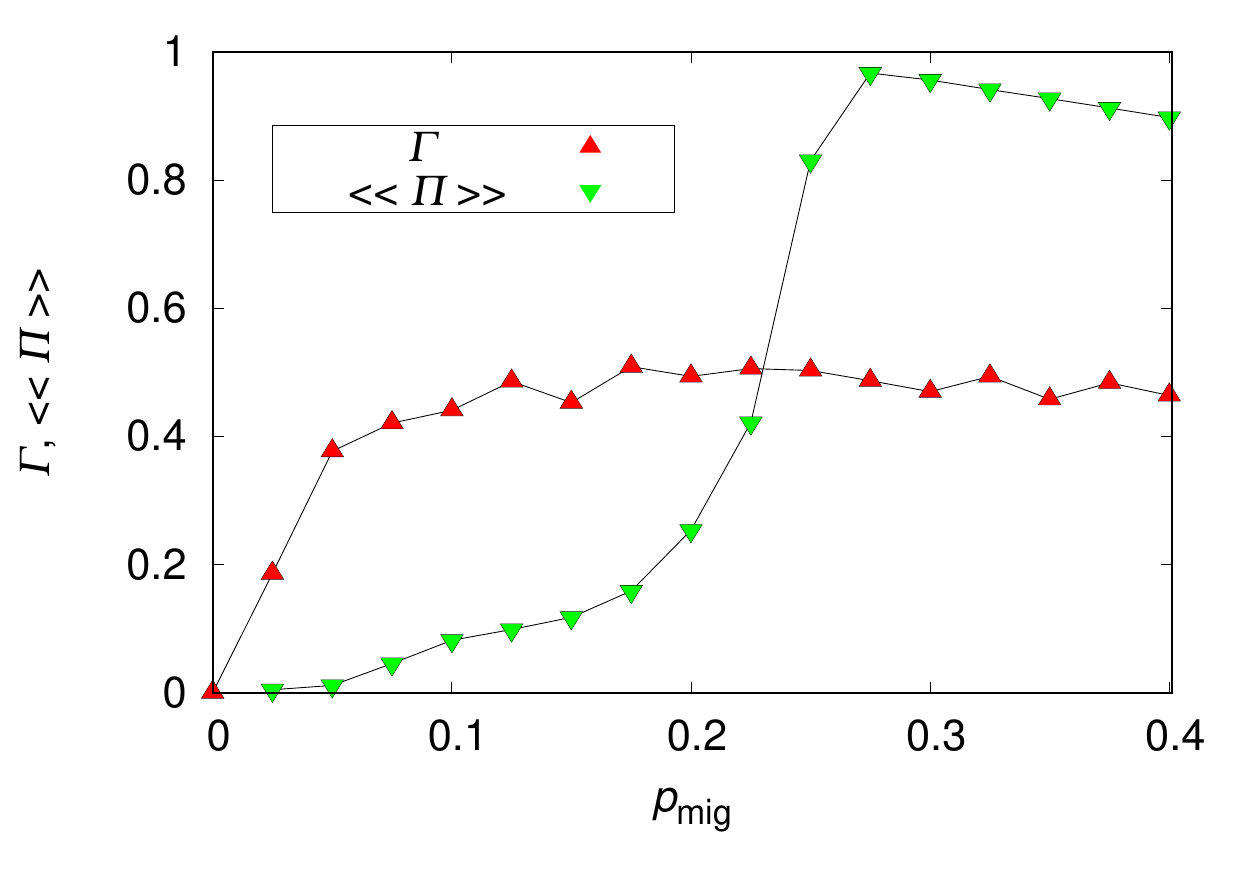}} 
 \subfigure{\includegraphics[width=0.48\textwidth]{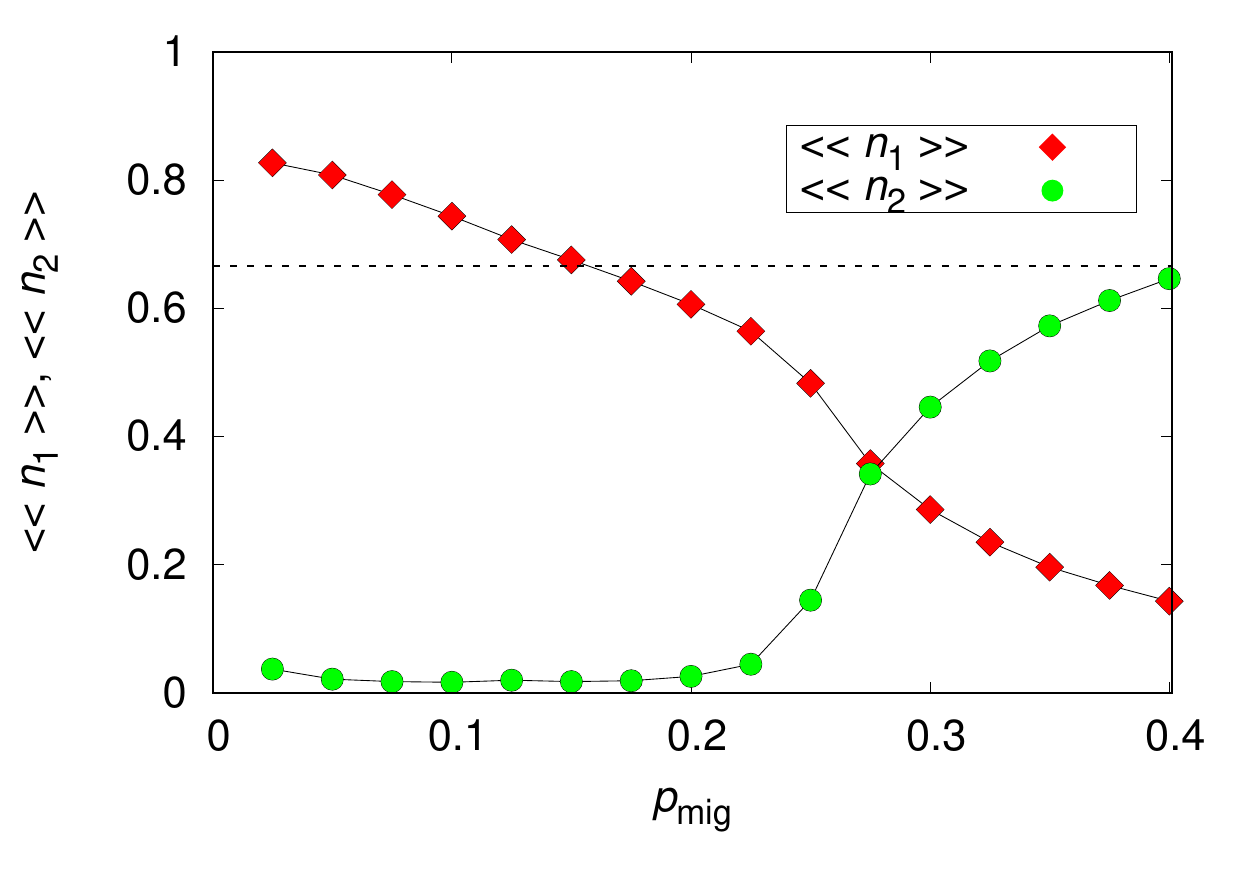}} 
\caption{Influence of the migration probability $p_{mig}$ on species coexistence for the uncorrelated environment. \textbf{Upper Panel:} Probability of coexistence in the metapopulation $\Gamma$ and probability of coexistence within a patch  $\langle \langle \Pi \rangle \rangle $. \textbf{Lower Panel:}  Mean patch relative abundances of the nonplastic species $\langle \langle  n_1 \rangle \rangle$  and  of  the plastic species $\langle \langle  n_2 \rangle \rangle$. The parameters are $L=20$, $K_{max} = 100$, $c=0$, $\sigma_e^2 = 2$  and $\rho =0$. The lines connecting the symbols are guides to the eye. The dashed horizontal line is $\langle \langle  n_1 \rangle \rangle = 1/a_{21} = 2/3$.
 }  
\label{fig:6}  
\end{figure}

Figure \ref{fig:6} shows the effect of migration on species coexistence for an uncorrelated landscape ($\rho=0$). Increasing the migration probability $p_{mig}$  has an effect  similar to increasing the environment ruggedness. As pointed out in our study of the single-species dynamics, migration affects  the adaptation of species 1 but has little to none influence
on  the adaptation of species 2. Hence the increase of the abundance of species 2 with increasing $p_{mig}$  shown in the figure is  a result of the effect of migration on the abundance of species 1 which in turn  affects species 2 in the ecological competition stage. 

\begin{figure} 
 \subfigure{\includegraphics[width=0.48\textwidth]{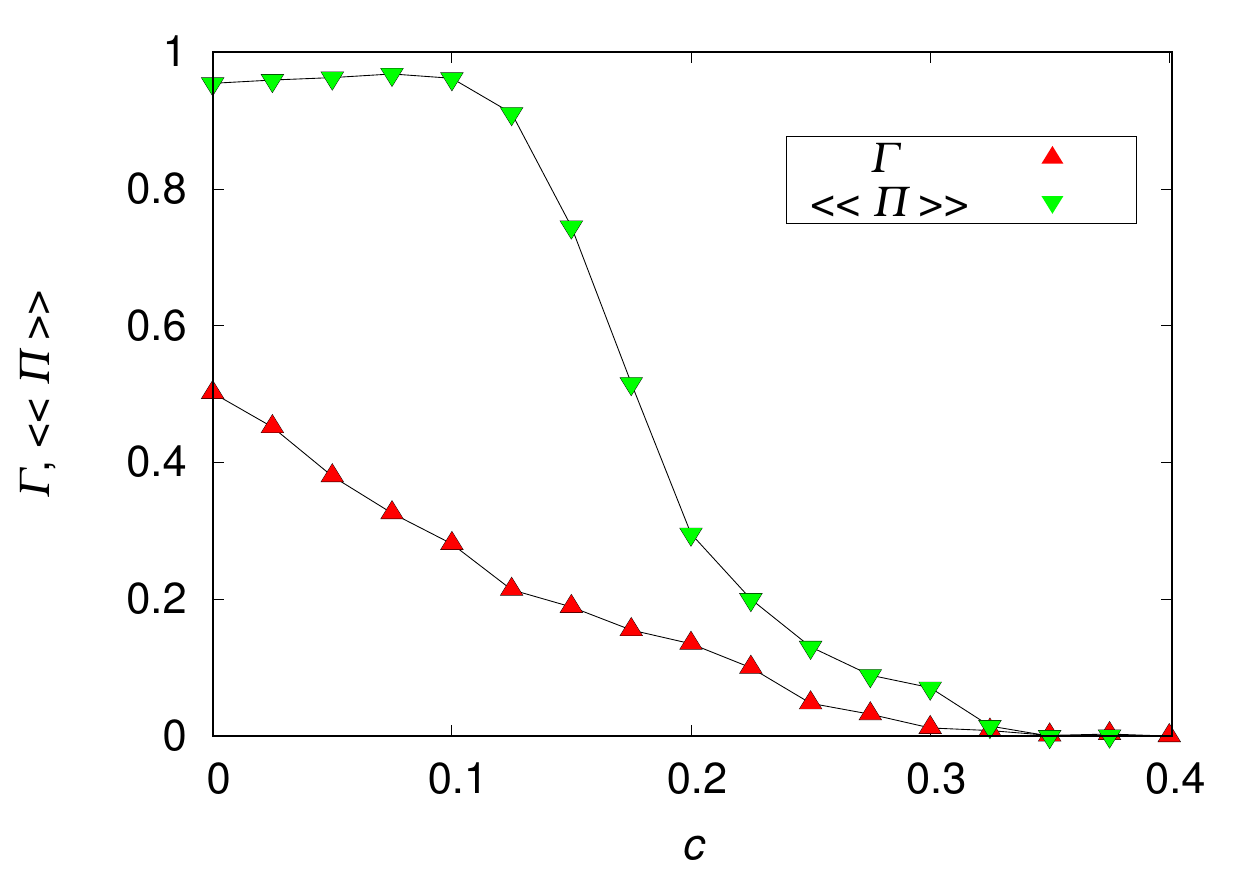}} 
 \subfigure{\includegraphics[width=0.48\textwidth]{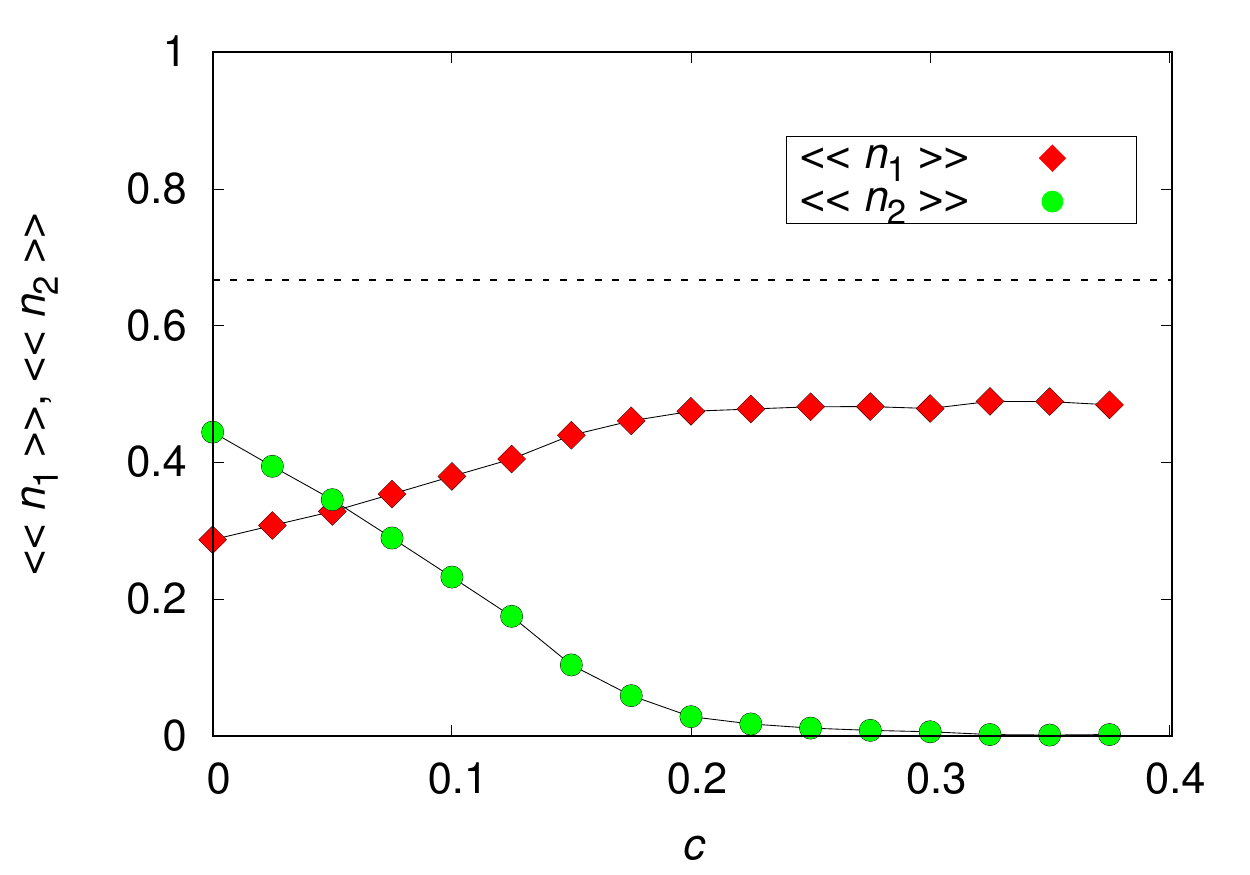}} 
\caption{Influence of the plasticity cost $c$ on species coexistence for the uncorrelated environment. \textbf{Upper Panel:} Probability of coexistence in the metapopulation $\Gamma$ and probability of coexistence within a patch $\langle \langle \Pi \rangle \rangle $. \textbf{Lower Panel:}  Mean patch relative abundances of the nonplastic species $\langle \langle  n_1 \rangle \rangle$ and  of  the plastic species $\langle \langle  n_2 \rangle \rangle $. The parameters are $L=20$, $K_{max} = 100$, $p_{mig}=0.3$, $\sigma_e^2 = 2$  and $\rho =0$. The lines connecting the symbols are guides to the eye. The dashed horizontal line is $\langle \langle  n_1 \rangle \rangle = 1/a_{21} = 2/3$. 
 }  
\label{fig:7}  
\end{figure}

The parameters $\sigma_e^2$, $\rho$ and $p_{mig}$ influence mainly the adaptation of the nonplastic species 1. The plasticity cost $c$, however,  affects the plastic species 2 only and figure \ref{fig:7} shows its effect on species coexistence. For the migration probability considered ($p_{mig} = 0.3$),  species 1 cannot prevent invasion (and, consequently, coexistence) but for large $c$ species 2 cannot take advantage of the maladaptation of species 1. We note that it is the presence of species 1 that drives species 2 to extinction, since species 2 alone can thrive  for large $c$ by turning off the plastic alleles (figure \ref{fig:3}). The data missing for $c=0.4$ is because none of the runs resulted in coexistence.

\subsubsection{Remarks on the simulation halting time, grid size, carrying capacity and recombination}\label{sec:param}

In our study, we assume that a  running time of  $2000$ generations is sufficient to proclaim that the metapopulation dynamics reached equilibrium and hence that coexistence was achieved.  Equilibrium population abundances are then evaluated by running the simulations for  additional 100 generations when the relevant quantities are stored for averaging purposes. In figure \ref{fig:S12} of the Supplementary Material we show the time dependence of the relative abundances of both species for typical runs that led to coexistence.  The results support our assumption that a halting time of $2000$ generations is adequate  to guarantee  the equilibration of the  metapopulation. Moreover,  the  dynamics reveals a most interesting feature of our model: the abundance of plastic species 2  increases  much faster than its rival's in the initial generations, so species 2 rapidly colonizes almost  the entire environment before it is partly or completely displaced by the nonplastic species 1 (see also figure \ref{fig:S9} of the Supplementary Material). 

Our analysis is restricted to a fixed grid size of linear length $L=20$ and patch carrying capacity $K_{max} = 100$, which results in a very large carrying capacity  for the  metapopulation (viz., $L^2 K_{max} = 40000$). Nevertheless, in the Supplementary Material we  present the  results for different choices of $L$ and $K_{max}$. In particular, we show that there is practically no difference between the results for $L=15$ and $L=20$ (figures  \ref{fig:S13} and \ref{fig:S14}), which indicates that our choice $L=20$  for the linear dimension of the grid gives a good approximation to the limit of an infinitely large grid.  The probability of coexistence $\Gamma$ and the fraction of patches that harbor the two species  $\langle \langle \Pi \rangle \rangle$ increase with patch’s carrying capacity $K_{max}$ (figure \ref{fig:S15}), but the mean relative abundances of both species  rapidly converge to their asymptotic values (figure \ref{fig:S16}), i.e., the values for 
$K_{max} \to \infty$. Since in the case of costless plasticity  it is the mean  relative abundance of species 1 that determines whether non-accidental coexistence can take place, these findings indicate that our choice of the grid size and patch carrying capacity probably describes very well the behavior of a  very large  population in a very large grid.

A limitation of our model  is the assumption of asexual reproduction. Nearly all invasive species are sexual and, in the case of plants, highly selfing or clonal which is not the same as being strictly asexual. However, the simulations of asexual populations are much faster and easier to implement and reproduce than for the  sexual populations, hence our option for that reproduction mode. In the Supplementary Material  we offer some results for sexual species (figure \ref{fig:S17}). Recombination favors the non-plastic species 1 in the competition with  the plastic species 2. In addition, for low and high mutation probabilities the sexual populations reach equilibrium faster than the asexual populations. But, as expected,  the main conclusion of the paper is not affected by the reproduction mode: there is a regime of accidental coexistence that happens for low migration probabilities that is due to the existence of  patches that have too extreme environments for the nonplastic species,  and a regime of robust coexistence that happens for high migration probabilities, where  the species coexist  within most patches. 

\subsubsection{Simple argument for coexistence}\label{sec:mec}

Although our extensive simulations point  rather unequivocally to the possibility of coexistence of the two species in a heterogeneous environment, here we offer analytical  evidence for that finding. The aim is  not only to dismiss  suspicion that the observed coexistence is an artifact of our simulations but to complement the simulation results. Since  the species at extinction risk  -- the plastic species  2 -- can thrive very well when alone in the patchy environment, the key to coexistence is  the ecological  competition stage (subsection \ref{sec:comp}), so let us  look at it more carefully.

First and foremost, we note that equations (\ref{N1}) and (\ref{N2}) are not recursion equations. In fact, the quantities $N_{1i}$ and $N_{2i}$ that appear in their right-hand sides  are the numbers of survivors of each species in patch $i$ after viability selection, whereas the quantities $N'_{1i}$ and $N'_{2i}$ that appear in their left-hand sides  are the numbers of offspring they bring forth. But only a fraction of these offspring will survive the selection sieve (and hence  become adults)  and this culling effect  is not included in equations (\ref{N1}) and (\ref{N2}).
Let us assume that the metapopulation  is at equilibrium (see, e.g., figure \ref{fig:S12}). The number of survivors of both species $ N_{1i}^{eq}$ and $ N_{2i}^{eq}$ at a given patch $i$ must satisfy the condition
\begin{equation}\label{cond2}
\frac{a_{21} N_{1i}^{eq} + a_{22} N_{2i}^{eq}}{K_{max}}   < 1,
\end{equation}
for the survival of species 2, and the condition
\begin{equation}\label{cond1}
\frac{a_{11} N_{1i}^{eq} + a_{12} N_{2i}^{eq}}{K_{max}}   < 1,
\end{equation}
for the survival of species 1. These are necessary conditions for survival of each species in patch $i$ as they ensure that the number of offspring will be greater than the number of  survivors.  (We recall that the number of  survivors in a given generation is only a fraction of the number of offspring in the previous generation). Inequality (\ref{cond2}) can be rewritten as 
\begin{equation}\label{cond2b}
 a_{22} \frac{N_{2i}^{eq}}{K_{max}}   < 1 - a_{21} \frac{ N_{1i}^{eq}}{K_{max}},
\end{equation}
which makes evident the impossibility of  an equilibrium scenario where species 2 is present  in patch $i$ and $N_{1i}^{eq}/K_{max} > 1/a_{21}$. Hence, increase of the entry $a_{21}$ decreases the chances of survival of species 2 and hence of coexistence.
This is the reason we draw a line  at  $\langle \langle n_1 \rangle \rangle  = 1/a_{21} = 2/3$ in the graphs for the relative abundance of species 1: the line delimits the regions where non-accidental coexistence is possible. Note that  a similar analysis for inequality (\ref{cond1}) indicates  that species 1 is  extinct in patch $i$ if  $N_{2i}^{eq}/K_{max} > 1/a_{12} = 2$, a condition that is never satisfied in our simulations since $N_{2i}$ and $N_{1i}$ are less than $K_{max}$ by construction. Therefore, non-accidental coexistence is a possible outcome of the metapopulation dynamics, provided species 1 is  locally maladapted in most patches, which is indeed the case for relatively large migration probabilities and environment variances.

However, the lower panel of figure \ref{fig:7} exhibits a scenario where inequality (\ref{cond2}) is satisfied and yet species 2 is extinct. Hence condition (\ref{cond2}) is necessary  for  survival of species 2, but it is not sufficient.  In fact, a necessary and sufficient condition   is that the  production of offspring compensates the population decrease due to viability selection.  For instance, assume that the number of offspring  is twice the number of survivors, so  condition (\ref{cond2}) is satisfied, but that viability selection reduces the population to $1/4$ of its size. Starting with 100 survivors, we get 200 offspring, then 50 survivors, then 100 offspring, then 25 survivors, and so on until extinction. This is the situation depicted in the lower panel of figure \ref{fig:7} for  high plasticity costs.   A similar argument can explain the possibility of extinction of species 1  as well, despite the fact that  inequality (\ref{cond1}) is always satisfied.
Unfortunately, we cannot express this necessary and sufficient condition in a simple mathematical  formula because it involves the viability selection process and hence information on the individuals' phenotypes.  This point highlights that to take advantage of the unfitness of species 1 in the rugged environment, species 2 must be well-adapted to it, hence the relevance  of plasticity in our model.

Finally, we note that increase of  the parameter $r$ that governs the growth of  both species in equations  (\ref{N1}) and (\ref{N2}) can be disastrous to species 2. The reason is that, other things being equal,  $N_{1i}^{eq}$ increases with $r$ so that the condition  $N_{1i}^{eq}/K_{max} > 1/a_{21}$ that prevents the growth of species 2 can be more easily fulfilled. Of course, the  increase in the number of offspring of species 1 resulting from increasing $r$ can be compensated by increasing the environment variance $\sigma_e^2$, which reduces their chances of  survival.

\section{Discussion}

Our results challenge predictions from classical ecological theory by showing that a competitively superior species cannot always displace an inferior competitor in absence of niche differentiation and in a standard scenario of density- and frequency-independent viability selection.  This conclusion obviously assumes that the ecologically inferior species 2 displays high levels of adaptive phenotypic plasticity (``any plasticity that allows individuals to have higher fitness in the new environment than it would were it not plastic''; \citet[p.  396]{Ghalambor(2007)}) and that plasticity can evolve quickly, which means that it harbors abundant genetic variation.

 It has been conjectured that greater plasticity is a key mechanism underlying the success of invasive species \citep{Baker(1965)}, an idea that has some positive support in plants \citep{Davidson(2011)} although there are counterexamples \citep{Godoy(2011)}. These inconsistent findings could be explained because adaptive plasticity might be a transient state during the invasion of new environments and thereafter disappear due to selection on the intersection of the reaction norm and eventual reduction of the slope, a process often referred to as ``genetic assimilation''\citep{Lande(2009),Lande(2015)}. The problem with this scenario is that for genetic assimilation to happen a very long time seems to be required if plasticity costs are low \citep{Scheiner(2021)}. In our case, with noncostly phenotypic plasticity the adaptation of species 2 during the colonization stage happens through phenotypic plasticity, i.e., the contribution of the rigid loci in equation (\ref{Z}) to the adapted phenotype is negligible. In the Supplementary Material we  test this scenario by assuming that no further migration takes place after the colonization period and find that  genetic evolution remained largely irrelevant and no genetic assimilation was detected (see figure \ref{fig:S4} of the Supplementary Material). The reason is that the increase in average fitness was very slow to impose any selection on the intersection of the reaction norm (see figure \ref{fig:S5} of the Supplementary Material). However, a different result is observed with costly plasticity, where adaptation after the colonization period results in a strong selective pressure to silence the contribution of the plastic alleles; i.e., genetic assimilation (see figure \ref{fig:S6} of the Supplementary Material). In any case, whether or not an initial greater plasticity during the colonization process confers higher fitness is more contentious, though  \citet{Davidson(2011)} consider that it is plausible
 
Perhaps more controversial is the model's assumption that there is always plenty of genetic variation for plasticity so that populations will be able to adequately track the environment more closely. For instance, there seems to be limited ability for plasticity in thermal tolerance of ectotherms (over 90\% of all animals), which should rely on behavioral thermoregulation to avoid overheating risk \citep{Gunderson(2015)}; see also \citet{Sunday(2014), Arnold(2019)}. Although at spatial scales there is ample information on the genetic evolution of latitudinal clines for thermal-related traits (e.g., \citet{Hoffmann(2002),Sgro(2010),Wallace(2014),Castaneda(2015)}) widely distributed Drosophila species do not seem to show higher plasticity for thermal tolerance than those from restricted areas, being their distributions more closely linked to species-specific differences in thermal tolerance limits \citep{Overgaard(2011)}. However, these conclusions are problematic because they were based on inferences that might grossly underestimate the population consequences of thermal plasticity. Thus, \citet{Rezende(2020)} have uncovered a dramatic effect of thermal acclimation in Drosophila, with warm-acclimated flies being able to increase the window for reproduction by nearly one month from mid-spring to early summer when compared with their cold-acclimated counterparts. In summary, answers to the important question of why adaptive plasticity is not more commonly observed should consider the heritability of plasticity (generally lower than trait heritability; \citet{Scheiner(1993)}, the interactions among different traits (e.g., temperature-dependent trade-offs between fitness traits; \citet{Svensson(2020)}, the reliability of habitat-specific cues \citep{Tufto(2000)}, and ecological constraints \citep{Valladares(2007),Scheiner(2013),Snell-Rood(2021)}.

We have  focused in the situation where both species can simultaneously expand their range, which might not be an unrealistic scenario as range expansions have always occurred in the history of most species \citep{Excoffier(2009)}, and we are currently witnessing how species' range edges are expanding polewards in response to global warming \citep{Mason(2015)}. 
The important message here is that a successful invading species does not necessarily need to be ecologically superior to the resident one, it only needs to display some level of not much costly adaptive phenotypic plasticity under environmental conditions that usually vary across space and over time \citep{Yeh(2004),Richards(2006)}. Since empirical evidence indicates that costs of plasticity are infrequent or small \citep{Murren(2015)}, the former conclusion seems to be robust.

Finally, we can only speculate about the empirical relevance of our model. A recent empirical study reports that plasticity can enhance species coexistence by swiftly changing species' traits in response to a shift in the competitive environment, which was however assumed to be constant \citep{Hess(2022)}.  It might be interesting to comment on Amarasekare's work on parasitoid coexistence in a spatially structured host–multiparasitoid community \citep{Amarasekare(2000a),Amarasekare(2000b)}. The two parasitoid species she studied show asymmetric competition in the laboratory with one species being potentially capable of displacing the other, but both species can coexist in some metapopulations even though the two parasitoids have overlapping niches and compete for a shared limiting resource. She tested whether coexistence could happen via a trade-off between competitive ability and a higher dispersal of the inferior competitor, which could find patches where the superior competitor was absent. Her data showed that this was not the case, but pointed to local interactions as, e.g., density-dependent processes that could ameliorate antagonistic interactions in her study system. However, she did not estimate whether the fitness of egg parasitoids in the patches was differentially altered in the two species depending on the environmental conditions (e.g., temperature) at which individuals developed \citep{Boivin(2010)}. In other words, could phenotypic plasticity have played any role in explaining Amarasekare's findings? We do not know, but perhaps this is a hypothesis that has some merit.

\bigskip

\acknowledgments
J. F. Fontanari was supported in part by Grant No.\ 2020/03041-3,  Fun\-da\-\c{c}\~ao de Amparo \`a Pesquisa do Estado de S\~ao Paulo (FAPESP) and by Grant No.\  305620/2021-5, Conselho Nacional de Desenvolvimento 
Cient\'{\i}\-fi\-co e Tecnol\'ogico  (CNPq). M. Matos is financed through the cE3c Unit FCT funding project UIDB/BIA/00329/2020. M. Santos is funded by grants PID2021-127107NB-I00 from Ministerio de Ciencia e Innovaci\'on (Spain), 2021 SGR 00526 from Generalitat de Catalunya, and the Distinguished Guest Scientists Fellowship Programme of the Hungarian Academy of Sciences (https://mta.hu)
This work benefited from discussions and insightful comments 
from Erol Ak\c{c}ay, Benjamin M. Bolker, Harold P. de Vladar,  E\"ors Szathm\'ary, Sam Scheiner, and an anonymous reviewer.

\bibliographystyle{Frontiers-Harvard} 


\clearpage

\newpage

\onecolumngrid
\begin{center}
  \textbf{\large SUPPLEMENTARY MATERIAL}\\[.2cm]
\end{center}


\setcounter{equation}{0}
\setcounter{figure}{0}
\setcounter{table}{0}
\setcounter{page}{1}
\setcounter{section}{0}
\renewcommand{\theequation}{S\arabic{equation}}
\renewcommand{\thefigure}{S\arabic{figure}}
\renewcommand{\thesection}{S\arabic{section}}
\renewcommand{\citenumfont}[1]{S#1}

\section{Adaptation of the nonplastic species}\label{sec:S1}

Here we offer a summary of the adaptation process of the nonplastic species in the case it is left alone to colonize the patchy environment (i.e., we set $b=c=N_{2i}=0$ for all $i$).  In particular, we  focus on the sum of the  nonplastic allelic values $S_r  \equiv \sum_{k=1}^{m_r} R_k$, which determines the  phenotype $Z_i$ of an individual  of species 1  at patch $i$  [equation (\ref{Z}) of the main text]. An important point to note is that neither $S_r$ nor $Z_i$ are  targets of selection: the target of selection is the  gaussian fitness $W_i$ [equation (\ref{W}) of the main text] that determines the probability of survival of the individuals. In that sense, the mean relative abundance $\langle n_1 \rangle$, viz., the number of surviving individuals  in a patch divided by the carrying capacity $K_{max}$  and averaged over all patches, presented in the main text is the relevant quantity to assess the adaptation of the species to  the patchy environment. Here we focus on the mean fitness as well, which has not been considered in the main text.  
  
As the optimum value of $S_r$ depends on the patch,  a possibility is to consider the ratio between $S_r$ and the optimum phenotype at that patch, i.e., $S_r /E_i $.  One may expect that an individual  well-adapted to its patch  is characterized by $ S_r/E_i   \approx 1$ and that the departure of this ratio  from  unity signals poor adaptation. This is not so. For instance, consider a patch for which $E_i = 0.001$ and an individual whose sum of the  nonplastic allelic values is $S_r = 0.1$. We recall that $E_i \sim N(0,\sigma_e^2)$ so patches characterized by small values of $E_i$ are  not rare. For this particular individual we have $S_r/E_i = 100 $ but the probability it survives the viability selection sieve is  very high, viz., $W_i  = \exp ( - 0.099^2/2 ) \approx 0.995$. 

\begin{figure}[H]
 \subfigure{\includegraphics[width=0.48\textwidth]{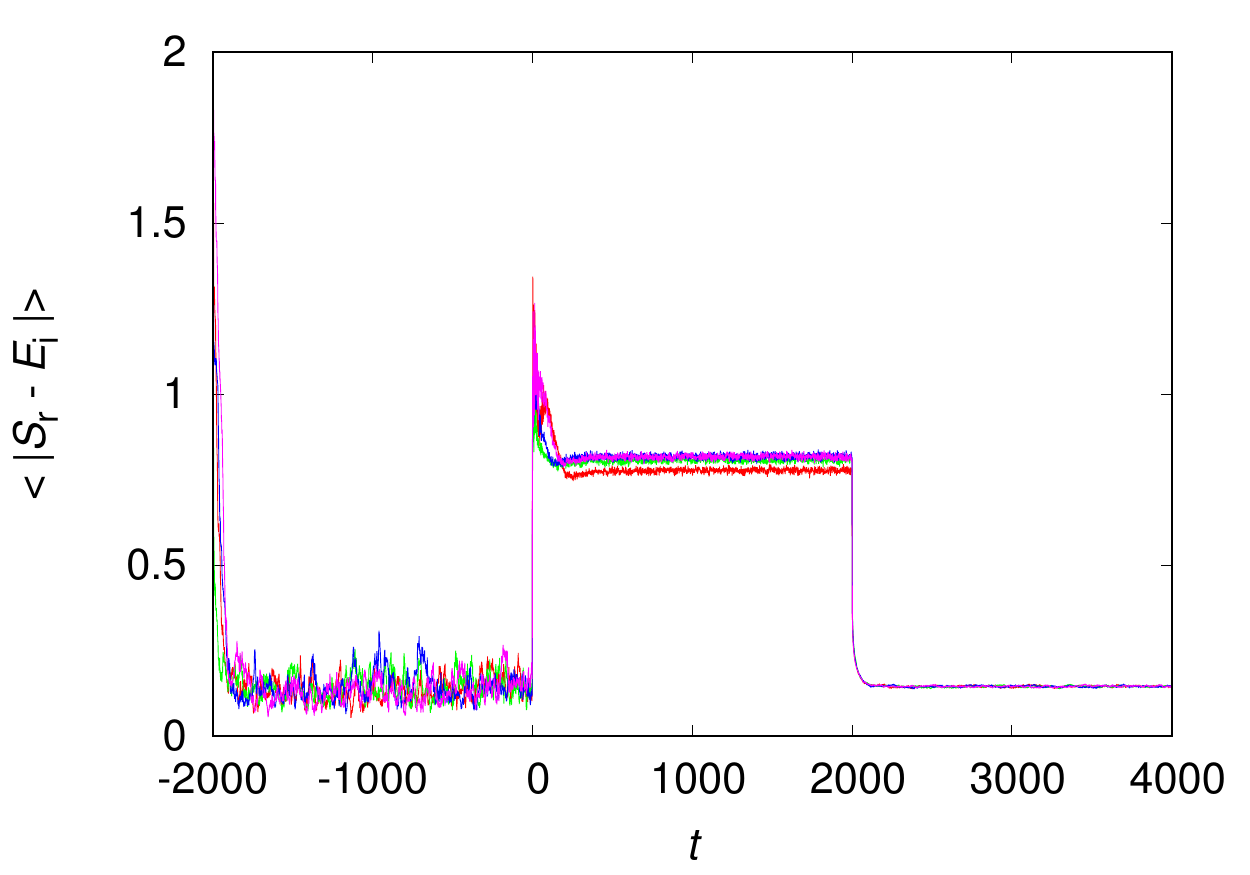}} 
 \subfigure{\includegraphics[width=0.48\textwidth]{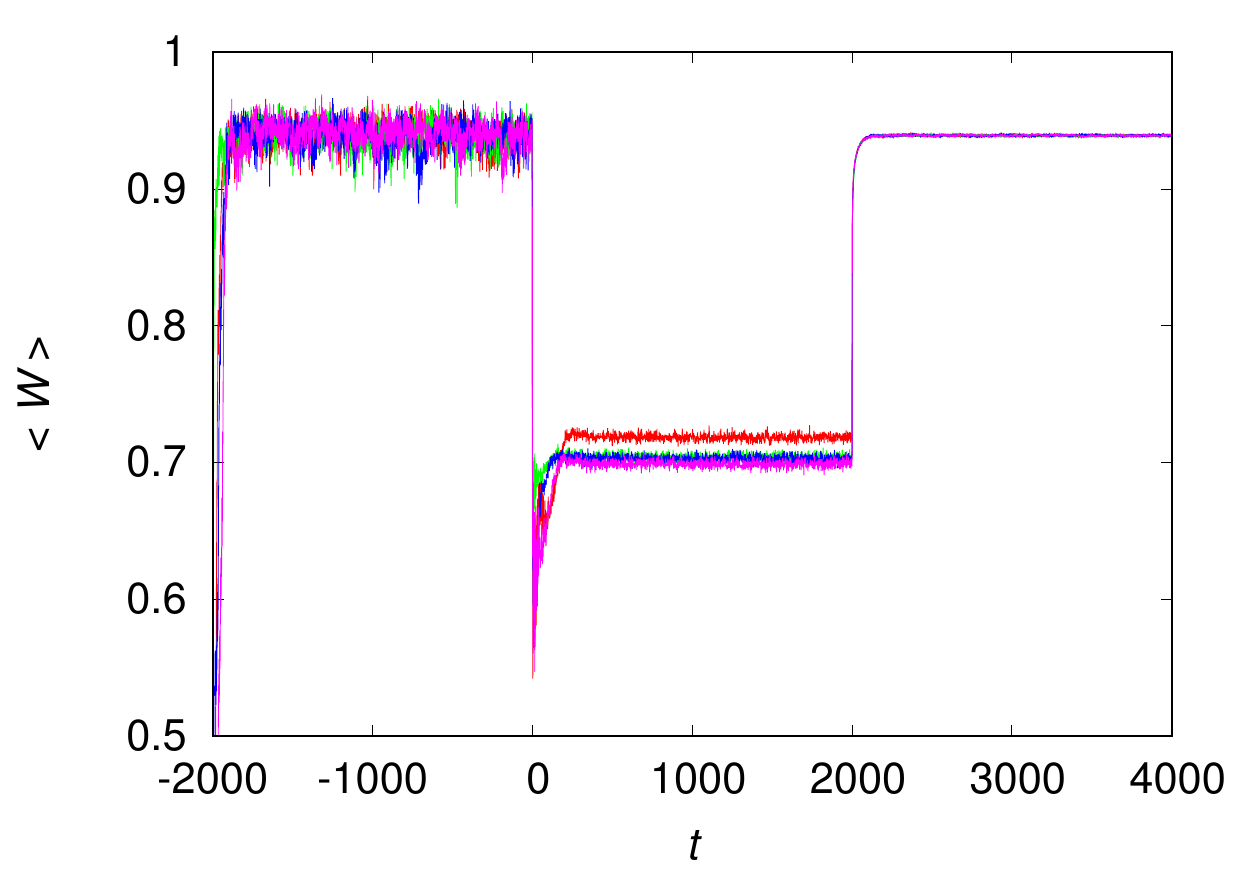}} 
\caption{Time evolution of the nonplastic species for four independent runs with $p_{mig} = 0.4$. \textbf{Left Panel:} Mean deviation of the sum of the nonplastic allele  values from the optimum phenotype $\langle \mid S_r - E_i \mid \rangle$. \textbf{Right Panel:}  Mean fitness of the nonplastic species $\langle W \rangle$.   The equilibration period in the seed patch occurs for $t \in [-2000,0)$ and the  colonization of the empty patches occurs for $t \in [0,2000)$. We set $p_{mig}=0$ at $t=2000$.  The parameters are $L=20$, $K_{max} = 100$,  $c=0$, $b=0$,  $\sigma_e^2 =2$ and $\rho = 0$. 
 }  
\label{fig:S1}  
\end{figure}

A  more suitable measure of the adaptation of an individual to its patch is  $\mid S_r - E_i \mid $: the closer  to zero this quantity is, the greater the odds that the individual survives the viability selection sieve.  
 Figure  \ref{fig:S1} shows the time evolution of  $\langle \mid S_r - E_i \mid \rangle $  as well as of the mean fitness $\langle W \rangle$ measured {\it before} the selection sieve for four independent runs. We recall that each run corresponds to a different environment. Here the single bracket notation stands for the average over the individuals in a patch and over all patches.  In addition, in figure  \ref{fig:S2} we show $\langle S_r/ E_i  \rangle $ to illustrate the inadequacy of this measure. 
 
\begin{figure}
 \includegraphics[width=0.48\textwidth]{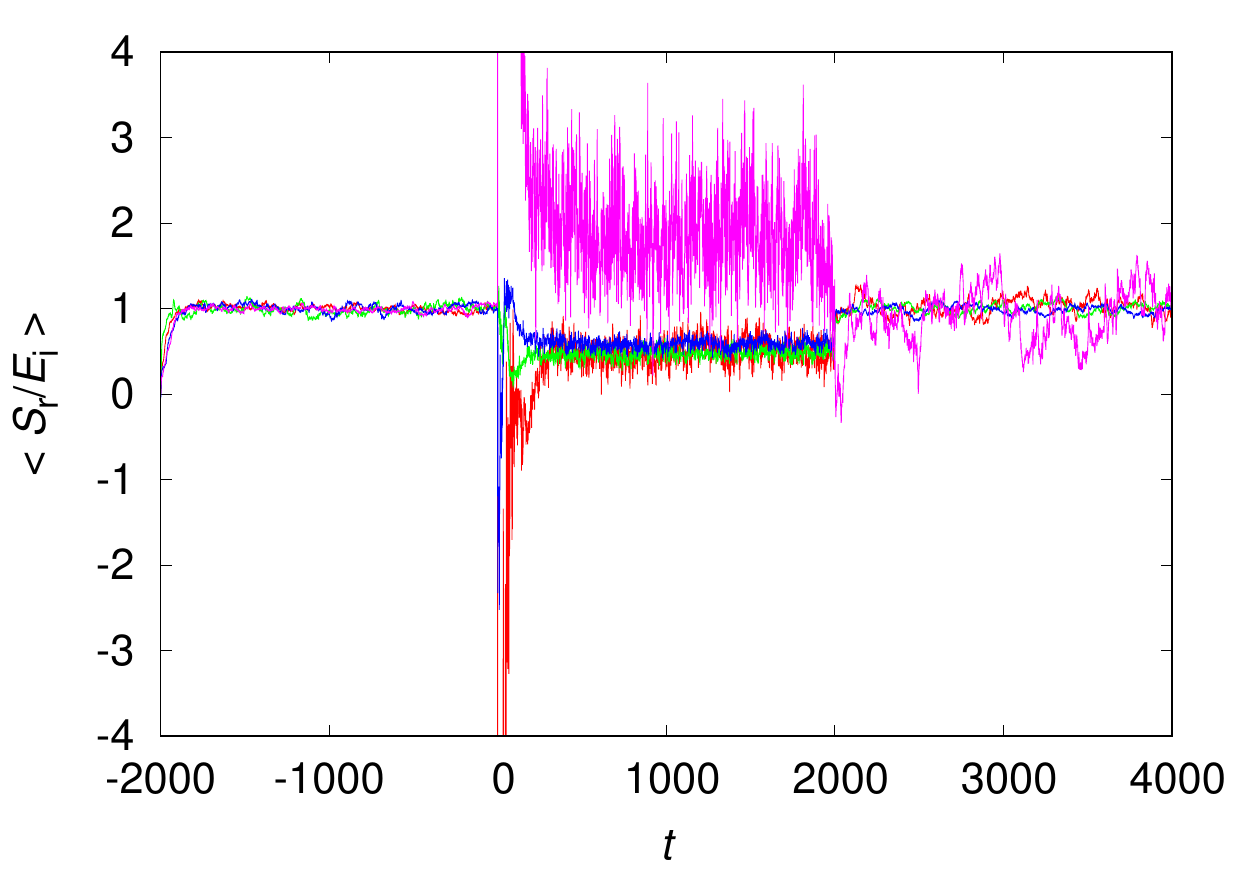} 
\caption{Time evolution of the ratio   $\langle S_r/ E_i  \rangle $   for the  four independent runs shown in figure \ref{fig:S1}. The y-scale was limited to the range $[-4,4]$ for the sake of visualization, but the actual range of   variation of this ratio  is $[-50,330]$. 
          The parameters are $L=20$, $K_{max} = 100$, $p_{mig} = 0.4$, $c=0$, $b=0$,  $\sigma_e^2 =2$ and $\rho = 0$. 
 }  
\label{fig:S2}  
\end{figure}

 To better illustrate the dynamics of adaptation, figures \ref{fig:S1} and \ref{fig:S2} exhibit the stage of equilibration that occurs for $t \in [-2000,0)$ in the seed patch  and  the colonization of the empty patches that occurs for $t \in [0, 2000)$ in the case that $p_{mig} > 0$. In addition, to prove that migration is the culprit for the poor adaptation  of species 1 in the heterogeneous environment, we set $p_{mig}=0$ for  $t \in [2000, 4000)$: the fact that the now isolated populations  quickly adapt to their local environments, as expressed by the increase of the mean fitness,  shows that adaptation is in fact taking place in our model. 
For a particular run, we note that the large fluctuations observed in the equilibration period in figure \ref{fig:S1}  are due to the small population size, which is on the order of $K_{max}=100$. When the population is allowed to colonize the entire grid the population is on the order of $K_{max}L^2=40000$ and the fluctuations  are negligible within a run.

\begin{figure}
 \subfigure{\includegraphics[width=0.48\textwidth]{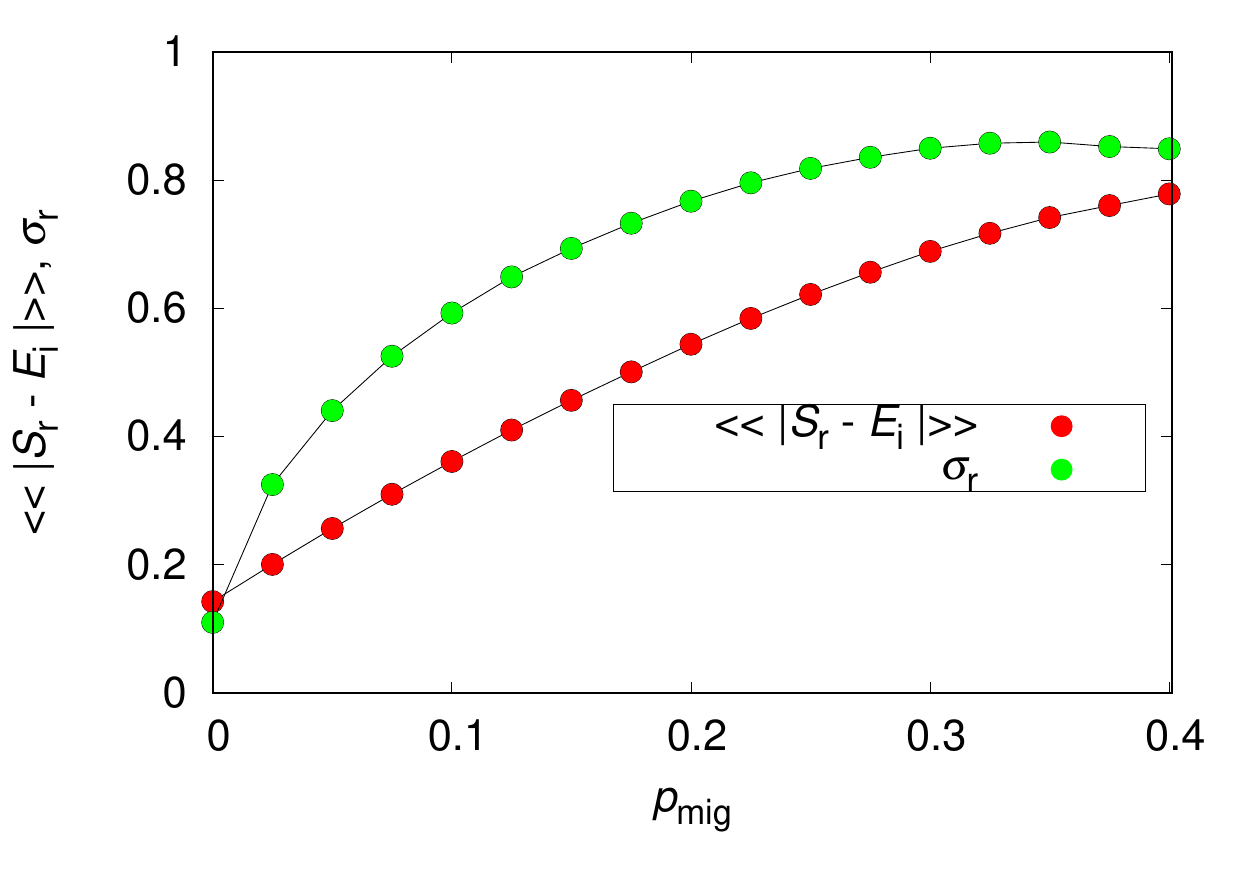}} 
 \subfigure{\includegraphics[width=0.48\textwidth]{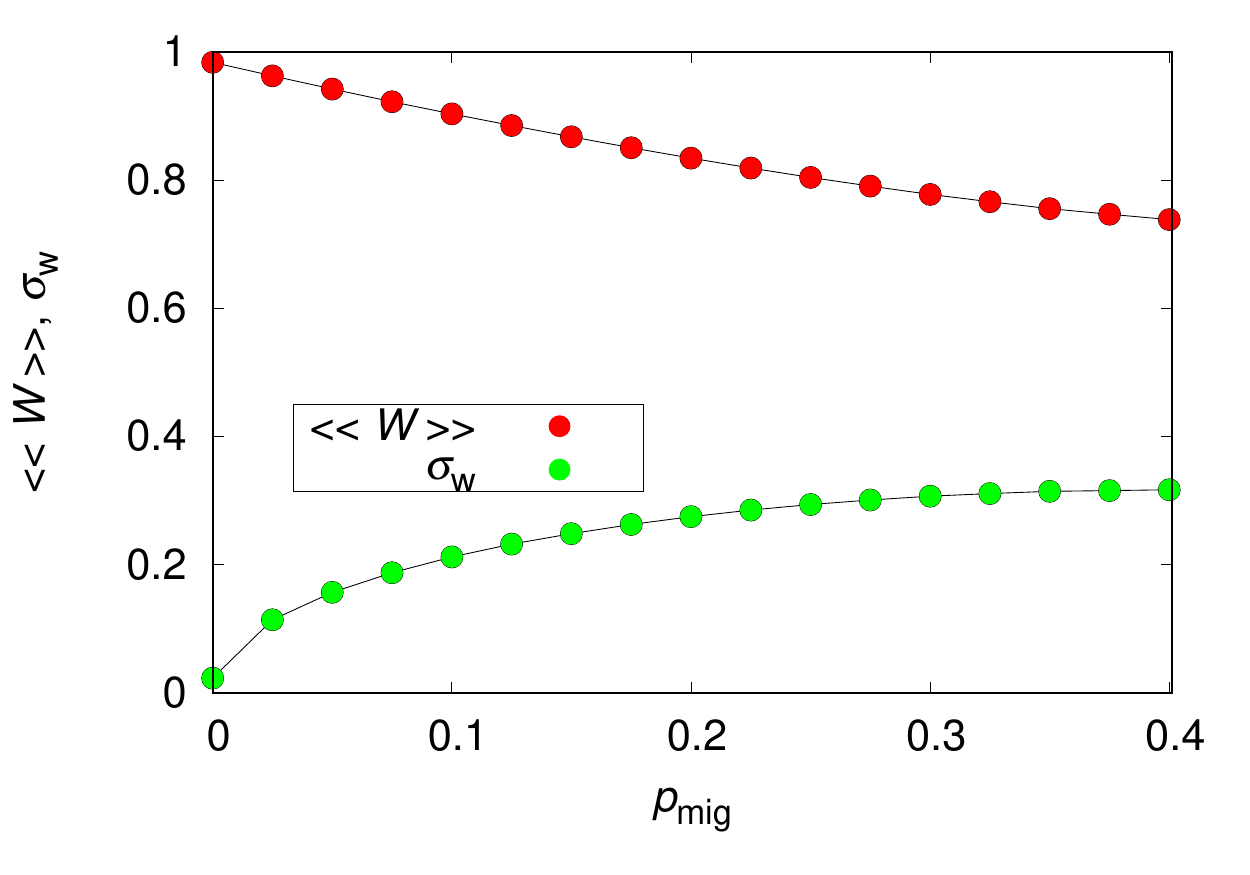}} 
\caption{Influence of the migration probability on the equilibrium of the nonplastic species.  \textbf{Left Panel:} Mean deviation of the sum of the nonplastic allele  values from the optimal phenotype $\langle \langle  \mid S_r - E_i \mid \rangle \rangle$ and the corresponding standard deviation $\sigma_r$. \textbf{Right Panel:} Mean fitness  of the nonplastic species $\langle \langle W \rangle \rangle$ and the corresponding standard deviation $\sigma_w$.   
          The parameters are $L=20$, $K_{max} = 100$, $c=0$, $b=0$,  $\sigma_e^2 =2$ and $\rho = 0$.  
 }  
\label{fig:S3}  
\end{figure}

In Figure \ref{fig:S3} we show the means and the  standard deviations (i.e., the square root of the  variance) of the random variables $\mid S_r - E_i \mid $ and $ W_i $ at equilibrium.
As in the main text, we record the values of these variables in the last 100 generations of the colonization stage for each one of the 1000 independent runs. 
These random variables are the values of the deviation of the sum of the nonplastic allele  values from the optimum phenotype and the fitness of a particular individual in the metapopulation. The statistical  ensemble  used to calculate the  moments of these variables   comprises the $10^5$  samples from all the individuals in the metapopulation, whose number is on the order of $K_{max} L^2 = 40000$. Hence the statistical ensemble  has about  $4 \times 10^9$ samples of the random variables $\mid S_r - E_i \mid $ and $ W_i $ from which we estimate the mean and the standard deviations exhibited in figure \ref{fig:S3}. In this figure, the  double brackets notation stands for the average over  individuals in each patch, over  patches, over the last 100 generations of the colonization phase and over runs.

It is instructive to use our estimate of the standard deviation  $\sigma_w$ to offer a rough approximation of  the size of the fluctuations observed in figure  \ref{fig:S1}. Let us consider the fitness of individual $j_i$ at patch $i$, which we will denote by $W_{j_i,i}$ for the present argument. The quantity shown in that figure (i.e.,  $\langle W  \rangle $), is the average of the fitness over individuals and patches, 
\begin{equation}\label{S1}
\langle W  \rangle \approx \frac{1}{L^2} \sum_{i=1}^{L^2} \frac{1}{K_{max}}\sum_{j_i=1}^{K_{max}} W_{j_i,i},
\end{equation}
where for the sake of simplicity we have assumed that all $L^2$ patches are at maximal occupancy $K_{max}$. Of course, $\langle W  \rangle$ is also a random variable with mean $\langle \langle W  \rangle \rangle$ and variance approximately $\sigma_w^2/(K_{max} L^2)$. Here we have  assumed that the fitness of the individuals are statistically independent variables, which is clearly not correct but it is fine for our order of magnitude calculation. The mean size of the fluctuations of the random variable $\langle W  \rangle$ are then on the order of $\sigma_w/\sqrt{K_{max} L^2} \approx 0.3/200 = 0.0015$, since $\sigma_w \approx 0.3$ for $p_{mig} = 0.4$. The true mean size of the fluctuations is greater than this estimate because the patches are not maximally occupied  and the individual fitness are not all independent, so there are effectively less than $K_{max} L^2 $ independent terms in the sum (\ref{S1}). This digression is useful because a similar argument holds for the estimate of the size of the fluctuations (error bars) of  $\langle \langle W \rangle \rangle$ or of any other averaged quantity considered in the paper. However, because the size of the statistical ensemble is on the order of  $4 \times 10^9$ samples we can be confident that the size of the error bars  is much smaller than the  sizes of the symbols used in our figures, even if a fraction of those samples are not independent variables.

\section{Adaptation of the plastic species}\label{sec:S2}

We consider now the adaptation process of the plastic species in the case it is left alone to colonize the patchy environment (i.e., we set $b=1$ and $N_{1i}=0$ for $i$).  We  focus on the sum of the  nonplastic allelic values $S_r  \equiv \sum_{k=1}^{m_r} R_k$ and on the sum of the plastic allelic values $S_p  \equiv \sum_{k=1}^{m_p} P_k$  which determine the  phenotype $Z_i$ of an individual  of species 2  at patch $i$  [equation (\ref{Z}) of the main text]. In the equilibration stage that happens at the seed patch $i_s$, any combination of $S_r$ and $S_p$ such that $S_r + E_{i_s} S_p  \approx E_{i_s}$ will guarantee the survival of the individuals at patch $i_s$. However, in the colonization stage only the choice  $S_r \approx 0$ and $S_p \approx 1$ guarantees the survival of an individual in an arbitrary patch. Hence we will use  $S_r $ and $ S_p  $ as measures of adaptation.

\begin{figure}
 \subfigure{\includegraphics[width=0.48\textwidth]{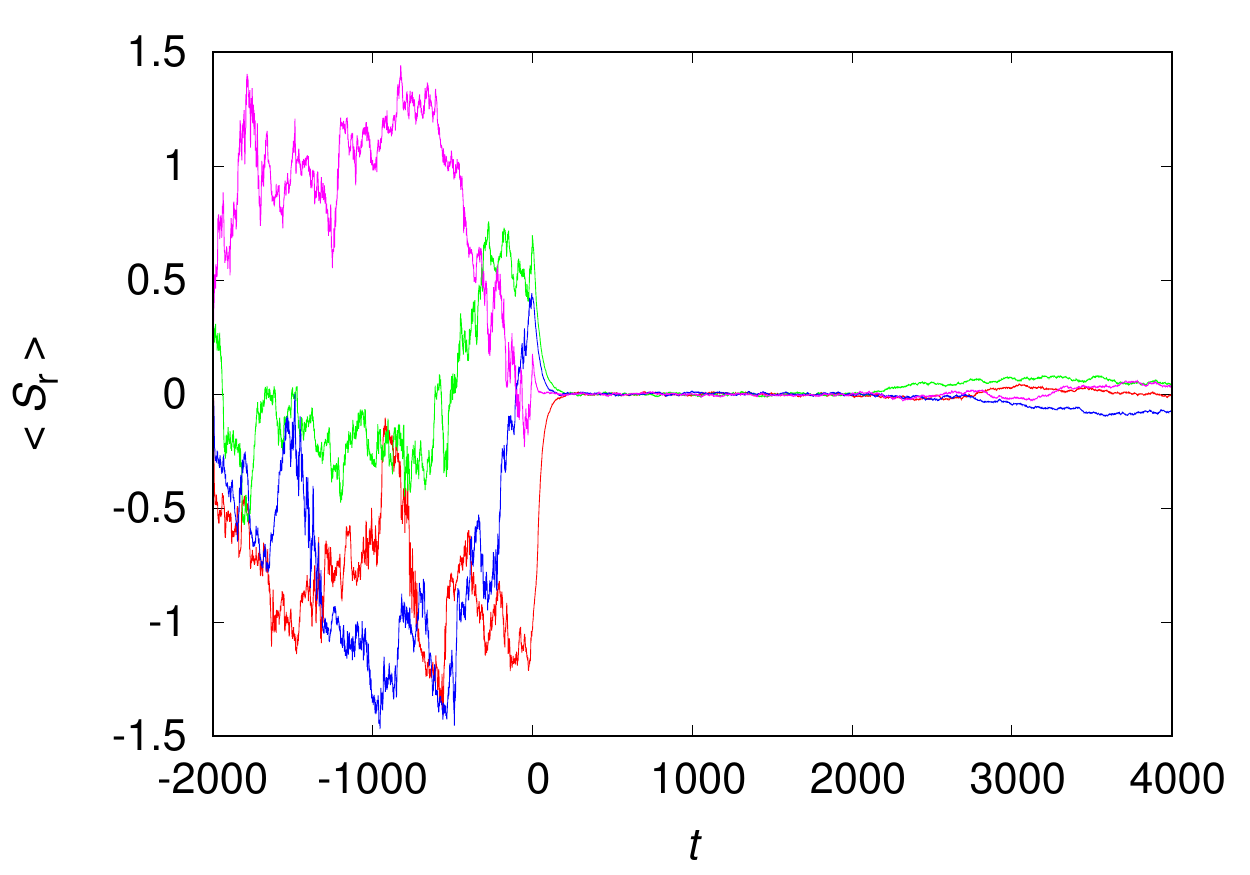}} 
 \subfigure{\includegraphics[width=0.48\textwidth]{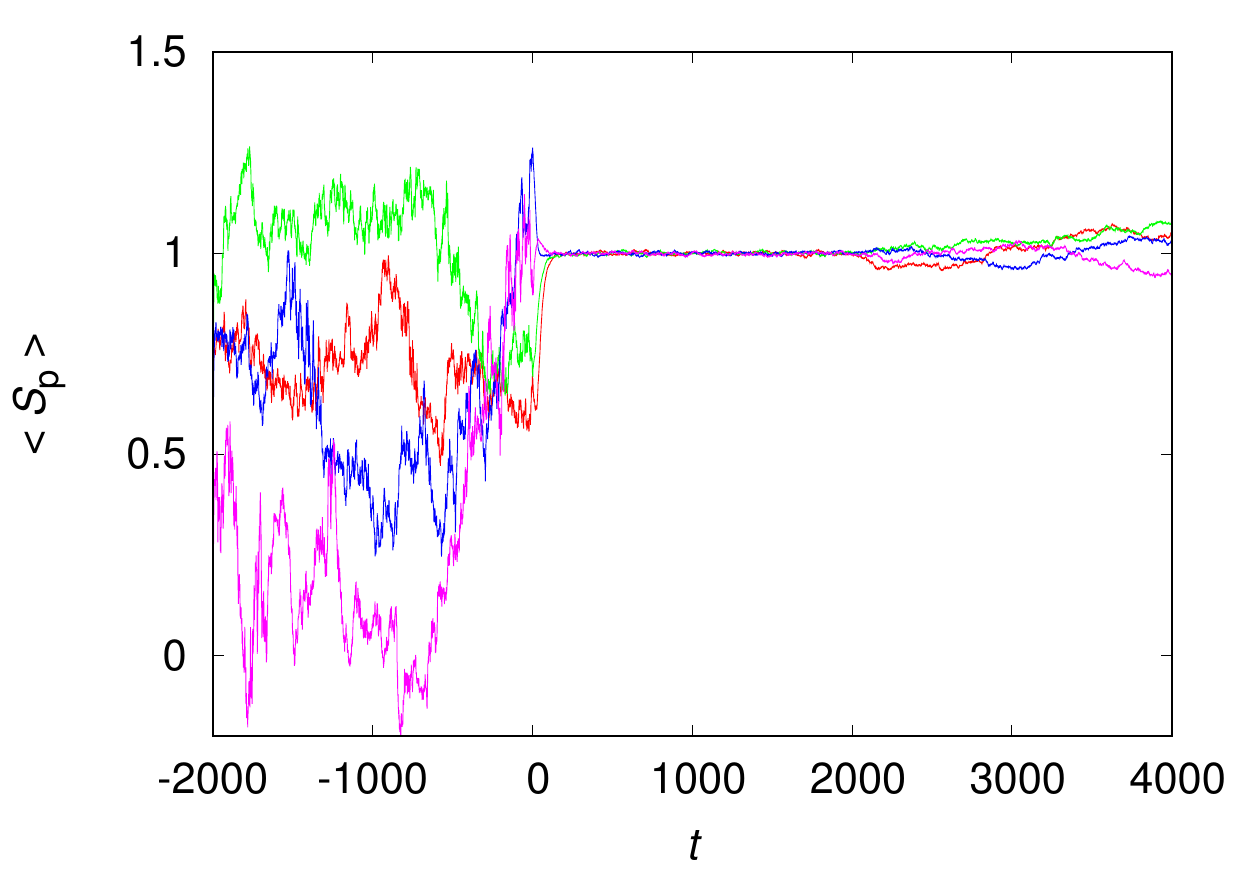}} 
\caption{Time evolution of the plastic species for four independent runs with $p_{mig} = 0.4$.  \textbf{Left Panel:} Mean sum of the nonplastic allele  values $\langle S_r \rangle$. \textbf{Right Panel:}  Mean sum of the plastic allele  values $\langle S_p \rangle$.  The equilibration period at the seed patch occurs for $t \in [-2000,0)$ and the  colonization of the empty patches occurs for $t \in [0,2000)$. We set $p_{mig}=0$ at $t=2000$. 
          The parameters are $L=20$, $K_{max} = 100$, $c=0$, $b=1$,  $\sigma_e^2 =2$ and $\rho = 0$.
 }  
\label{fig:S4}  
\end{figure}

\begin{figure}
 \includegraphics[width=0.48\textwidth]{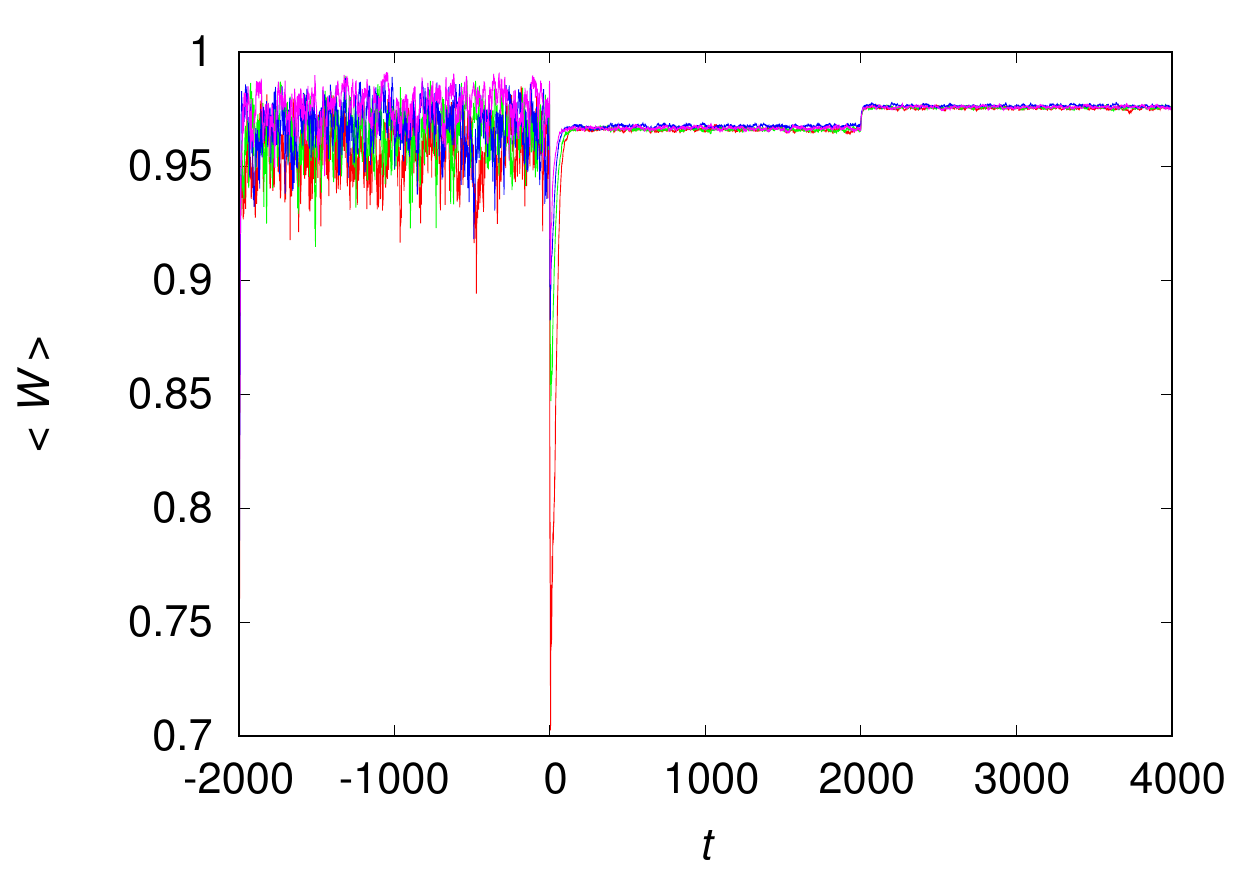} 
\caption{Time evolution of the mean fitness of the plastic species  for four independent runs with $p_{mig} = 0.4$. The equilibration period at the seed patch occurs for $t \in [-2000,0)$ and the  colonization of the empty patches occurs for $t \in [0,2000)$. We set $p_{mig}=0$ at $t=2000$. 
          The parameters are $L=20$, $K_{max} = 100$, $p_{mig} = 0.4$, $c=0$, $b=1$,  $\sigma_e^2 =2$ and $\rho = 0$. 
 }  
\label{fig:S5}  
\end{figure}

Figure  \ref{fig:S4} shows the time evolution  of $\langle S_r \rangle $ and $\langle S_p \rangle $  measured {\it before} the selection sieve for four independent runs. As before, the single bracket notation stands for the average over the individuals in a patch and over all patches. As expected, during the colonization stage there is a selective pressure to silence the nonplastic alleles and set the plastic alleles to their optimum values. It is interesting that when migration is turned off at $t=2000$ that selective pressure disappears and the alleles  begin to drift so as to improve the adaptation at the local patches, as shown in figure  \ref{fig:S5}, which exhibits the time dependence of the mean fitness of species 2.  Despite the large fluctuations on the plastic and nonplastic allele values during the equilibration phase, the population is well adapted to the seed patch, as indicated by the large values of the mean fitness.

\begin{figure}
 \subfigure{\includegraphics[width=0.48\textwidth]{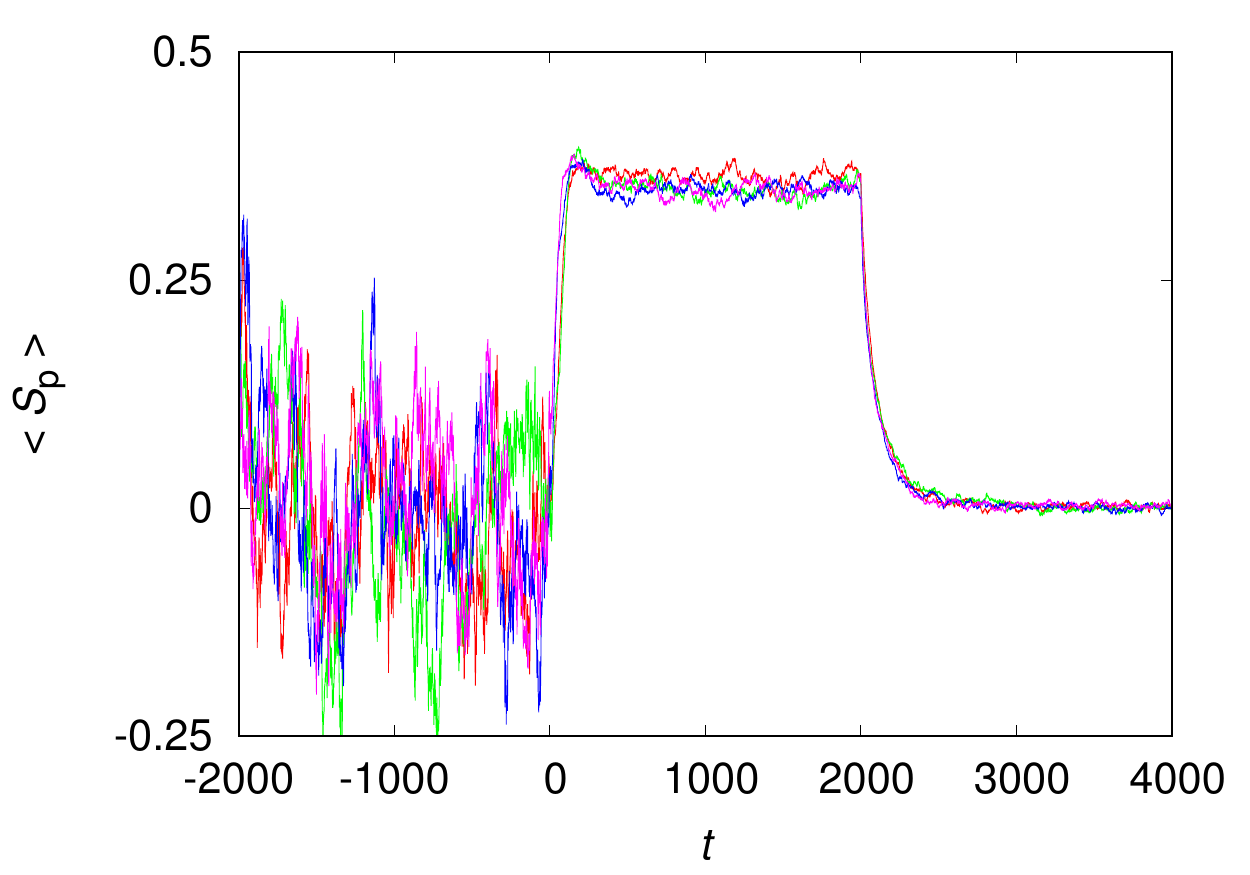}} 
 \subfigure{\includegraphics[width=0.48\textwidth]{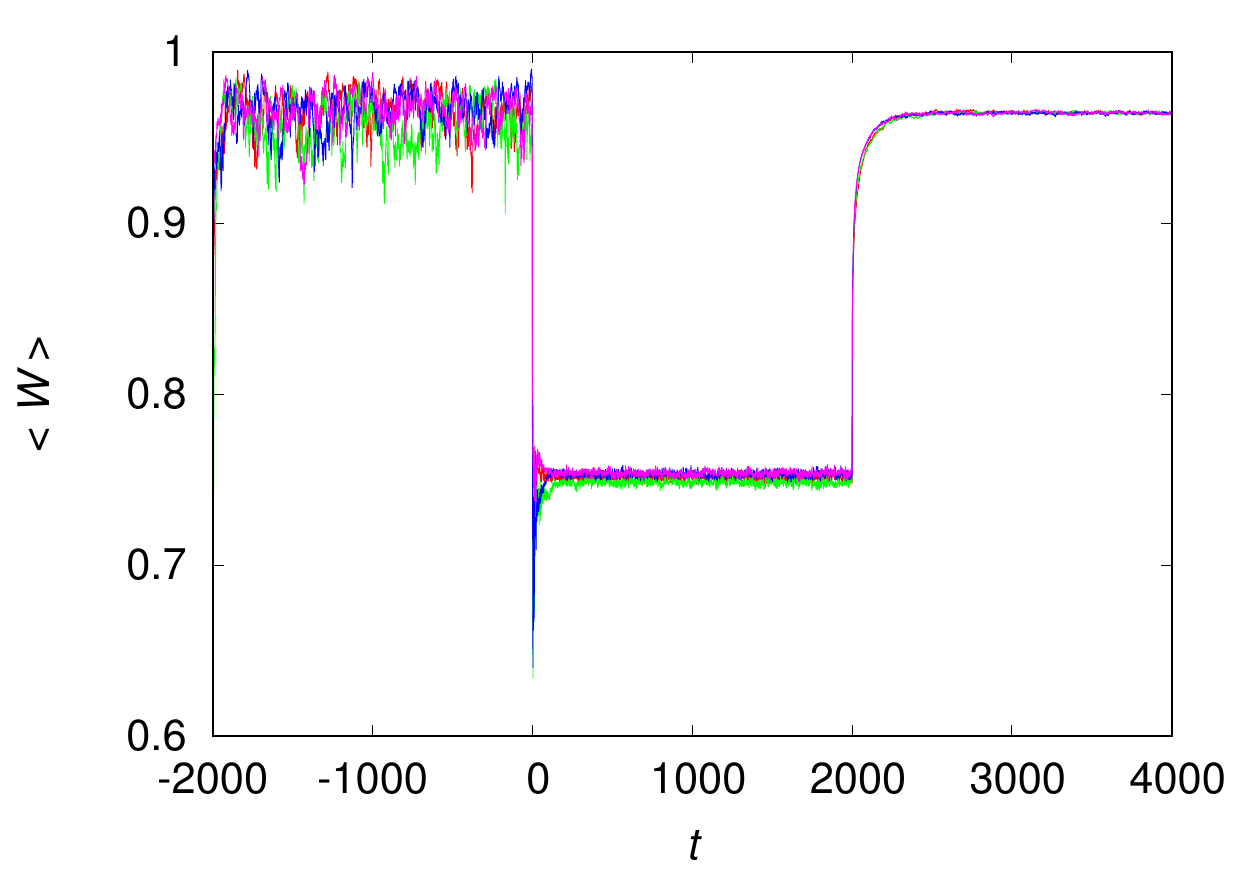}} 
\caption{Time evolution of the plastic species for four independent runs with $p_{mig} = 0.4$ and plasticity cost $c=1$.  \textbf{Left Panel:} Mean sum of the plastic allele  values $\langle S_p \rangle$. \textbf{Right Panel:}  Mean fitness $\langle W \rangle$.  The equilibration period at the seed patch occurs for $t \in [-2000,0)$ and the  colonization of the empty patches occurs for $t \in [0,2000)$. We set $p_{mig}=0$ at $t=2000$. 
          The parameters are $L=20$, $K_{max} = 100$, $b=1$,  $\sigma_e^2 =2$ and $\rho = 0$.
 }  
\label{fig:S6}  
\end{figure}

Figure  \ref{fig:S6} shows  the effect of the plasticity cost  $c$ on the values of the plastic  alleles and on the mean fitness for four independent runs.  The optimal strategy for an  isolated   population  (as  happens during  the equilibration period  or after $t=2000$ when migration is not allowed) is to silence the plastic alleles $P_k$ and this is exactly what we observe in the figure. In the case that migration is allowed there is a trade-off between the advantage and the cost  of plasticity, so $\langle S_p \rangle $ takes on an intermediate value between $0$ and $1$. The mean value of the nonplastic alleles $\langle S_r \rangle  $ is not informative so we do not present it here. In fact, since   $S_r \approx E_i$  within each patch and recalling that $E_i \sim N(0,\sigma_e^2)$  we have $\langle S_r \rangle \approx 0 $ when we average over patches. Note that  only  the nonplastic  alleles contribute to adaptation when the populations are isolated in the patches.

\begin{figure}
 \subfigure{\includegraphics[width=0.48\textwidth]{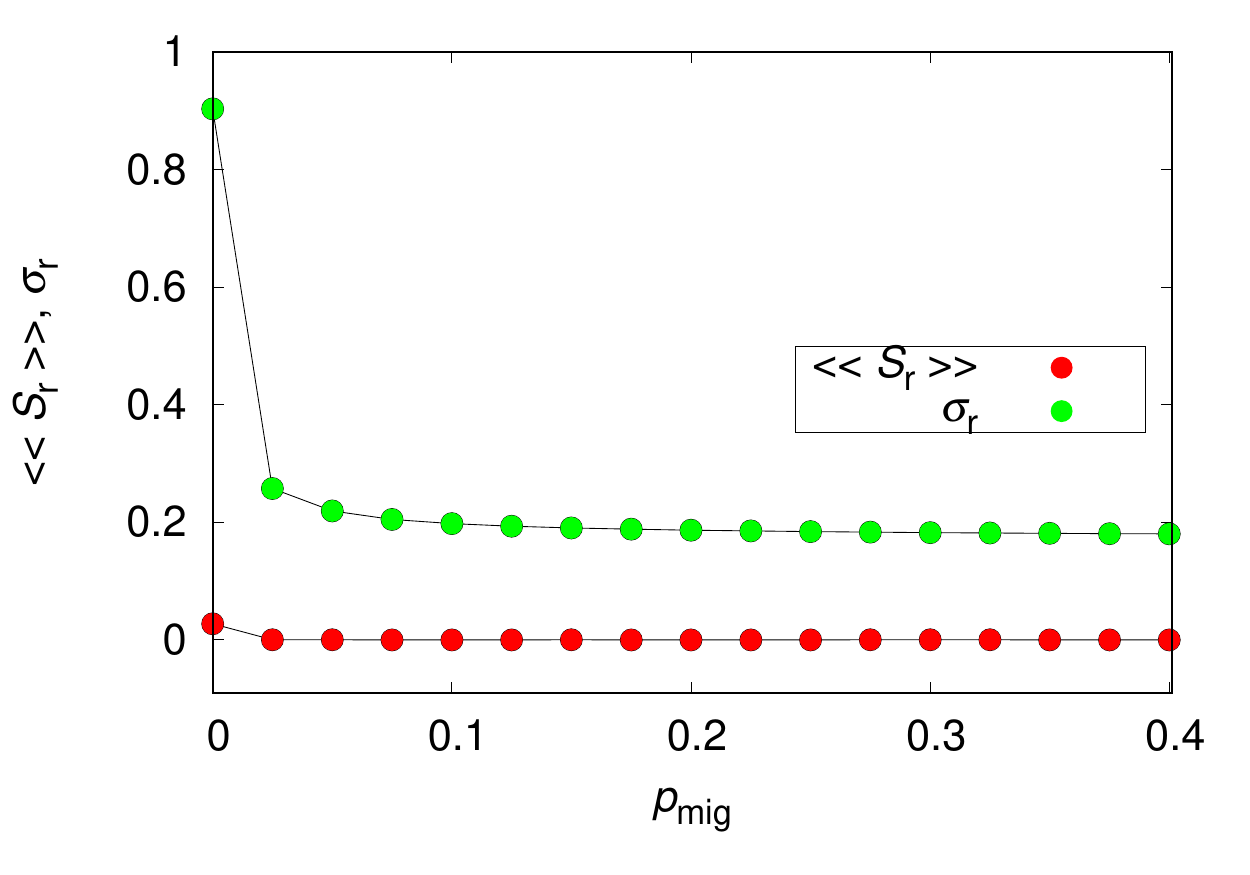}} 
 \subfigure{\includegraphics[width=0.48\textwidth]{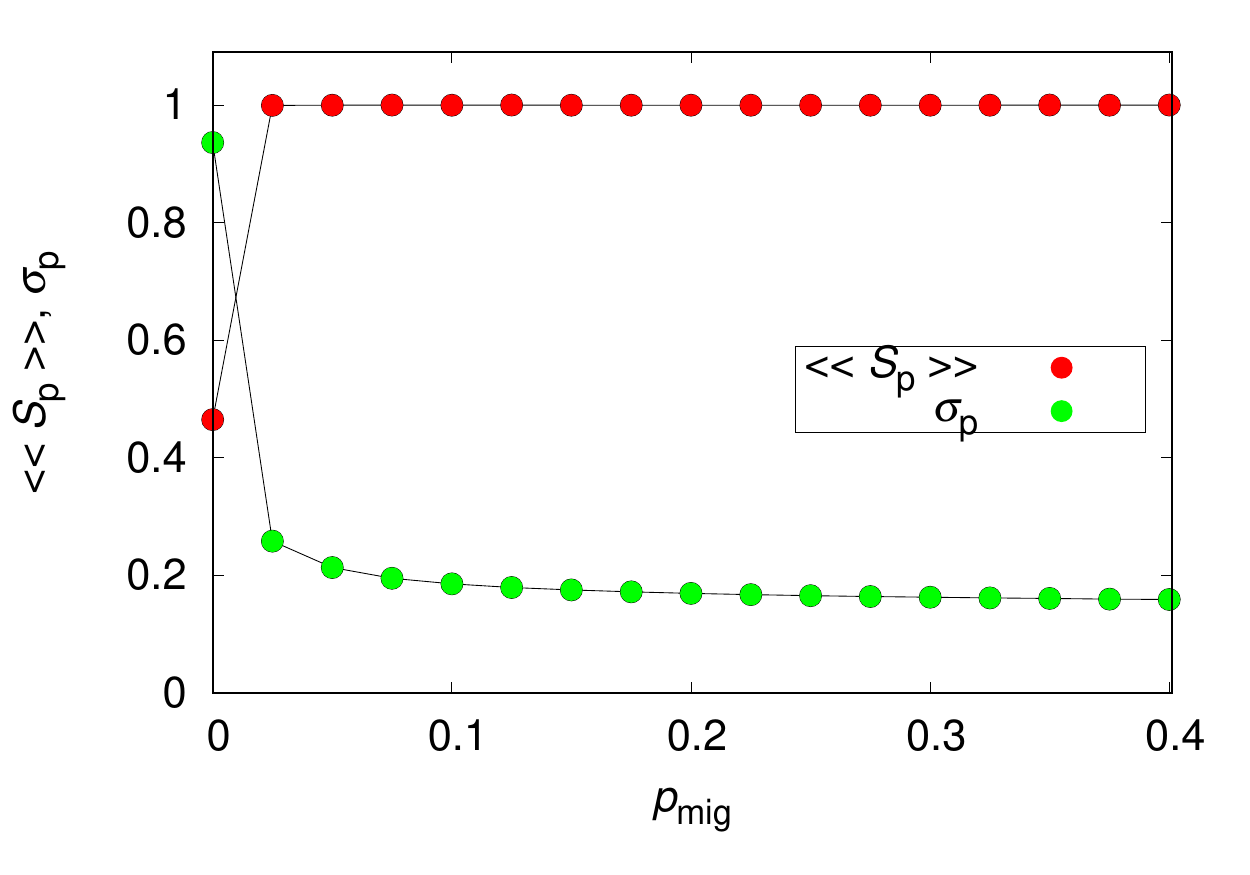}} 
\caption{Influence of the migration probability on the equilibrium of the plastic species.  \textbf{Left Panel:} Mean sum of the nonplastic allele  values  $\langle \langle S_r \rangle \rangle$ and the corresponding standard deviation $\sigma_r$. \textbf{Right Panel:} Mean sum of the plastic allele  values  $\langle \langle  S_p \rangle \rangle$ and the corresponding standard deviation $\sigma_p$.  
          The parameters are $L=20$, $K_{max} = 100$, $c=0$, $b=1$,  $\sigma_e^2 =2$ and $\rho = 0$.
 }  
\label{fig:S7}  
\end{figure}

In figure  \ref{fig:S7} we show the dependence of $\langle \langle S_r \rangle \rangle $ and $\langle \langle S_p \rangle \rangle $ and the corresponding standard deviations $\sigma_r$ and $\sigma_p$   on the migration probability at equilibrium. As in the analysis of the adaptation of the  nonplastic species, to produce this figure we average the sums of the allele values over all individuals,  over the last 100 generations of the colonization stage and over 1000 runs. In agreement with the expectation about the characteristics of a plastic species, the sums of the allele values and their standard deviations  are practically unaffected by changes in the migration probability $p_{mig}>0$. The same  conclusion holds true for the fitness, as shown in figure \ref{fig:S8}. 

\begin{figure}
 \includegraphics[width=0.48\textwidth]{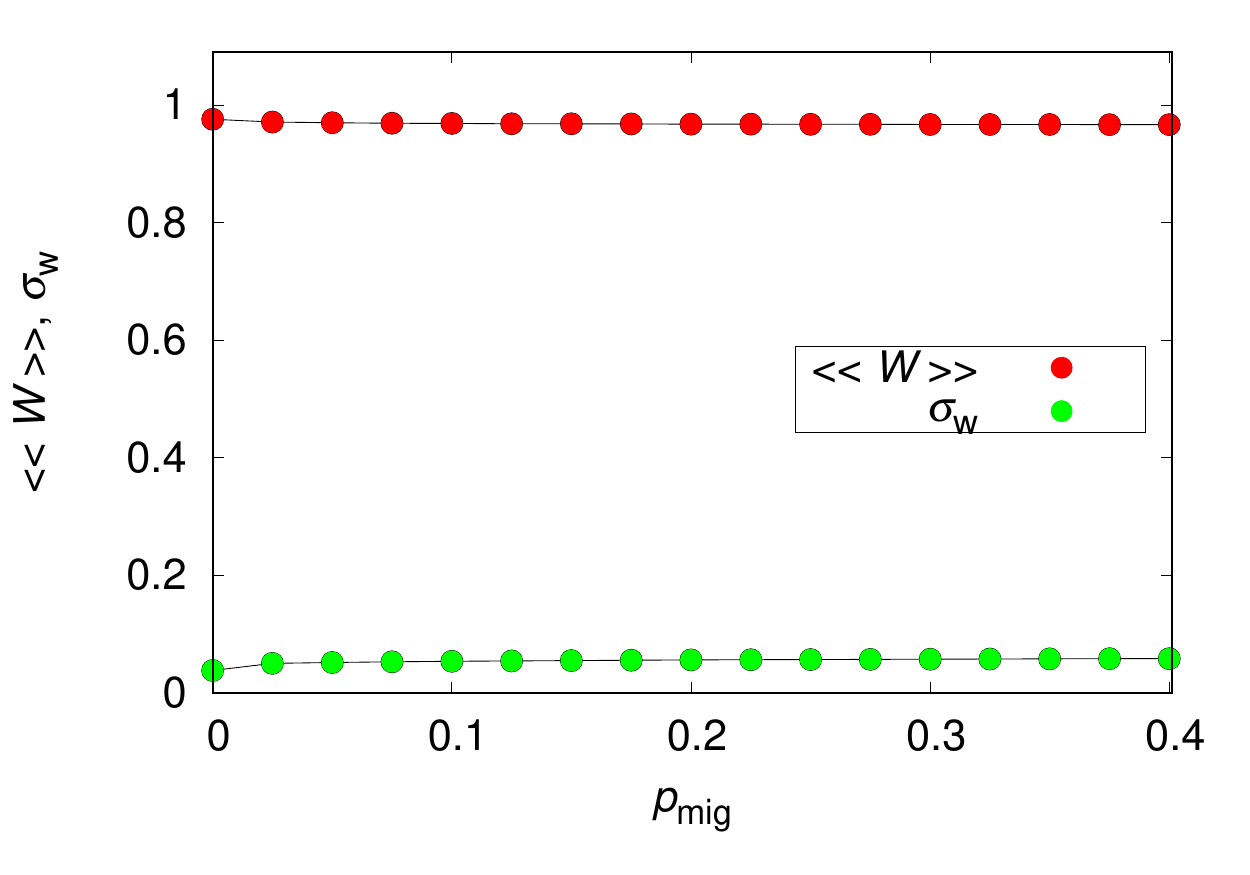} 
\caption{Influence of the migration probability $p_{mig}$  on the mean $\langle \langle W \rangle \rangle$  and standard deviation $\sigma_w$ of the fitness for the plastic species at equilibrium. The parameters are $L=20$, $K_{max} = 100$,  $c=0$, $b=1$,  $\sigma_e^2 =2$ and $\rho = 0$.}
\label{fig:S8}  
\end{figure}

As mentioned before, the size of the error bars of our estimates of the mean values of the quantities that characterize the plastic species at equilibrium are negligible due to the large size of the statistical ensemble used to calculate the averages.

\section{Spatial distribution of the competing species}\label{sec:S3}

Here we offer a brief discussion on the spatial distribution of the competing species in the patchy environment. The  main point is to show that there is no  spatial organization process governing the distribution of the species in the patches. In addition, we use  snapshots of the relative abundances of the species in the patches to explain and argue in favor of the quantities used to characterize the equilibrium of the metapopulation dynamics in the main text.

Figure \ref{fig:S9} exhibits an instance of  equilibrium accidental coexistence for $p_{mig}= 0.2$ in which  species 1 and 2  coexist in the metapopulation but not within patches.    The seed patch $i_s$ is located at the center of the grid and at $t=0$  both species are well adapted  to the seed patch environment   with densities $n_{1i_s}=n_{2i_s} = 0.5$.  All other patches are empty. We recall that the relative abundance of a species in a given patch is  the number of individuals of that species in the patch divided by the carrying capacity $K_{max}$. The relative abundance values are shown in  a color scale in the figure and are always less than 1.

\begin{figure}
 \subfigure{\includegraphics[width=0.4\textwidth]{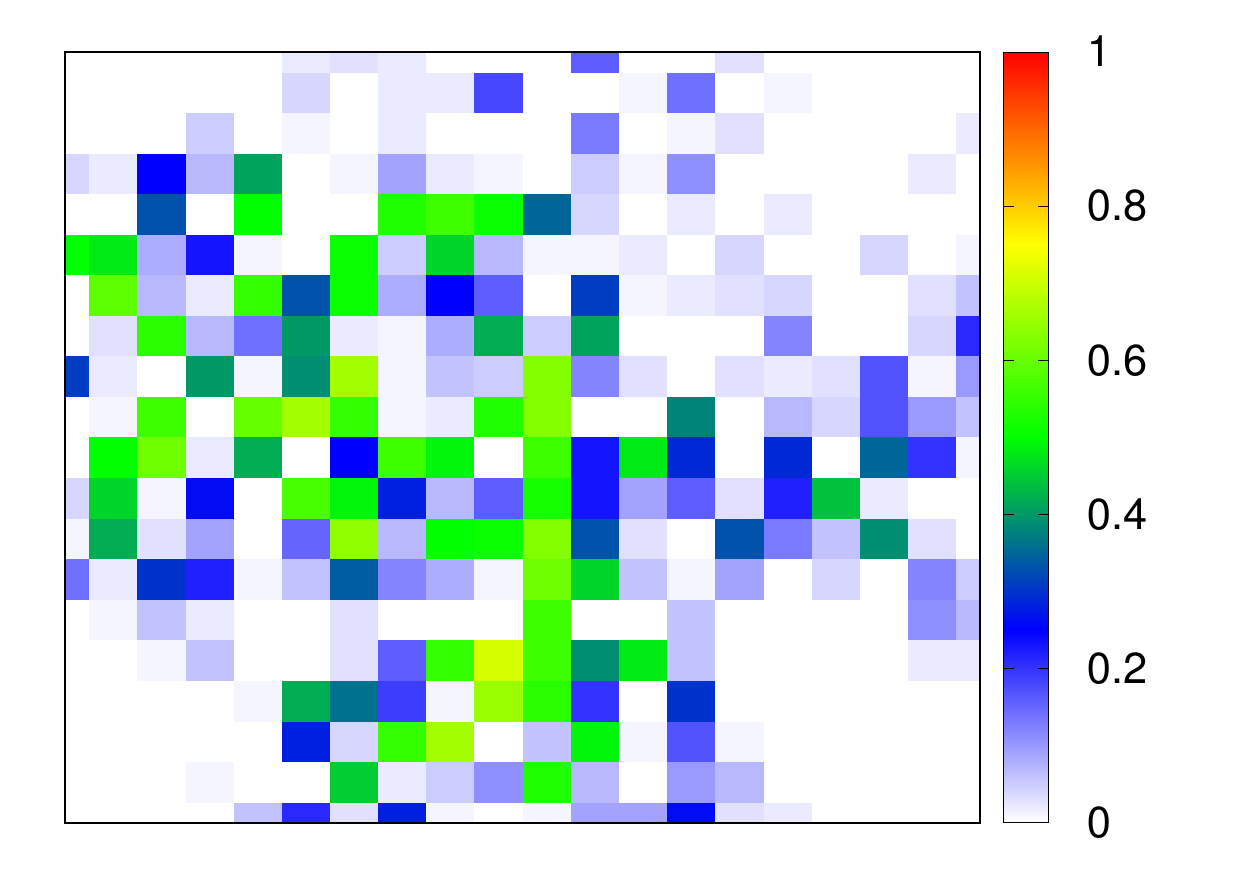}} 
 \subfigure{\includegraphics[width=0.4\textwidth]{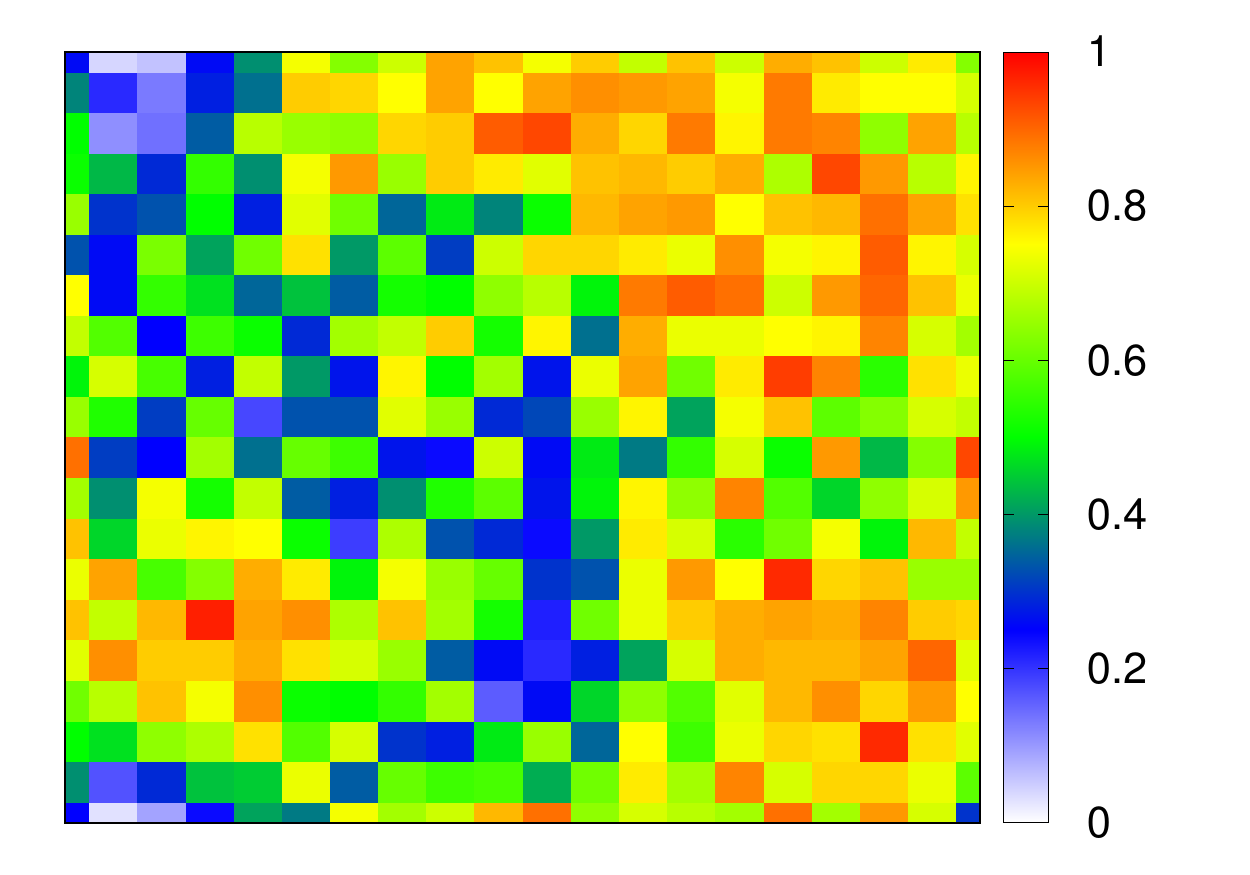}}  
 \subfigure{\includegraphics[width=0.4\textwidth]{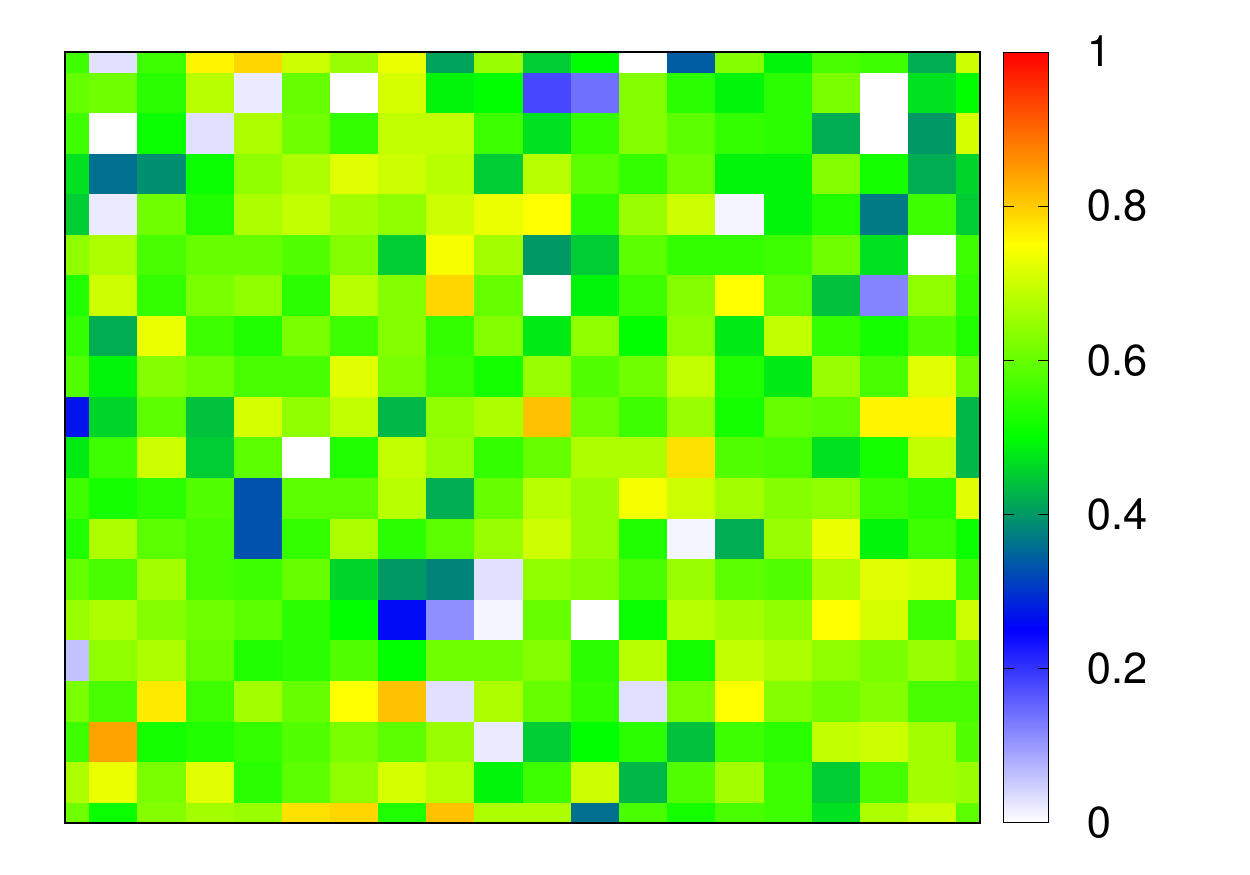}} 
 \subfigure{\includegraphics[width=0.4\textwidth]{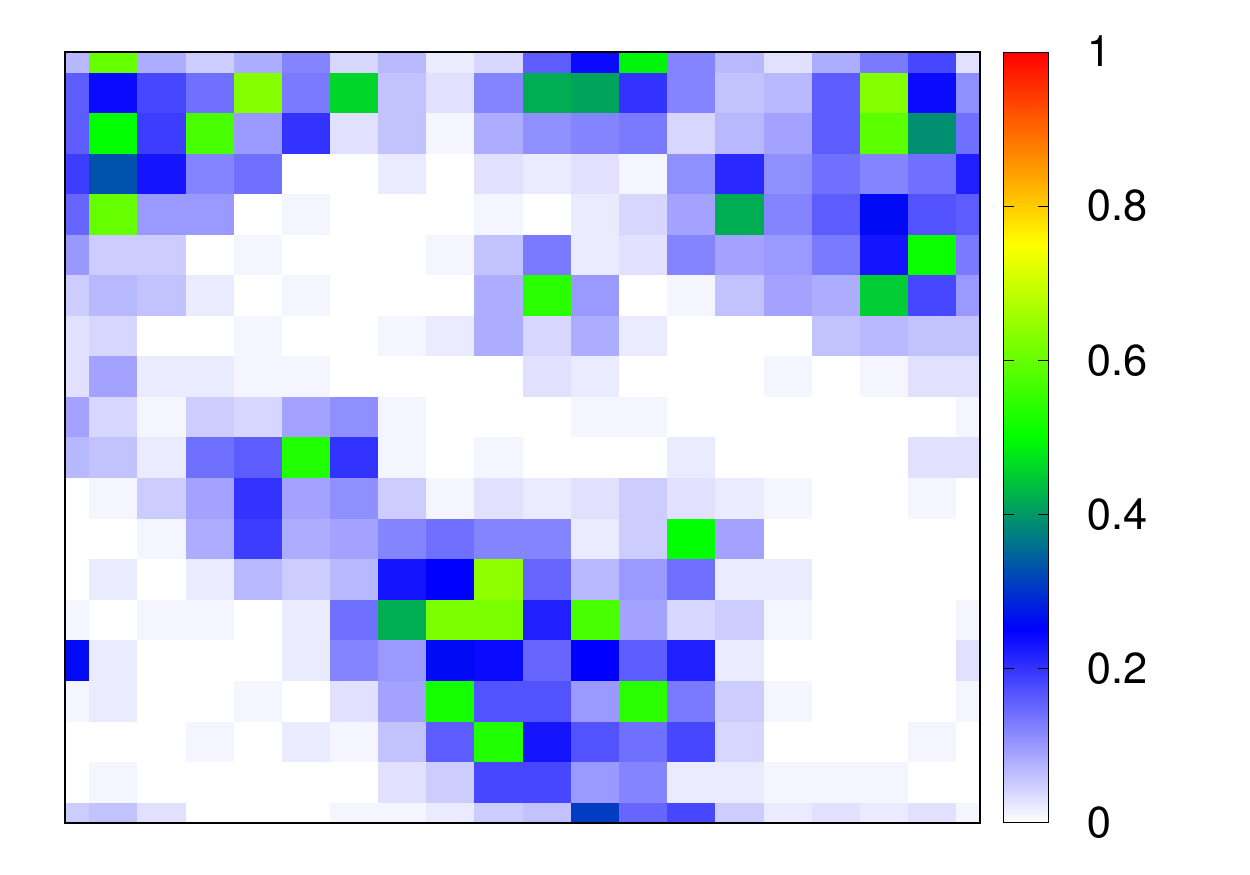}} 
 \subfigure{\includegraphics[width=0.4\textwidth]{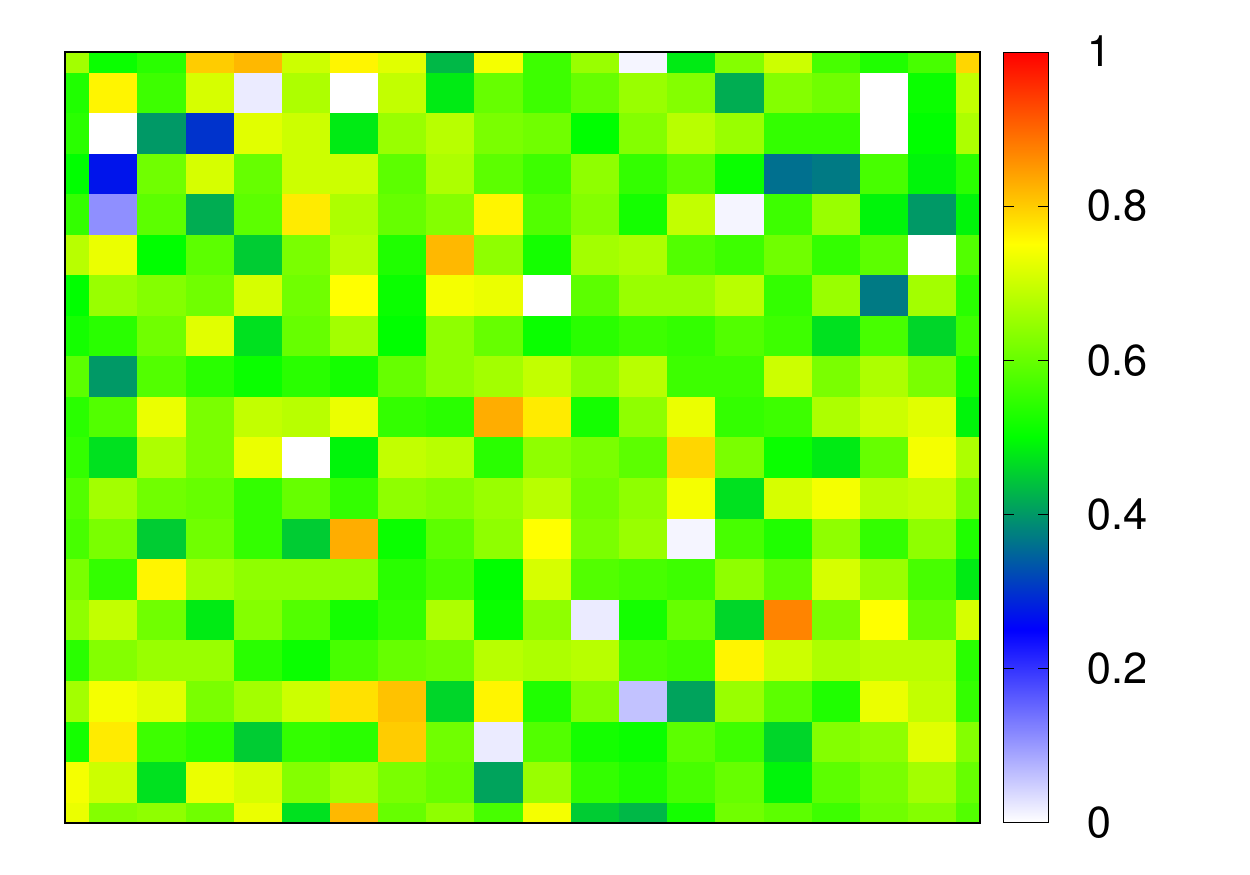}} 
  \subfigure{\includegraphics[width=0.4\textwidth]{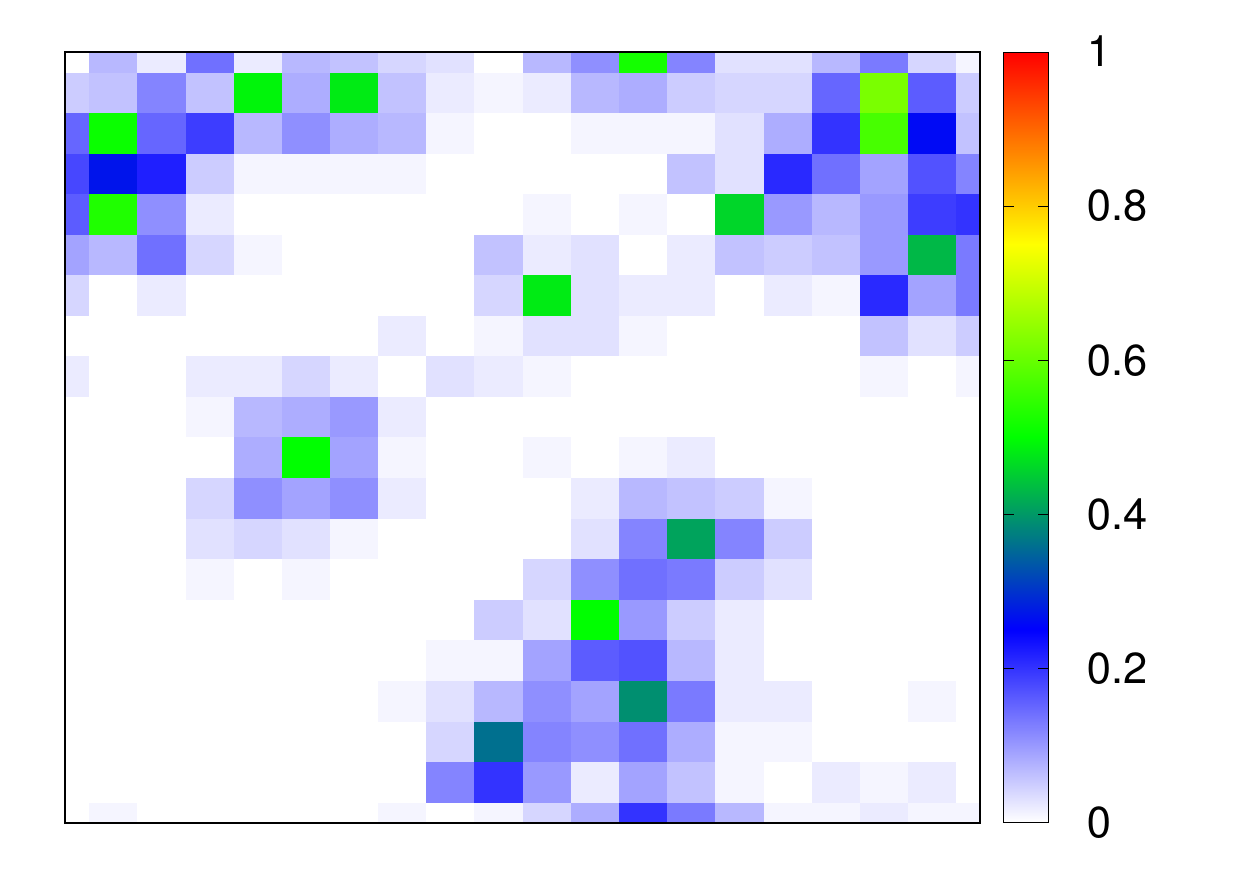}} 
   \subfigure{\includegraphics[width=0.4\textwidth]{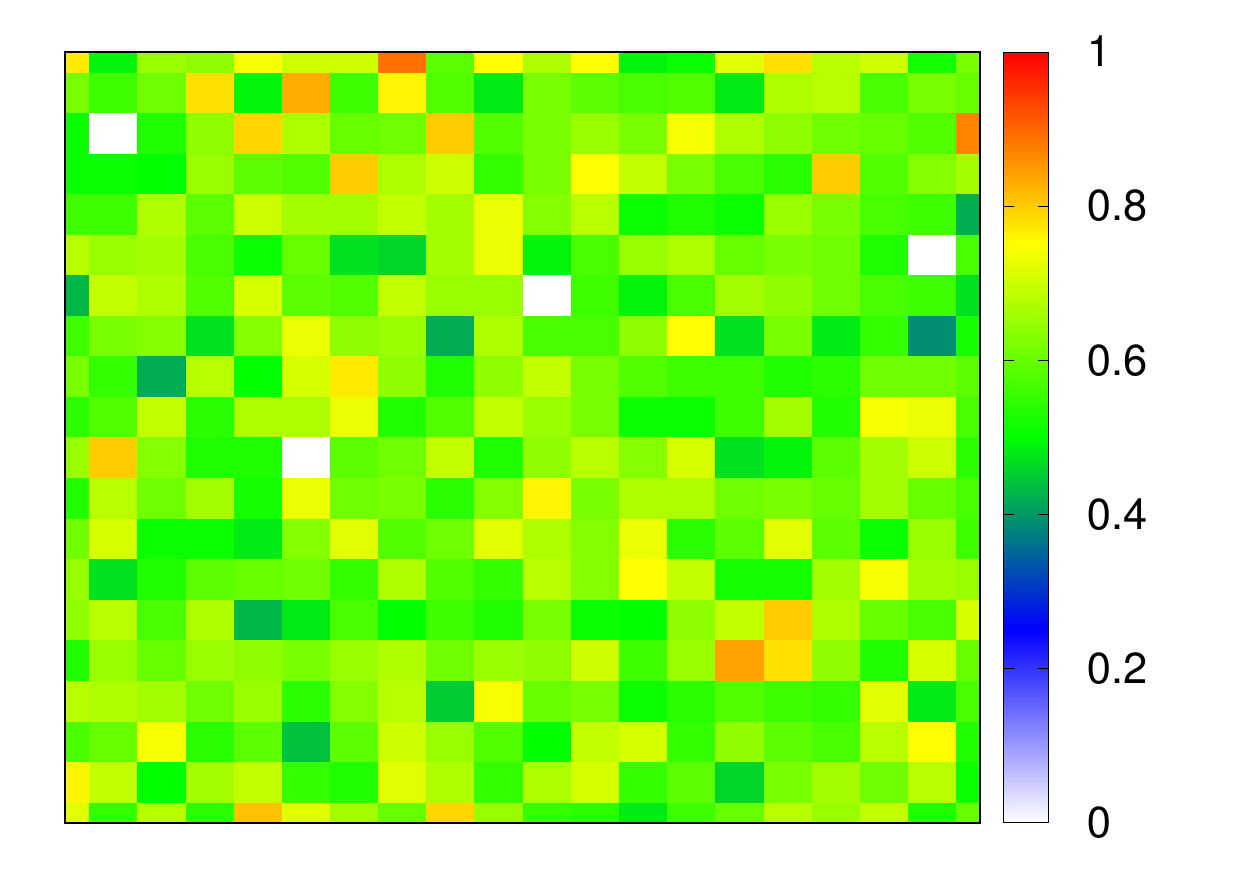}} 
    \subfigure{\includegraphics[width=0.4\textwidth]{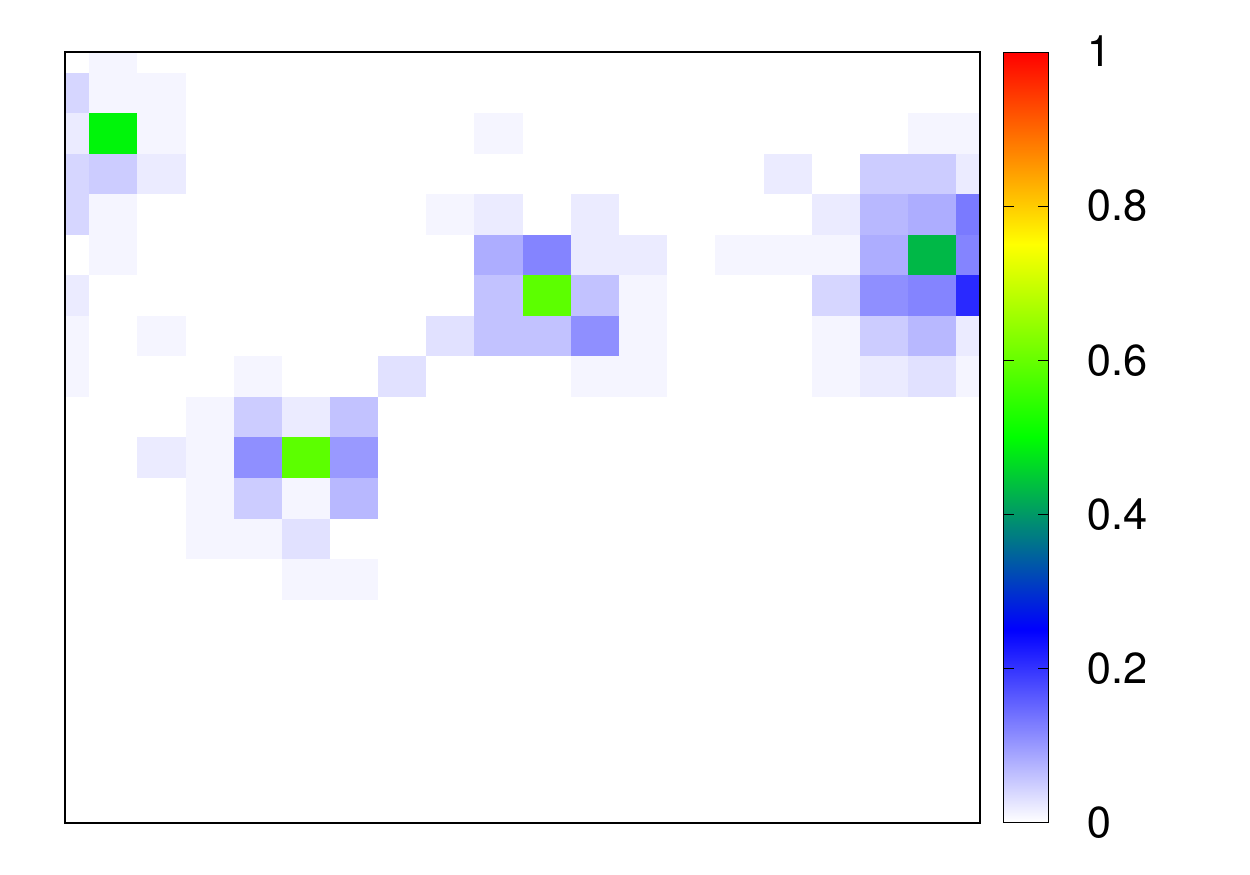}} 
\caption{Snapshots of the relative abundances  for $p_{mig}=0.2$. The rows show the grid configurations  at (top to bottom)  $t=50, 500, 1000$ and $10000$. The first column shows $n_{1i}$ and the second column shows  $n_{2i}$.
          The parameters are $L=20$, $K_{max} = 100$, $c=0$, $b=1$,  $\sigma_e^2 =2$ and $\rho = 0$.
 }  
\label{fig:S9}  
\end{figure}

The fact that the relative abundances are less than 1 supports our claim that the  hard upper bound on the number of individuals used in the modeling of the ecological competition (see Section \ref{sec:comp} of the main text) does not affect the dynamics. It is interesting that the plastic species rapidly colonizes all patches whereas the nonplastic species struggles to adapt to the environment of  each new visited  patch, as shown in the panels for $t=50$. Eventually, species 1 manages to displace species 2 from all but four patches, which are characterized by extreme values of the environmental values, as shown in the panels for $t=10000$. Those  patches are safe heavens  for species 2, which then  sends  a constant flow of  doomed migrants to their neighboring patches. We emphasize that there is no spatial organization in the distribution of species 2: it survives only in those patches that species 1 cannot colonize because of their extreme environments. Because the existence as well as the location of those patches are decided randomly at the setting of the environment,  we refer to this type of coexistence as accidental coexistence. 

Figures \ref{fig:S10} and  \ref{fig:S11} show instances of equilibrium robust coexistence for $p_{mig}= 0.3$ and $p_{mig}= 0.4$. Since the poor adapted species 1 cannot displace species 2 because $n_{1i} < 1/a_{21} = 2/3$ in almost all patches $i$, species 2  spreads all over the environment. In these cases, the patches where there is no coexistence are the patches where species 1 is absent.  At this point we can appreciate the relevance of  the measure $\langle \langle \Pi \rangle \rangle$ (i.e., the mean fraction of patches that carry both species)  to the understanding of the metapopulation equilibrium. In fact, for large migration probabilities $1-\langle \langle \Pi \rangle \rangle$  measures the fraction of patches not occupied by species 1 (i.e., the white patches in the left  panels of figures  \ref{fig:S10} and  \ref{fig:S11}).  For instance, the decrease of $\langle \langle \Pi \rangle \rangle$ with increasing $p_{mig}$ for $p_{mig} > 0.275$ shown in the upper panel of figure \ref{fig:6} of the main text quantifies the observed increase of  patches  lacking species 1 exhibited in the relative abundance snapshots. The low values of $\langle \langle \Pi \rangle \rangle$ for small migration probabilities indicates that (accidental) coexistence happens only in the neighborhood of the patches with extreme environments and in this case  $1-\langle \langle \Pi \rangle \rangle$ measures essentially the fraction of patches lacking species 2.

\begin{figure}
 \subfigure{\includegraphics[width=0.45\textwidth]{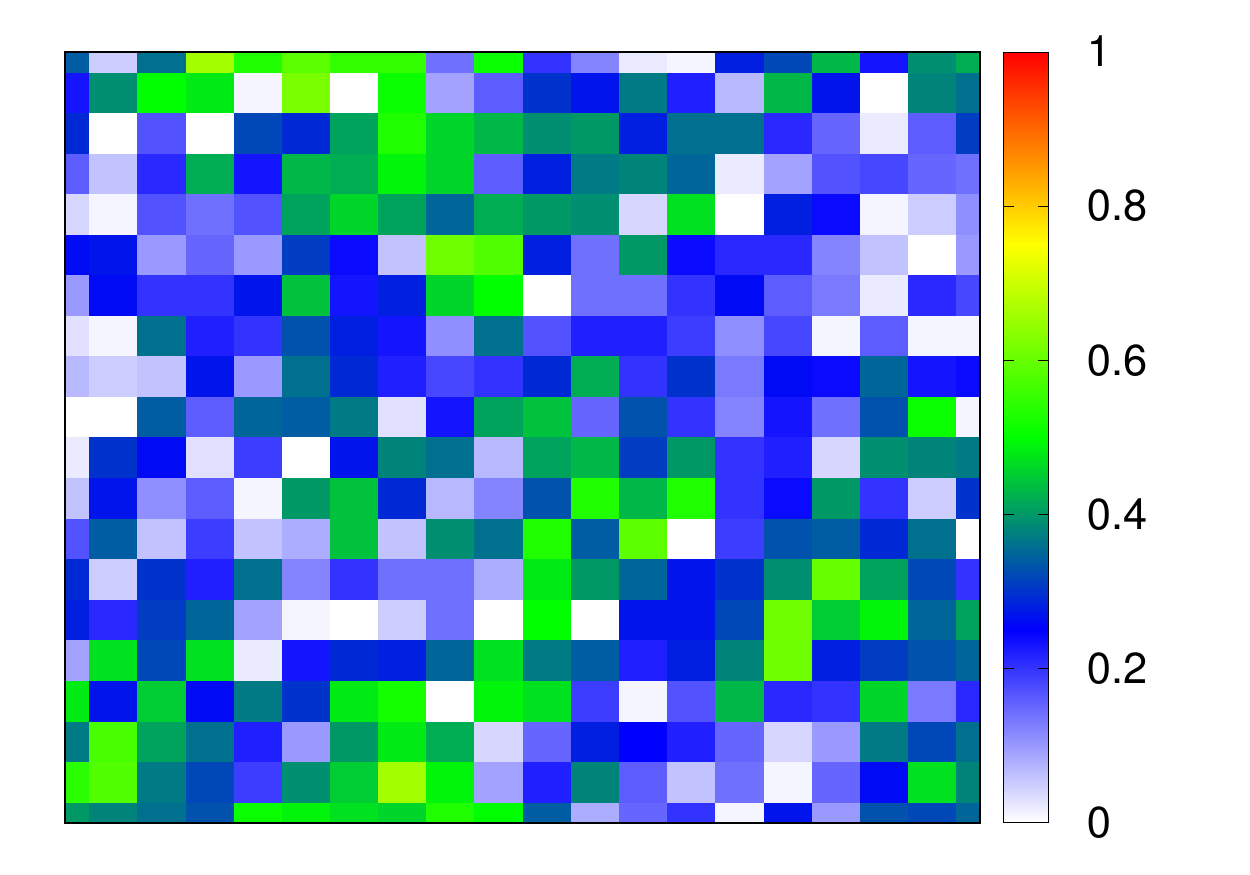}} 
 \subfigure{\includegraphics[width=0.45\textwidth]{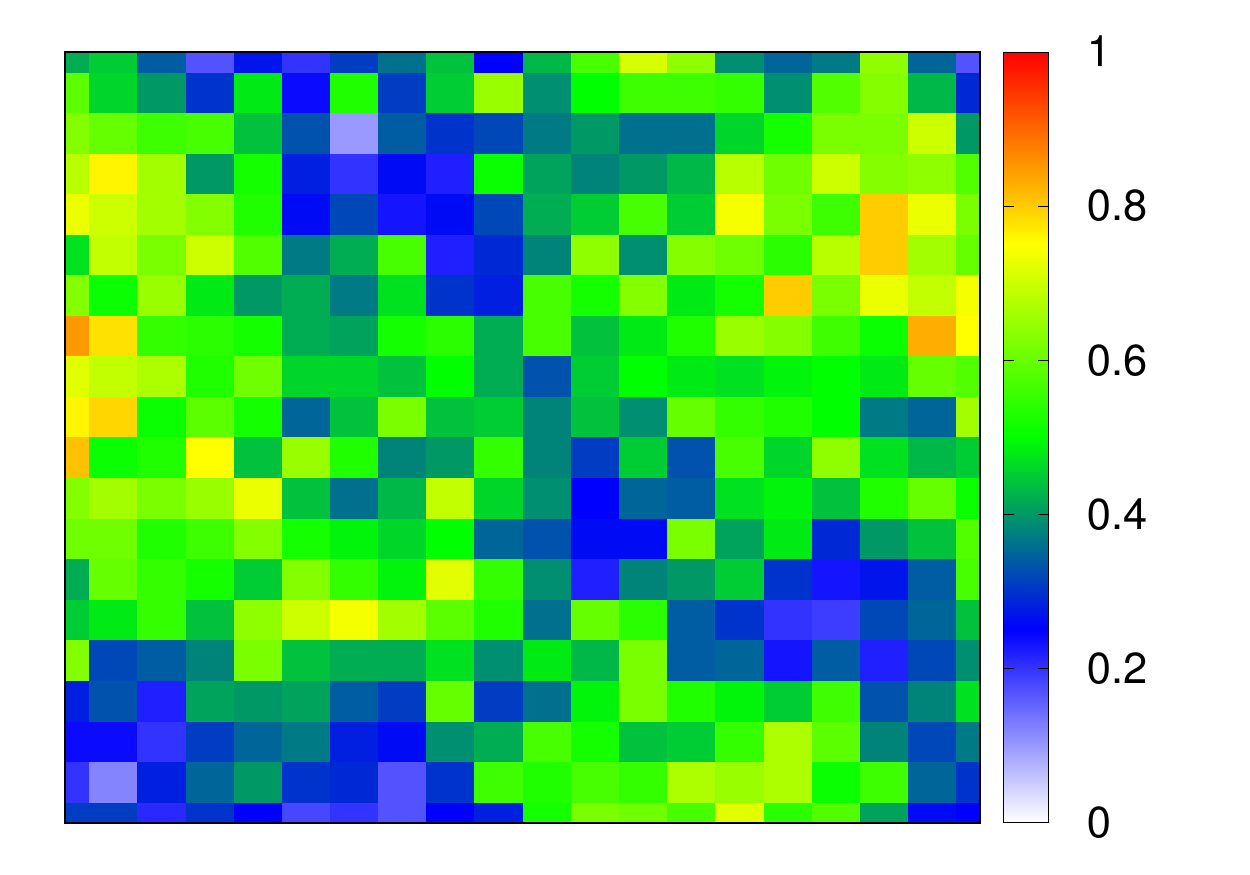}} 
\caption{Snapshots of the density of both species in the patches for $p_{mig}=0.3$.  \textbf{Left Panel:} Species 1 at $t=10000$. \textbf{Right Panel:}  Species 2 at $t=10000$.  
          The parameters are $L=20$, $K_{max} = 100$, $c=0$, $b=1$,  $\sigma_e^2 =2$ and $\rho = 0$.
 }  
\label{fig:S10}  
\end{figure}

\begin{figure}
 \subfigure{\includegraphics[width=0.45\textwidth]{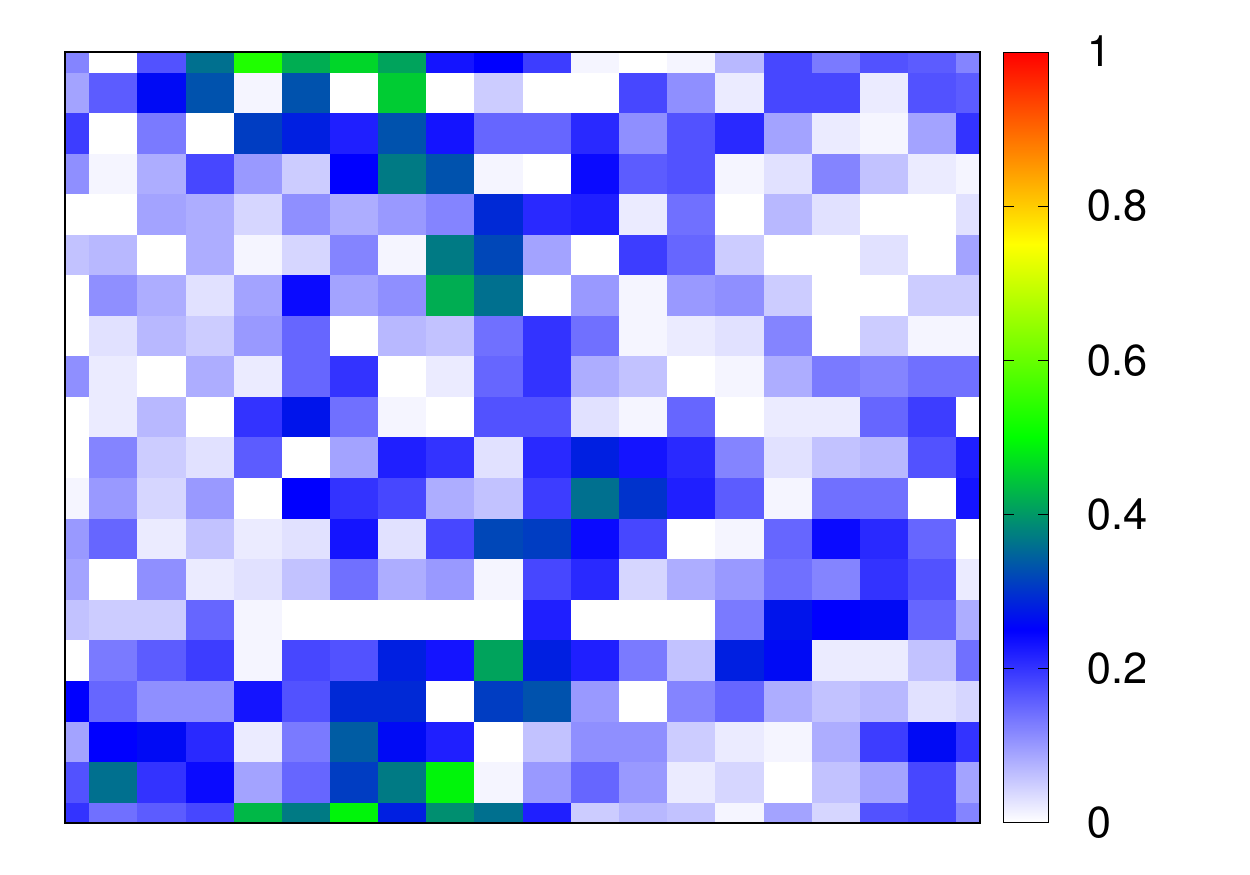}} 
 \subfigure{\includegraphics[width=0.45\textwidth]{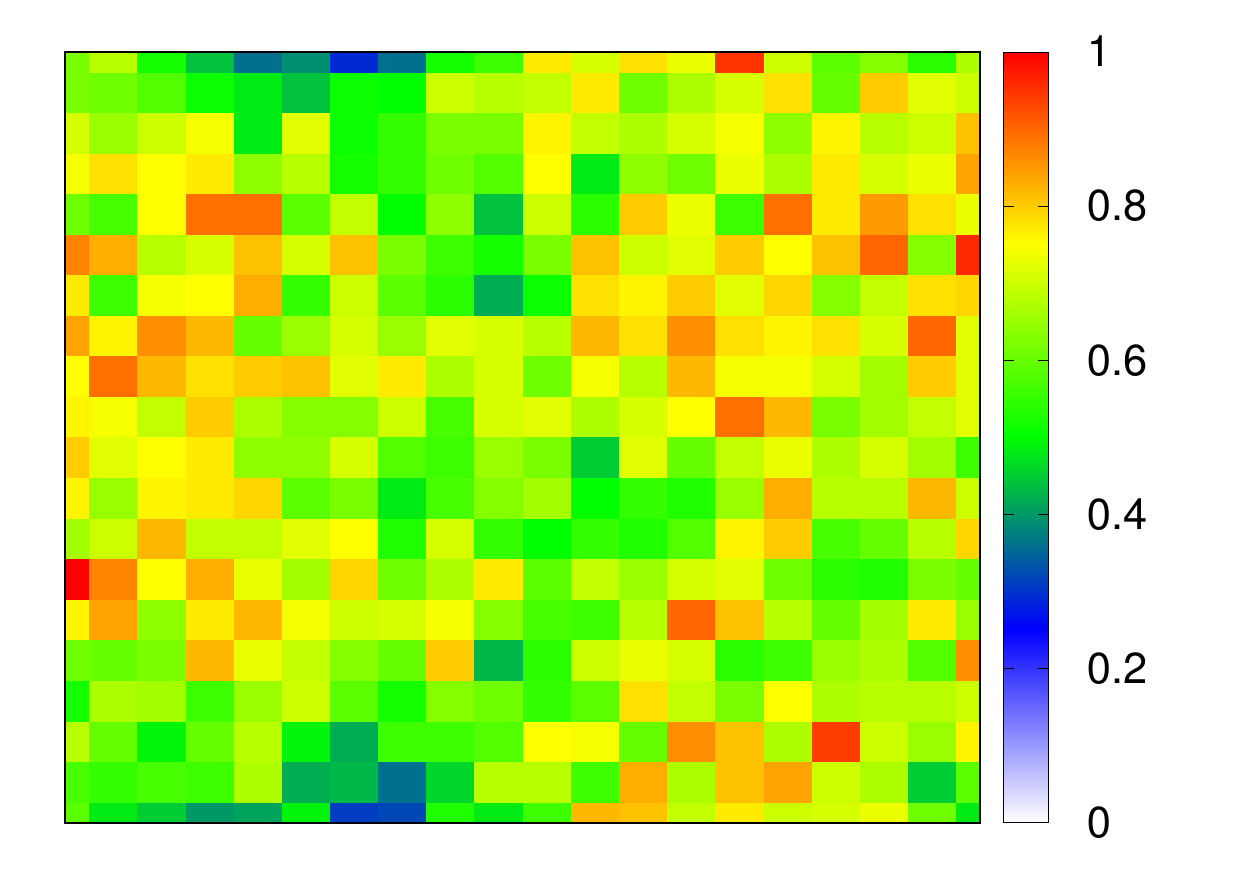}} 
\caption{Snapshots of the density of both species in the patches for $p_{mig}=0.4$.  \textbf{Left Panel:}  Species 1 at $t=10000$. \textbf{Right Panel:}  Species 2 at $t=10000$. 
          The parameters are $L=20$, $K_{max} = 100$, $c=0$, $b=1$,  $\sigma_e^2 =2$ and $\rho = 0$. 
 }  
\label{fig:S11}  
\end{figure}

The other quantities we introduced in the main text to characterize  the metapopulation at equilibrium are the mean patch relative abundances of both species $\langle \langle n_1 \rangle \rangle$ and $\langle \langle  n_2 \rangle \rangle$. The double brackets notation stands for the average of the relative abundances over all patches of the grid, as well as over runs and over the last 100 generations of each run. So those quantities represent the overall  abundances of the species in the grid.   The decrease of $\langle  n_1 \rangle $ and the increase of $\langle  n_2 \rangle$ with increasing $p_{mig}$ observed in the metapopulation snapshots for $t=10000$ are quantified in the lower panel of figure \ref{fig:6} in the main text. Here the single bracket notation stands for the average over patches only.

The interpretation of the measures used in the main text, viz. $\langle \langle \Pi \rangle \rangle$, $\langle \langle n_1 \rangle \rangle$ and $\langle  \langle n_2 \rangle \rangle$,  is now clear with the aid of the grid snapshots. It is also clear that those measures offer a very detailed  characterization of the coexistence observed in equilibrium regime of the metapopulation dynamics.

\section{Temporal evolution of the competing species}\label{sec:4}

Complementing the microscopic information about the spatial distribution of the species in the patchy environment presented in section \ref{sec:S3}, here we offer a brief appraisal of the temporal evolution of the abundances of the competing species averaged over all patches.  Figure \ref{fig:S12} shows the relative species abundances for typical runs that led to species coexistence. The run for $p_{mig}=0.1$ shows an instance of accidental coexistence (species 2 was still present in the metapopulation  up to $t=10^5$), whereas the run for $p_{mig}=0.3$ illustrates an instance of robust coexistence. It is clear that in both runs, but more conspicuously  in the run that led to non-accidental coexistence, the 2000 generations upper limit  offers a reliable guarantee that the dynamics  is in the equilibrium regime.  We note that  the analysis presented in the main text focused only on the equilibrium properties of the metapopulation  (e.g., species abundances and fraction of patches harboring the two species), which we measured in  the generation window $t \in [2000,2100]$. 

We recall that the relative abundance of, say, species 1 is obtained by adding the number of individuals of species 1 in each patch and then dividing the result by $K_{max}$ and $L^2$. This is the reason that the relative abundances at generation $t=0$ are very small (viz., 
$\langle  n_1 \rangle  = \langle n_2 \rangle   = 1/2L^2 $) and that a large relative abundance means that the species is spread all over the $L^2$ patches. We recall that here the single bracket notation stands for the average over patches only. Hence, figure  \ref{fig:S12} shows that  the plastic species 2 rapidly colonizes almost  the entire grid before it is completely or partially displaced by the nonplastic species 1. Of course, this result corroborates  the spatio-temporal  snapshots of the metapopulation dynamics illustrated in figure \ref{fig:S9}.  Interestingly, for patches distant from the seed patch, the actual competition scenario in our model is that where the nonplastic species 1 invades a resident population of plastic species 2. 

\begin{figure}
 \subfigure{\includegraphics[width=0.48\textwidth]{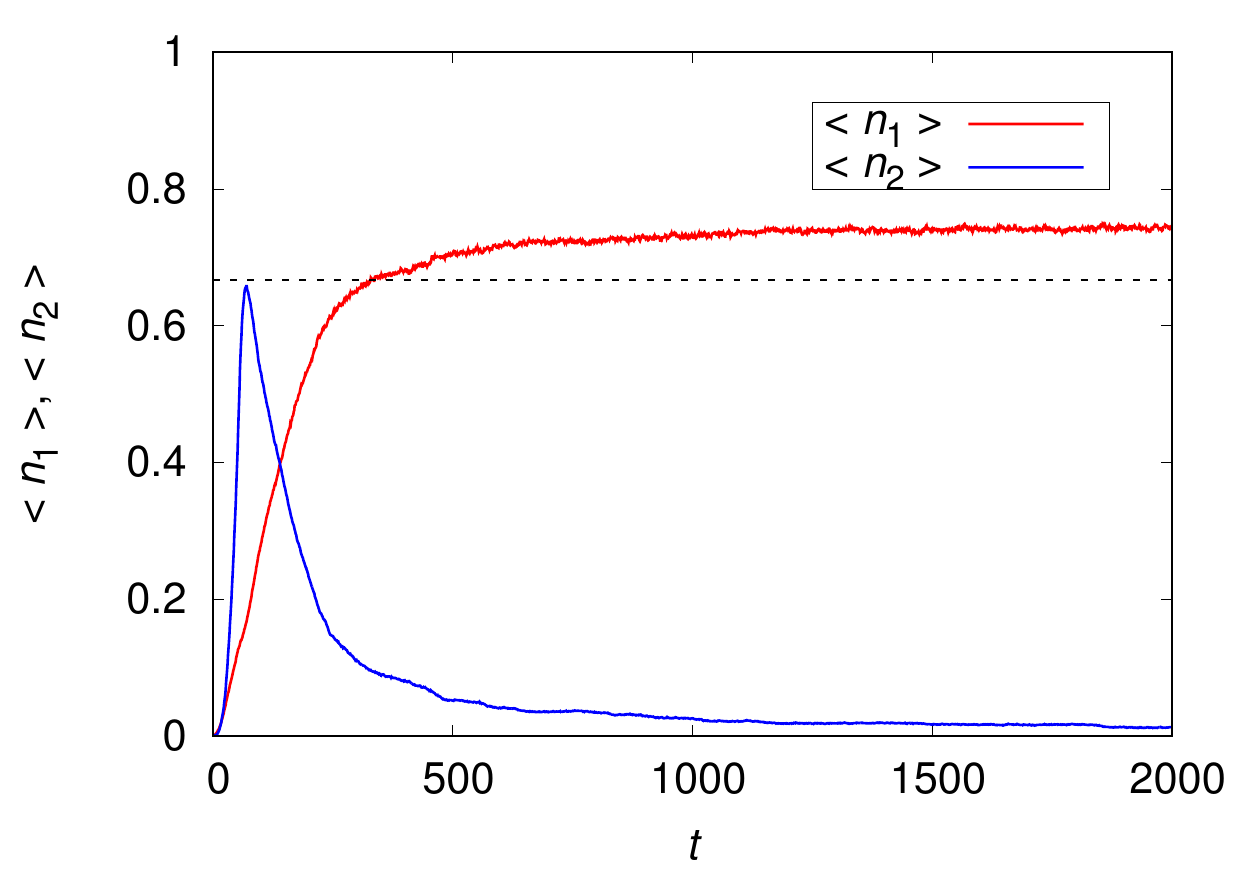}} 
 \subfigure{\includegraphics[width=0.48\textwidth]{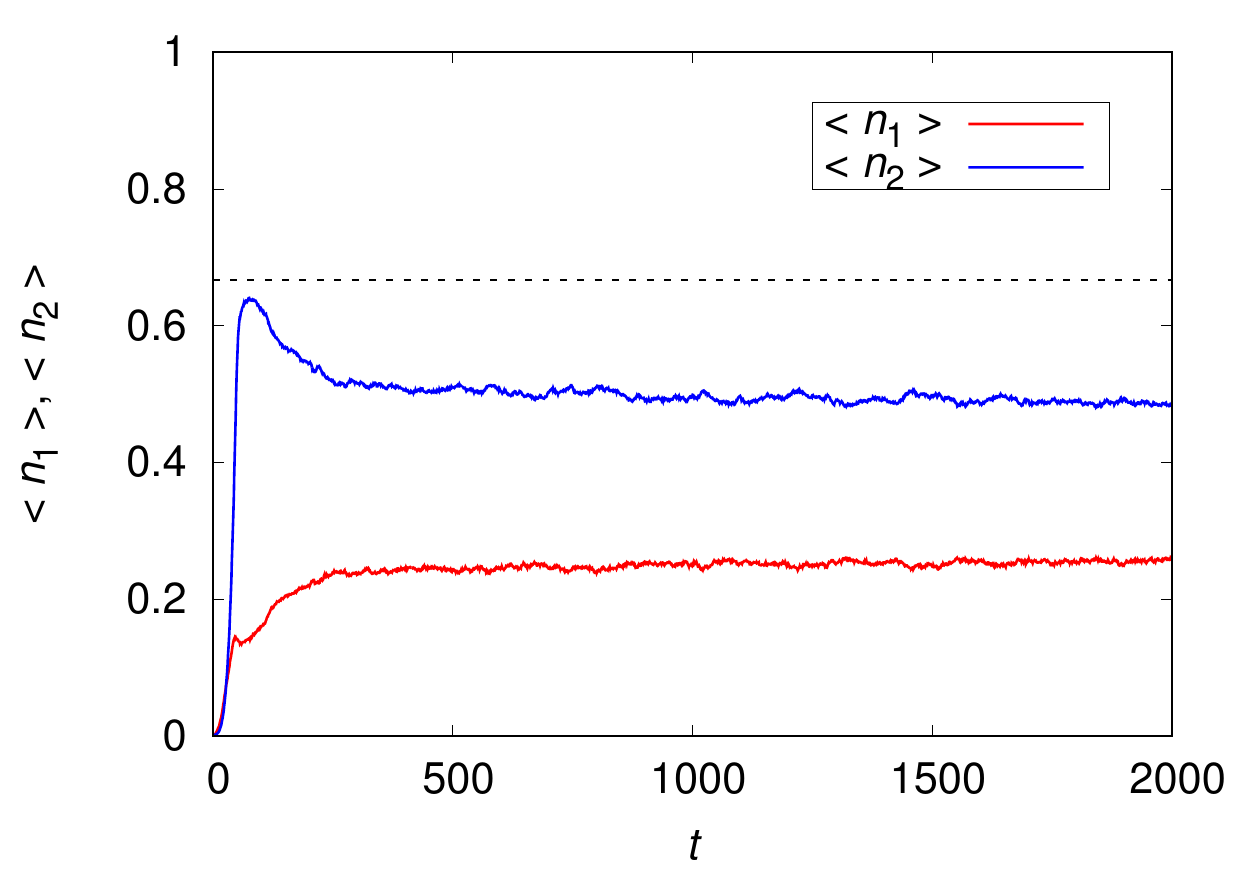}} 
\caption{Time evolution of the relative abundances of the nonplastic $\langle n_1 \rangle$ and plastic $ \langle n_2 \rangle  $ species for a single run that led to coexistence.  \textbf{Left Panel} $p_{mig} = 0.1$. \textbf{Right Panel:}  $p_{mig} = 0.3$. The parameters are $L=20$, $K_{max} = 100$, $c=0$,  $\sigma_e^2 =2$ and $\rho = 0$. The dashed horizontal line is $\langle n_1 \rangle = 1/a_{21} = 2/3$.
 }  
\label{fig:S12}  
\end{figure}

\section{Effect of the grid size}\label{sec:5}

Figures \ref{fig:S13} and \ref{fig:S14} show that the grid size has practically no influence on the two-species metapopulation dynamics, except for very small $L$. Hence the linear grid size $L=20$  used throughout the  paper gives a good approximation to the limit of an infinitely large grid.

\begin{figure}
 \subfigure{\includegraphics[width=0.48\textwidth]{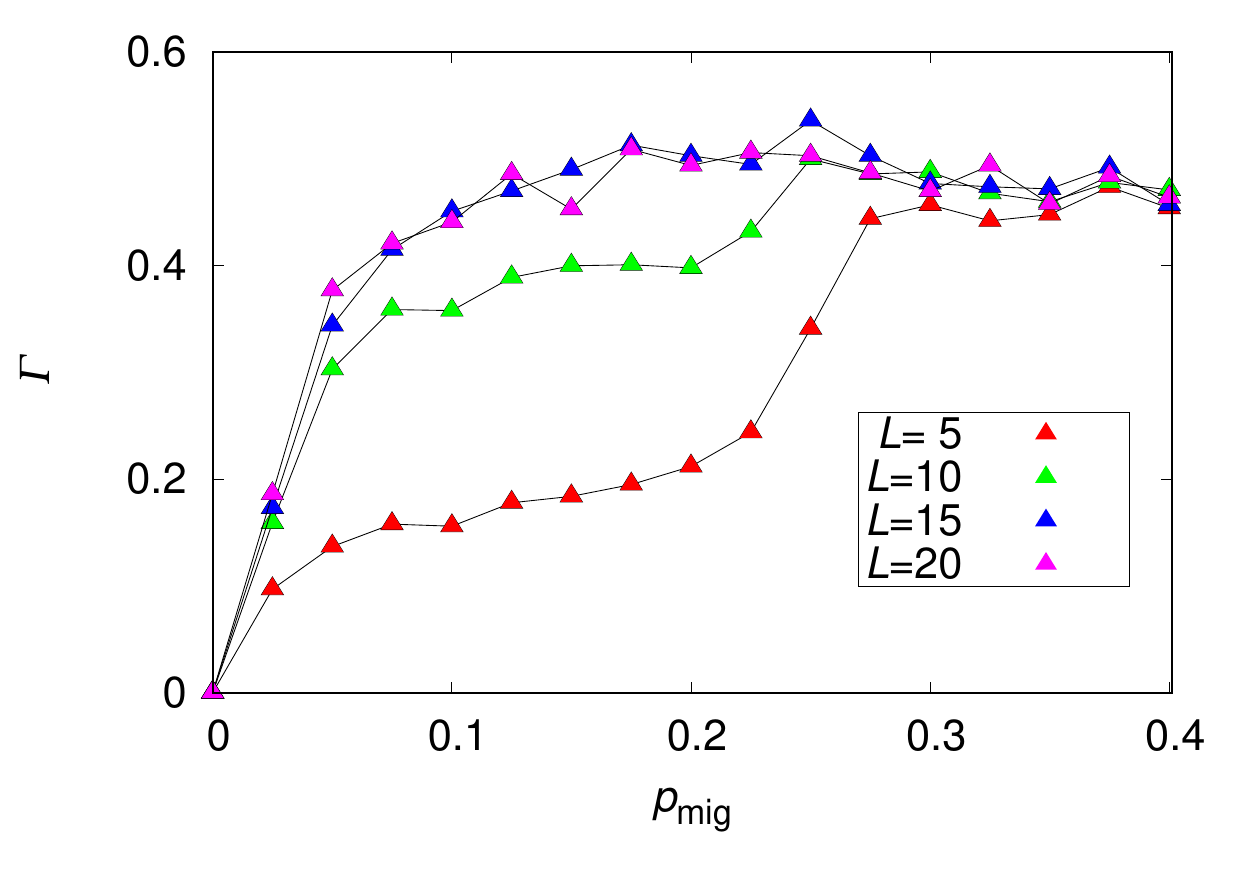}} 
 \subfigure{\includegraphics[width=0.48\textwidth]{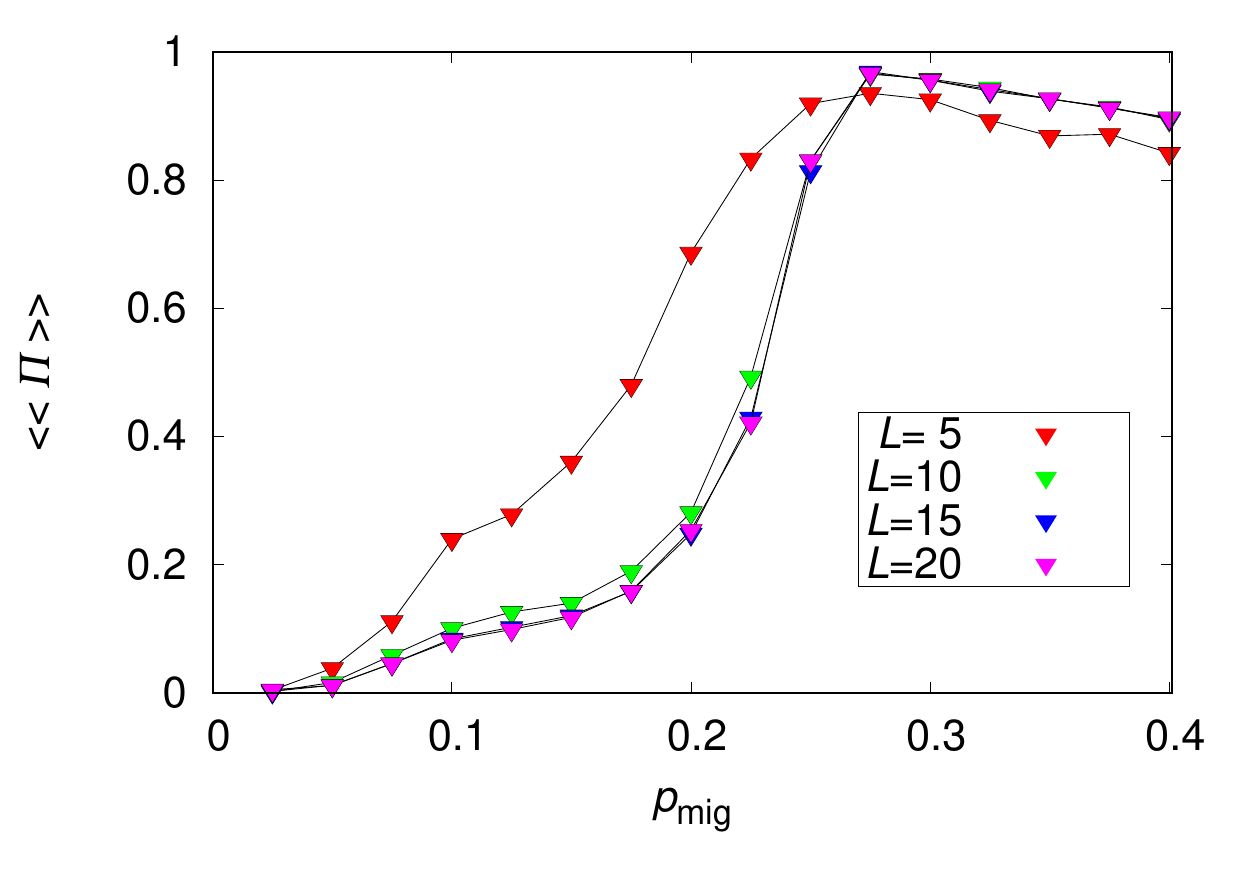}} 
\caption{Influence of  the migration probability  on species coexistence for grids of linear size $L=5, 10, 15$ and $20$, as indicated.   \textbf{Left Panel:} Fraction of runs that led to species coexistence. \textbf{Right Panel:} Fraction of patches where there is species coexistence. The parameters are $K_{max} = 100$, $c=0$,  $\sigma_e^2 =2$ and $\rho = 0$. The lines connecting the symbols are guides to the eye.
 }  
\label{fig:S13}  
\end{figure}

\begin{figure}
 \subfigure{\includegraphics[width=0.48\textwidth]{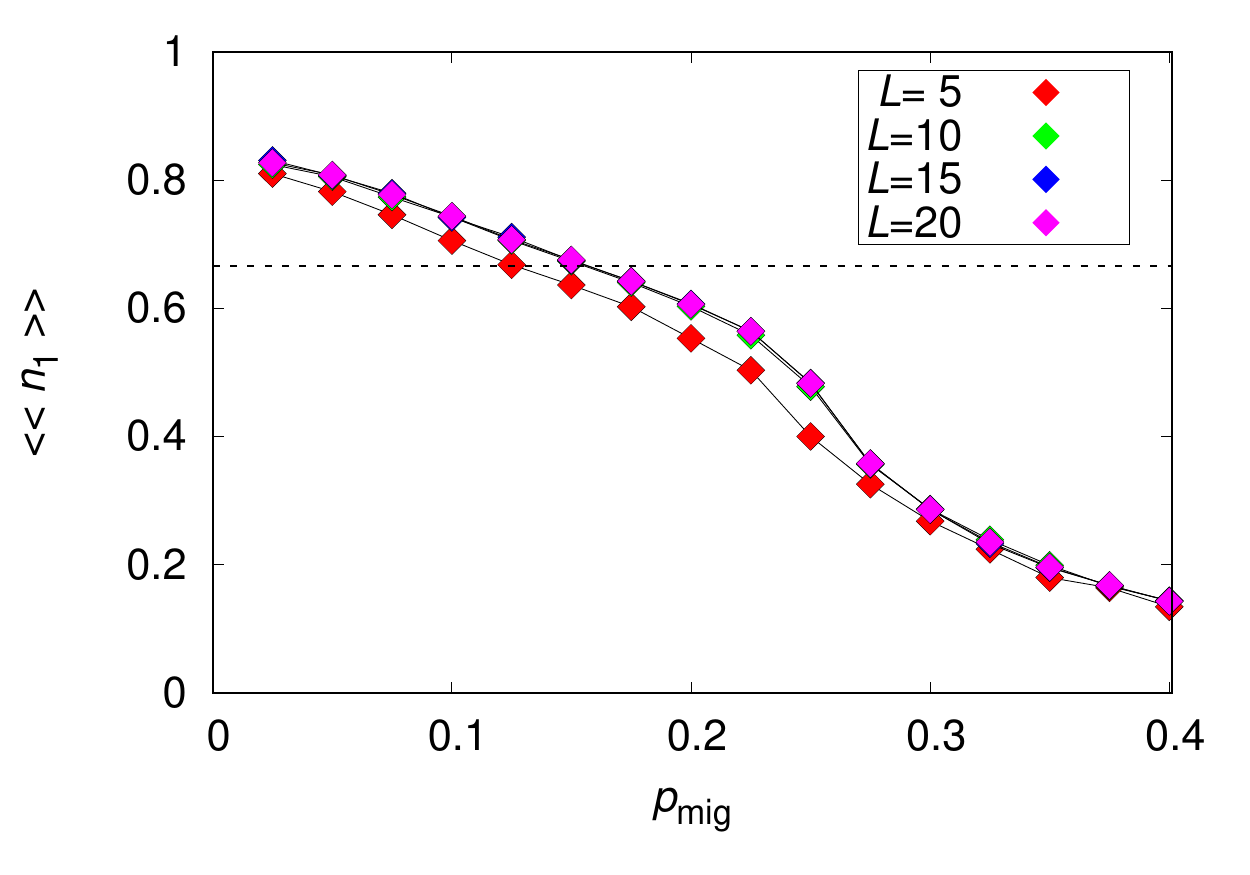}} 
 \subfigure{\includegraphics[width=0.48\textwidth]{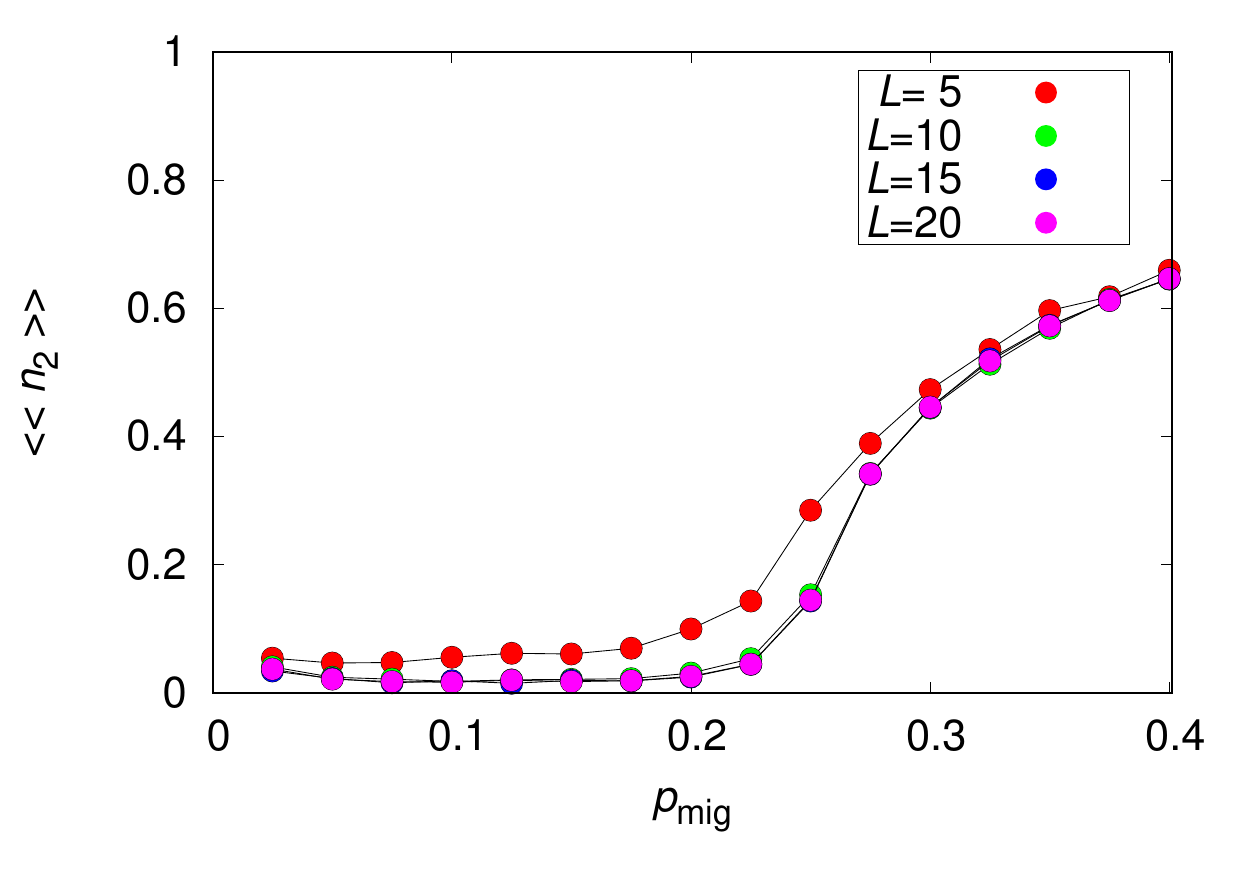}} 
\caption{Influence of  the migration probability  on the  mean  relative abundances for grids of linear size $L=5, 10, 15$ and $20$, as indicated. \textbf{Left Panel:} Nonplastic species 1. \textbf{Right Panel:} Plastic species 2.  The parameters are $K_{max} = 100$, $c=0$,  $\sigma_e^2 =2$ and $\rho = 0$. The lines connecting the symbols are guides to the eye. The dashed horizontal line is $\langle \langle  n_1 \rangle \rangle = 1/a_{21} = 2/3$.}
\label{fig:S14}  
\end{figure}

The coexistence observed for $p_{mig}$ such that $\langle \langle  n_1 \rangle \rangle > 1/a_{21} = 2/3$ is due to the existence of  patches with extreme environment values  that cannot be colonized by the nonplastic species 1. We refer to it as accidental coexistence. The odds that these extreme values appear increase with the number of patches, which explains the dependence on $L$  of the probability $\Gamma$ of finding the two species in the metapopulation at equilibrium for small $p_{mig}$.  In this line, we find that decreasing  $L$ with fixed $\sigma_e^2$  is equivalent to  decreasing  $\sigma_e^2$  with  $L$ fixed. More pointedly,  we find that $\Gamma$ is a function of the parameters combination $\sigma_e^2 L^{0.8}$ (data not shown). Now, given that accidental coexistence happens, i.e., that species 2 occupies an extreme patch, it will continuously send doomed migrants to that patch's  neighbors (see figure \ref{fig:S9}), which will count to the fraction of patches harboring the two species. If the total number of patches is small, the extreme patch and its neighbors can make a substantial contribution to $\langle \langle \Pi \rangle \rangle$, as observed in figure \ref{fig:S13}. We recall that the double brackets notation means an average over independent runs, over the last 100 generations of each run, and  over patches (in the case of species abundances).

\section{Effect of the  patch's carrying capacity}\label{sec:6}

Figures \ref{fig:S15} and \ref{fig:S16} show that the patch's carrying capacity $K_{max}$ has a significant influence on the probability of  coexistence but not on the mean  relative abundances of the two species, given coexistence.

\begin{figure}
 \subfigure{\includegraphics[width=0.48\textwidth]{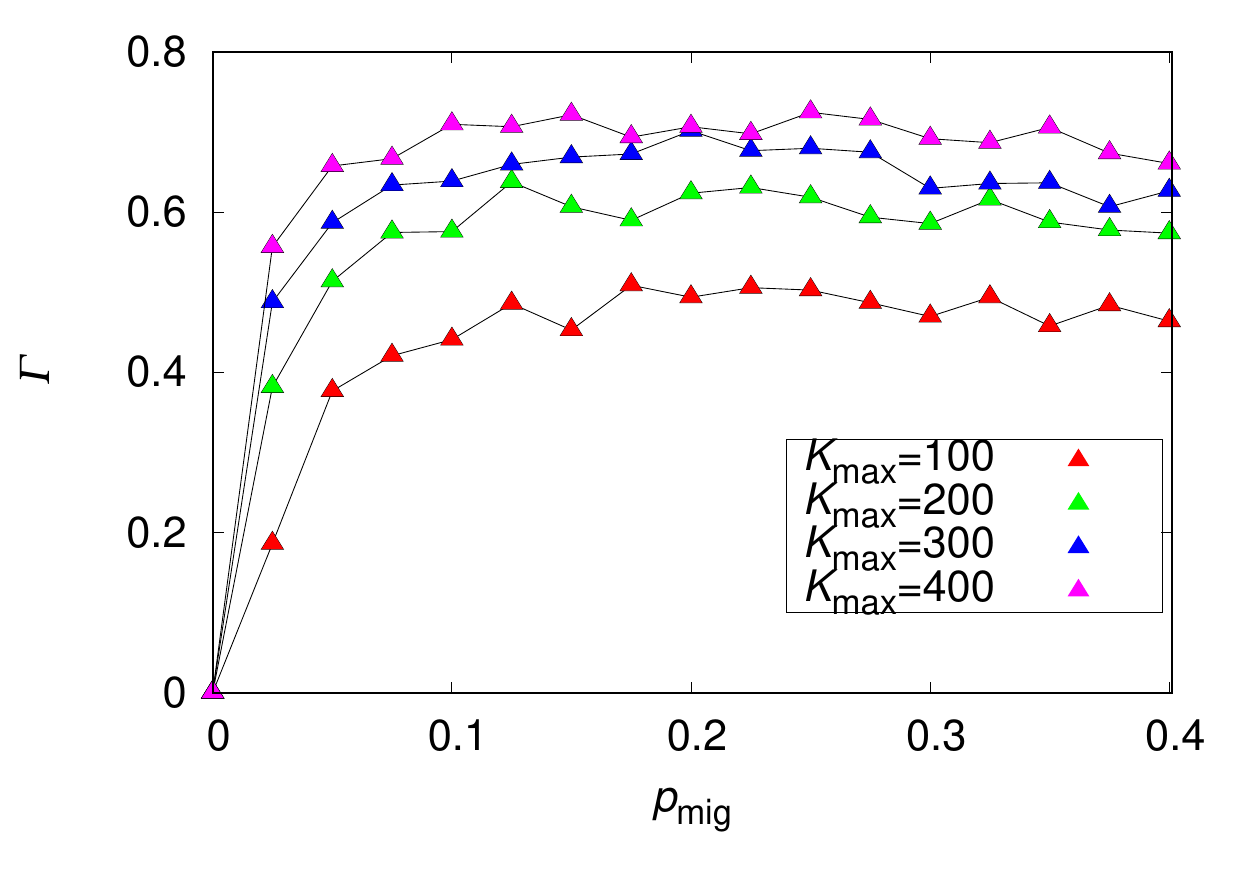}} 
 \subfigure{\includegraphics[width=0.48\textwidth]{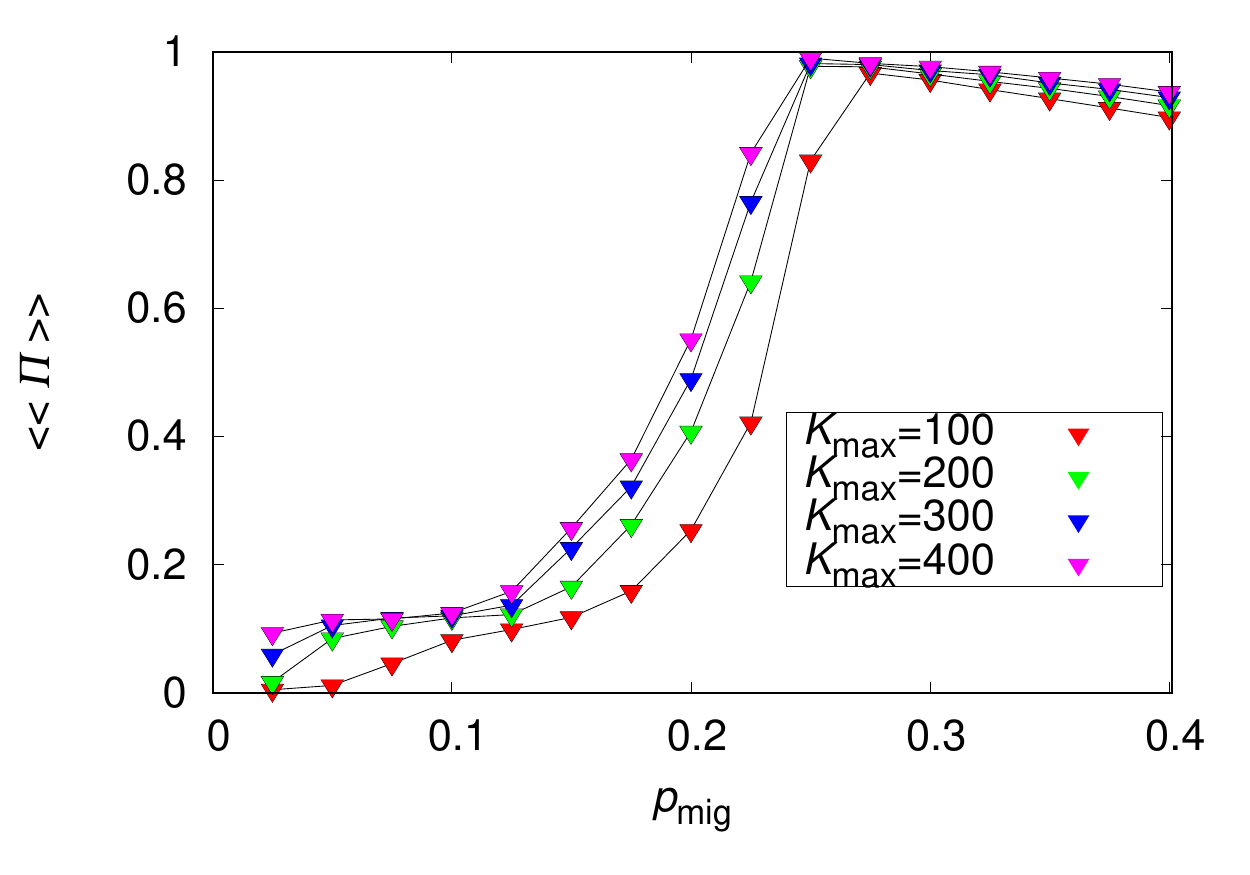}} 
\caption{Influence of  the migration probability  on species coexistence for patch's carrying capacity $K_{max} =100, 200, 300$ and $400$, as indicated. \textbf{Left Panel:} Fraction of runs that led to species coexistence. \textbf{Right Panel:} Fraction of patches where there is species coexistence. The parameters are $L= 20$, $c=0$,  $\sigma_e^2 =2$ and $\rho = 0$. The lines connecting the symbols are guides to the eye.}
\label{fig:S15}  
\end{figure}

As before, the explanation of the effect of $K_{max}$ on $\Gamma$ and $\langle \langle \Pi \rangle \rangle$ has to do with the existence of  patches with extreme environment values  that cannot be colonized by the nonplastic species 1. For $L=20$ it is almost certain that those patches exist in all runs. (We recall that we generate a new environment  for each run.) The key point is that  species 2 can be extinct before reaching those safe heavens, but the odds that this happens decreases with increasing $K_{max}$: the more individuals of species 2, the greater the odds that some of them will reach the extreme patches. In this sense, we expect that $\Gamma \to 1$ as $K_{max} \to \infty$ for $p_{mig} > 0$. Regarding the increase of $\langle \langle \Pi \rangle \rangle$ with increasing $K_{max}$, it can be explained by  the large number of doomed migrants (approximately $p_{mig} K_{max}$) that are  continuously  sent to the neighborhood of the extreme patches and that  are likely to reach way beyond  their nearest neighbors simply  because of their numerosity. 

\begin{figure}
 \subfigure{\includegraphics[width=0.48\textwidth]{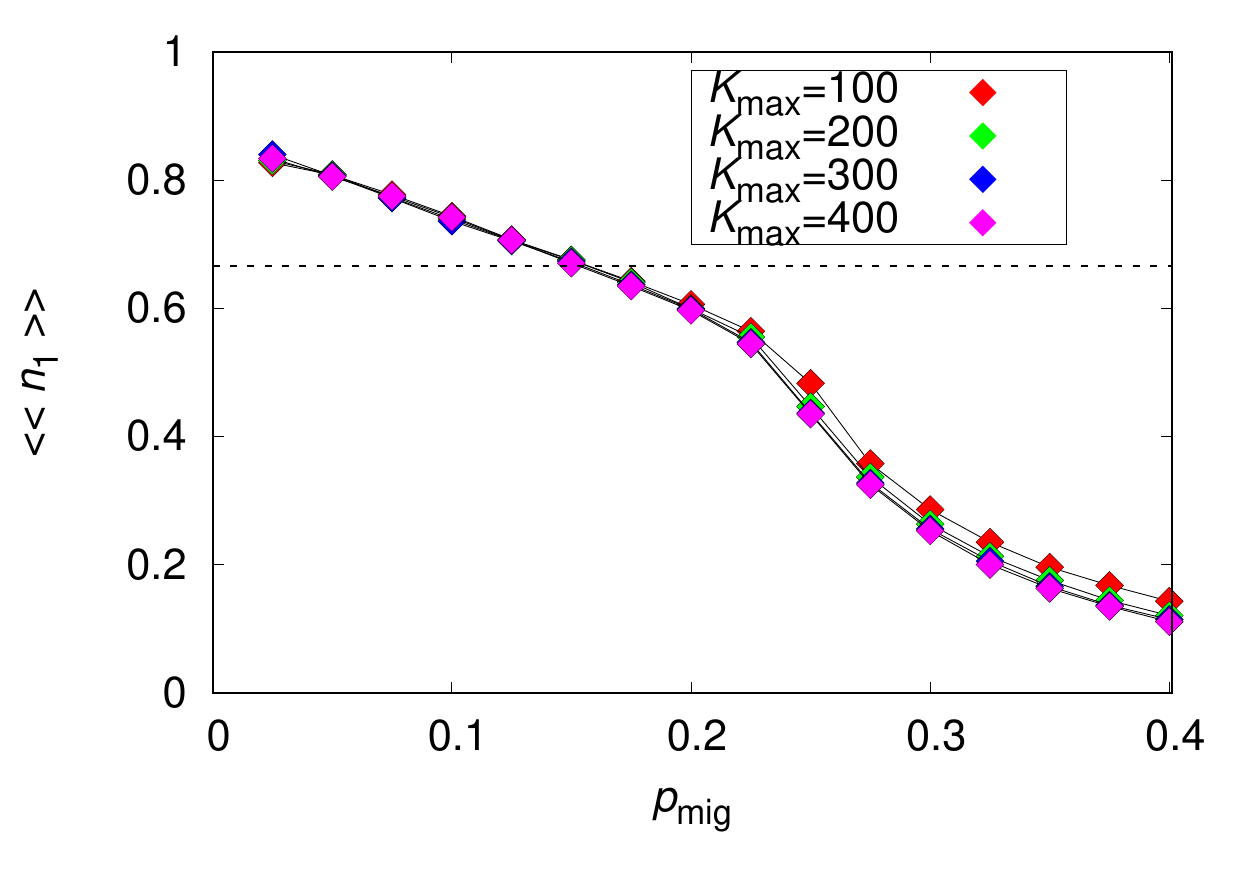}} 
 \subfigure{\includegraphics[width=0.48\textwidth]{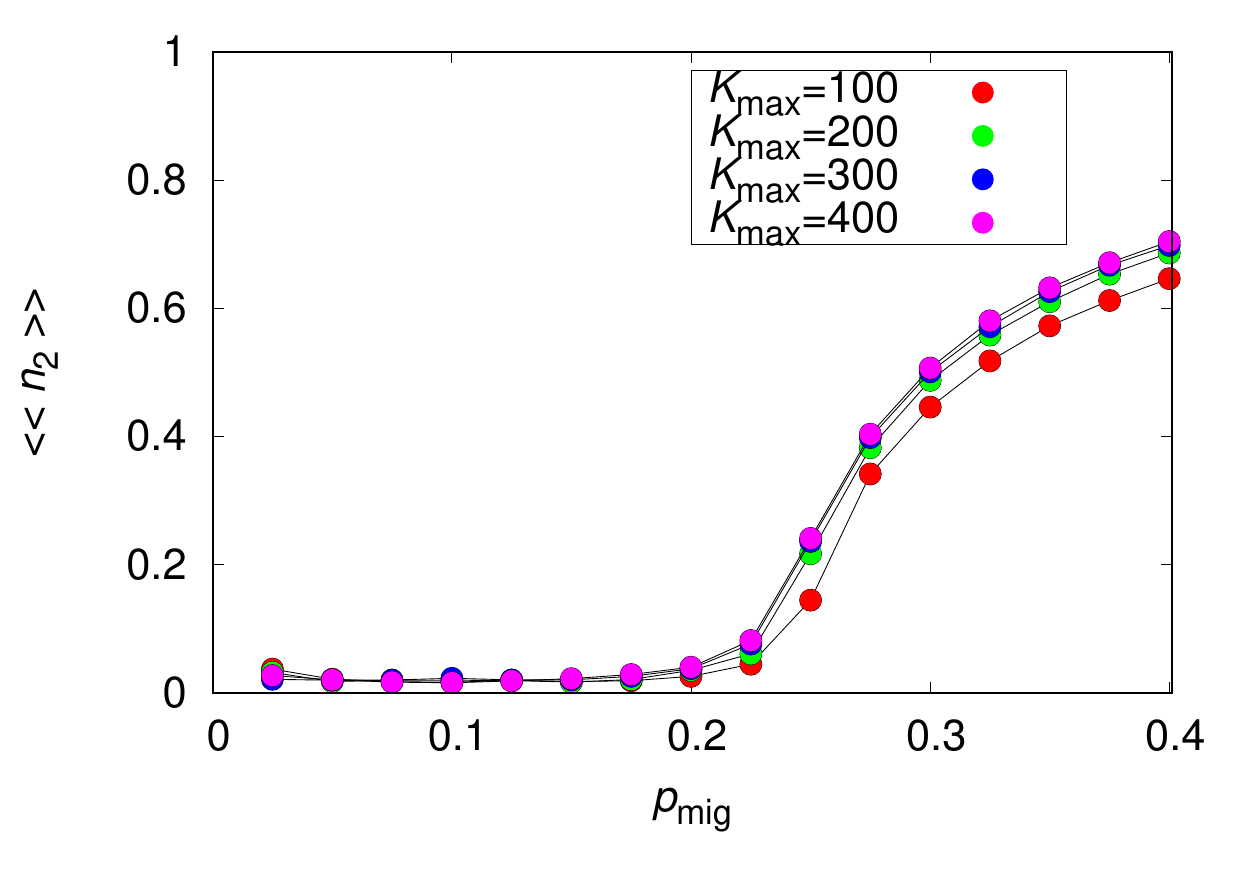}} 
\caption{Influence of  the migration probability  on the  mean  relative abundances for patch's carrying capacity $K_{max} =100, 200, 300$ and $400$, as indicated. \textbf{Left Panel:} Nonplastic species 1. \textbf{Right Panel:} Plastic species 2.  The parameters are $L = 20$, $c=0$,  $\sigma_e^2 =2$ and $\rho = 0$. The lines connecting the symbols are guides to the eye. The dashed horizontal line is $\langle \langle n_1 \rangle  \rangle = 1/a_{21} = 2/3$.}
\label{fig:S16}  
\end{figure}

In fact, figure \ref{fig:S16} confirms that, although certain, the   coexistence observed in the region for which $\langle \langle n_1 \rangle \rangle > 1/a_{21} = 2/3$ is largely irrelevant since  $\langle \langle n_2 \rangle \rangle \approx 0$ implies that species 2 is present in only a few patches. The rapid convergence to the asymptotic  relative abundances values with increasing $K_{max}$ indicates that the results for $K_{max}=100$ already offer a good approximation to the infinite population size limit.

\section{Effect of recombination}\label{sec:S7}

The results of the main text and of the previous sections of the Supplementary Material considered only asexually reproducing species, although we acknowledge that nearly all invasive species are sexual.  Our goal was to verify whether  plasticity  can offer advantage to a competitively inferior species in a scenario of mandatory  migration among heterogeneous patches. We expect  the mode of reproduction of the species to  play a minor role as compared with the migration rate and the environment heterogeneity, so we considered the asexual mode since it allows  optimizations of the code that greatly speed up the simulations. This speed up is necessary to study the equilibrium of the metapopulation dynamics as well as to carry out averages over a great number of  independent runs.  For instance, in the asexual  reproduction simulations  we  can ignore the loci and consider only the sum of the allele values  over all loci. Nevertheless, here we offer a brief analysis (single runs) of the effect of recombination on the competition between the plastic and nonplastic species.  As expected, the results  show that the conclusions drawn from the study of asexual populations hold true for sexual  reproducing  individuals as well. 

The sexual reproduction scheme is implemented as follows. Mating within each patch is random for each species: mating pairs are formed by randomly drawing the survivors with replacement, and each pair produces a single
offspring. The number of survivors of species 1 and 2 in patch $i$ are $N_{1i}$ and $N_{2i}$, respectively. The process is repeated $N'_{1i} $ and $N'_{2i}$ times according to equations (\ref{N1}) and (\ref{N2}) of the main text.  During reproduction, parental gametes mutate at  rate $u=5/1000$  per-locus for both  nonplastic and plastic loci. 
Following mutation, each allelic value  changes according to the continuum-of-alleles model  with Gaussian mutations with mean 0 and variance $1/100$ added to the existing allelic value, as in the asexual reproduction mode. Recombination of parental chromosomes occurs with probability one. The cross-over operator picks one internal point at random to form one haploid gamete by taking all alleles from one chromosome up to the crossover point, and all alleles from the other chromosome beyond the crossover point. The cross-over point is chosen so as to guarantee that the offspring is always a recombinant.

Figure \ref{fig:S17} shows the time dependence of the relative abundances of each species for four independent runs and different migration probabilities. The results indicate that recombination favors  the nonplastic species against the plastic species. For instance, the abundances of the plastic species at $t=2000$ are slightly lower  than the abundances shown in figure  \ref{fig:S14} for the asexual population at equilibrium. The opposite holds for  the  abundances of the nonplastic species.  In principle, this is expected since recombination reduces the time the nonplastic lineages need to adapt to their environments and so adaptation can happen before the lineage is disrupted by migration.  In fact, comparing the evolution of the sexual and asexual populations for $p_{mig}=0.1$ (see figure \ref{fig:S12}) we observe that the sexual population reaches the equilibrium situation, which implies the colonization of almost all patches, much faster than the asexual population. However, for $p_{mig}=0.3$ the sexual population takes a very long time to reach equilibrium. We recall that $p_{mig} = 0.3$ is close to  the point of transition between the regimes of accidental and  robust (i.e., non-accidental) coexistence, so perhaps the difficulty to reach equilibrium  is reminiscent of  the critical slowing down phenomenon of  phase transitions.

The important point is that the competition between the plastic and nonplastic sexual  species exhibits the same two regimes of coexistence observed in the study of the asexual  species, viz., a regime of accidental coexistence that happens for small migration probabilities and is due to the existence of  patches that have too extreme environments for the nonplastic species,  and a regime of robust coexistence that happens for large migration probabilities, where  the species coexist  within most patches.

\begin{figure}
 \subfigure{\includegraphics[width=0.4\textwidth]{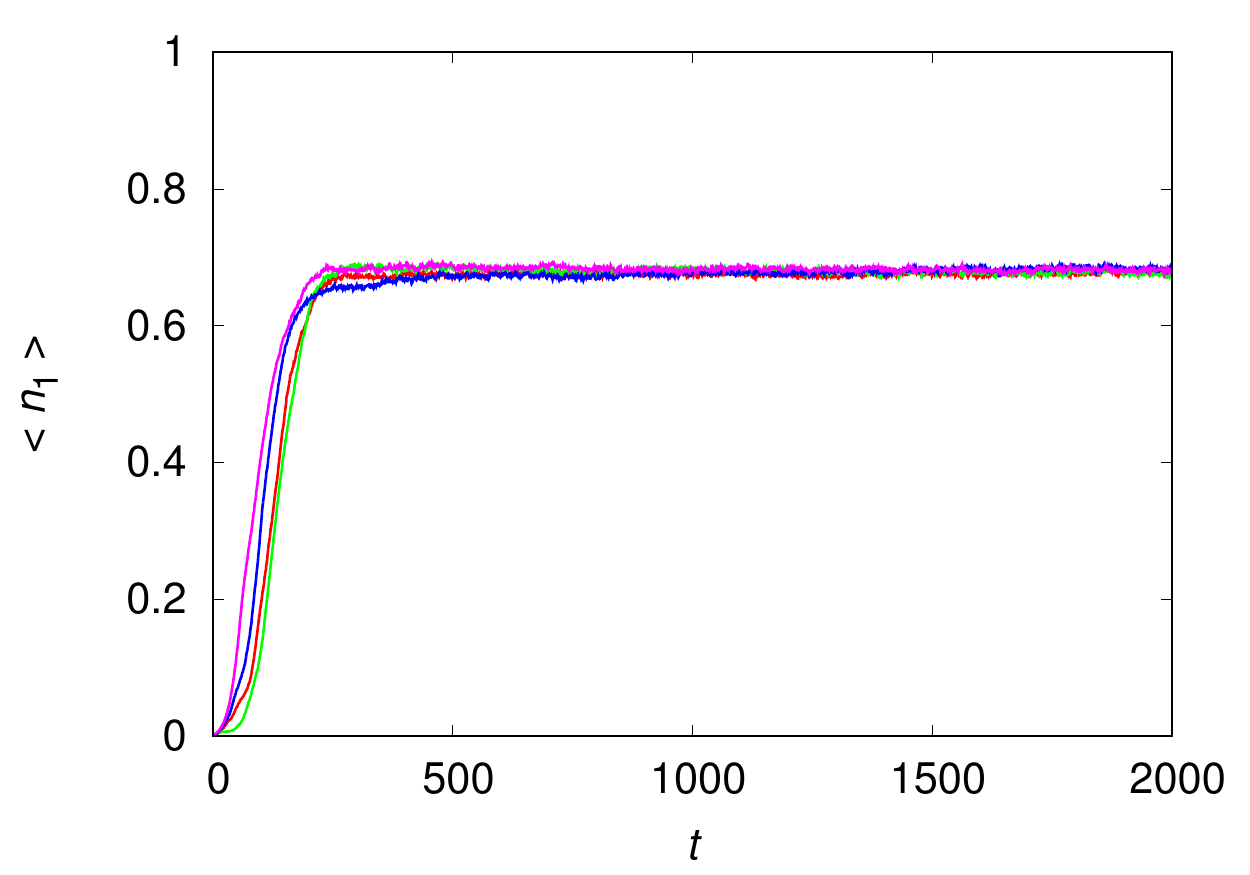}} 
 \subfigure{\includegraphics[width=0.4\textwidth]{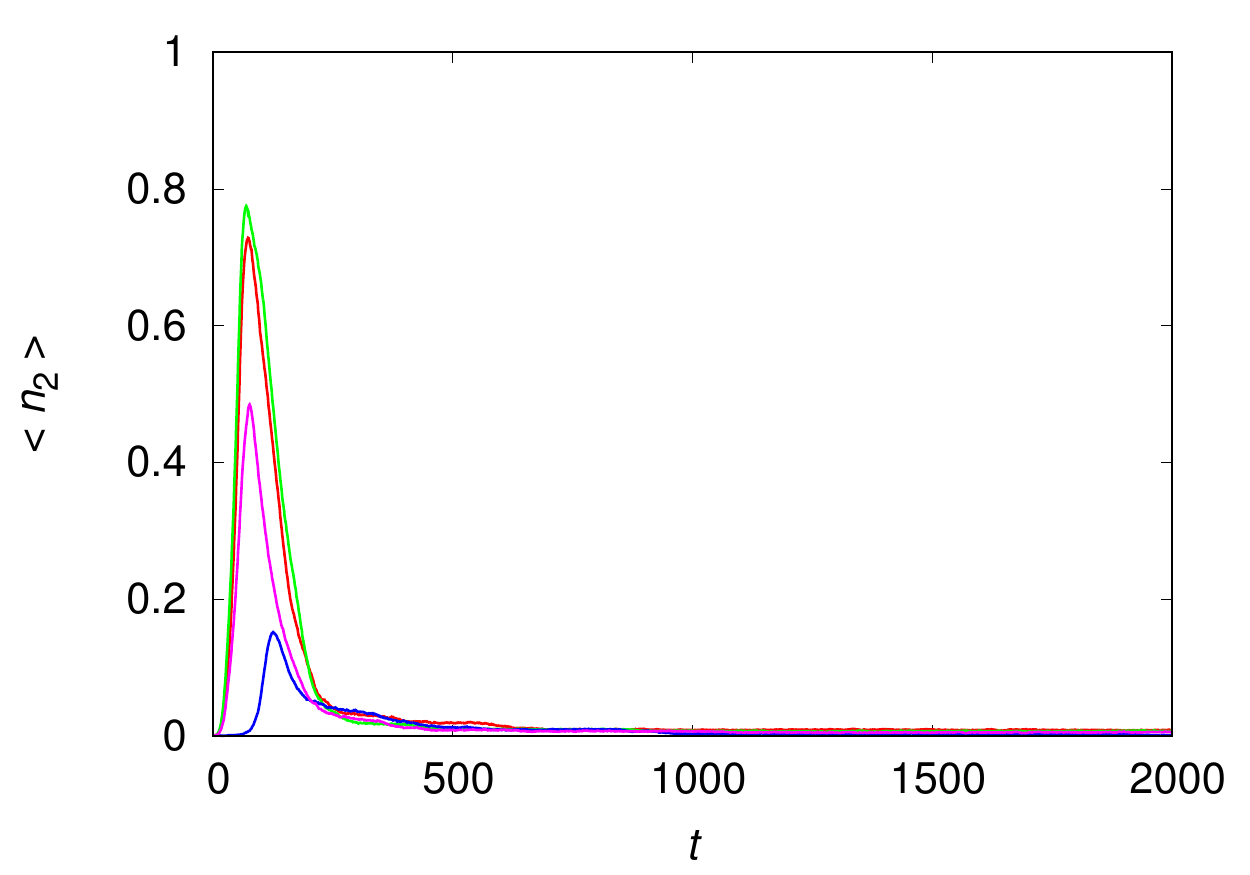}}  
 \subfigure{\includegraphics[width=0.4\textwidth]{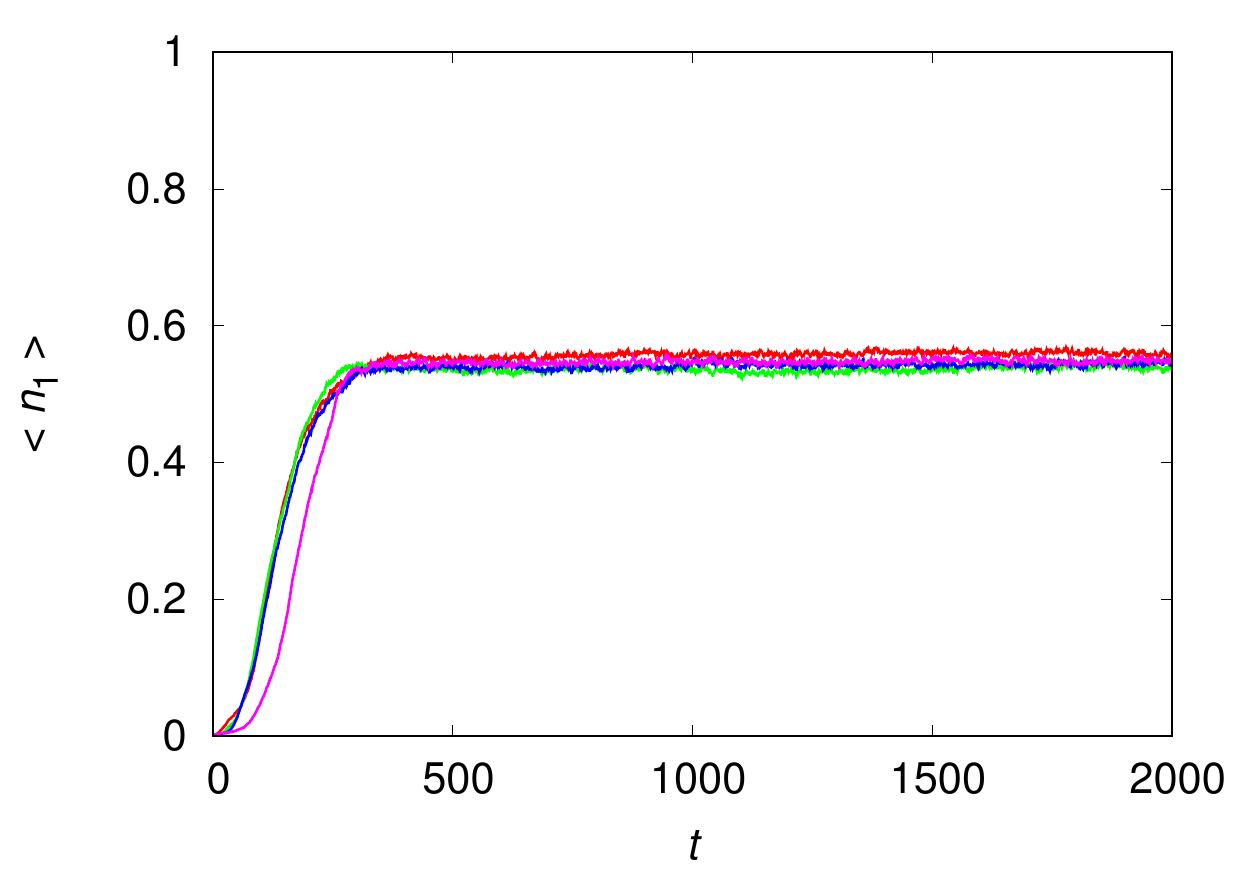}} 
 \subfigure{\includegraphics[width=0.4\textwidth]{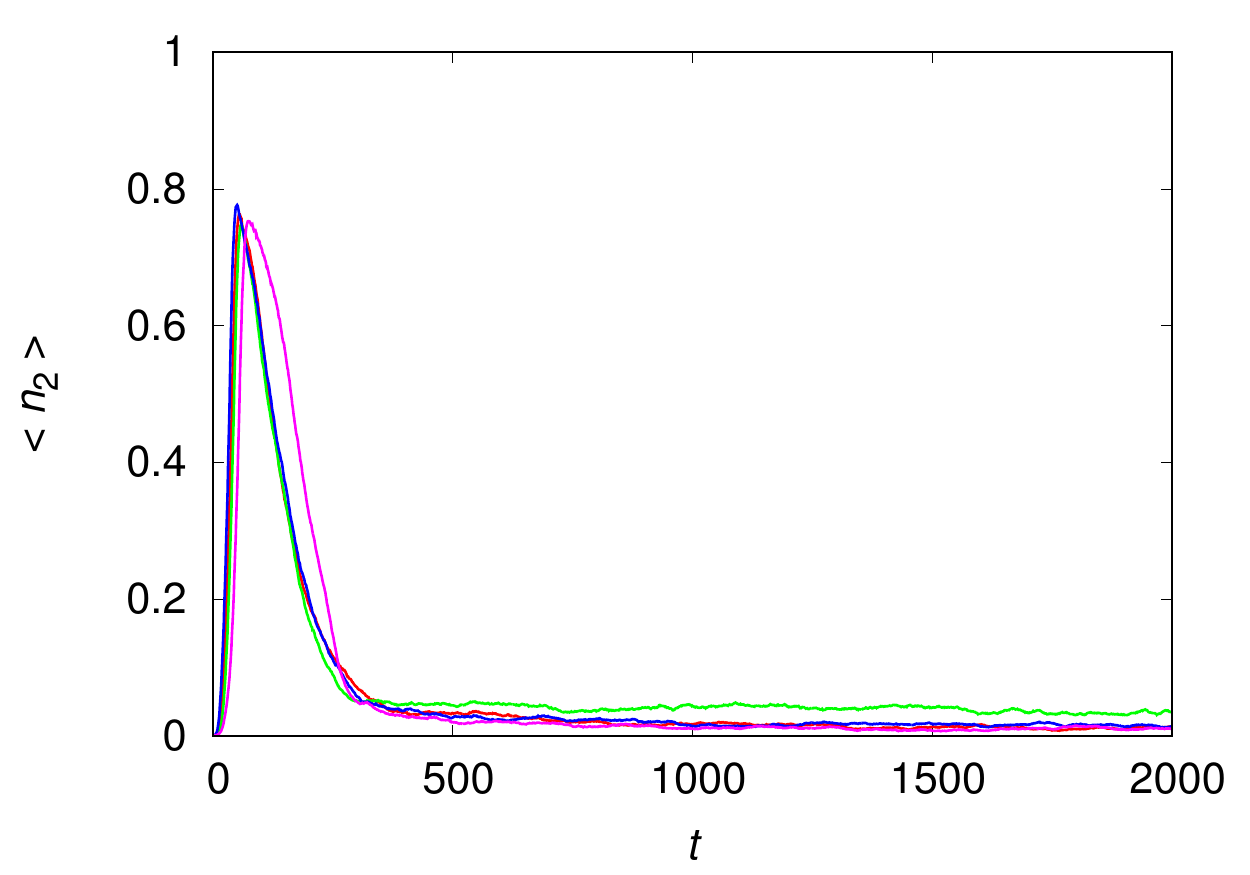}} 
 \subfigure{\includegraphics[width=0.4\textwidth]{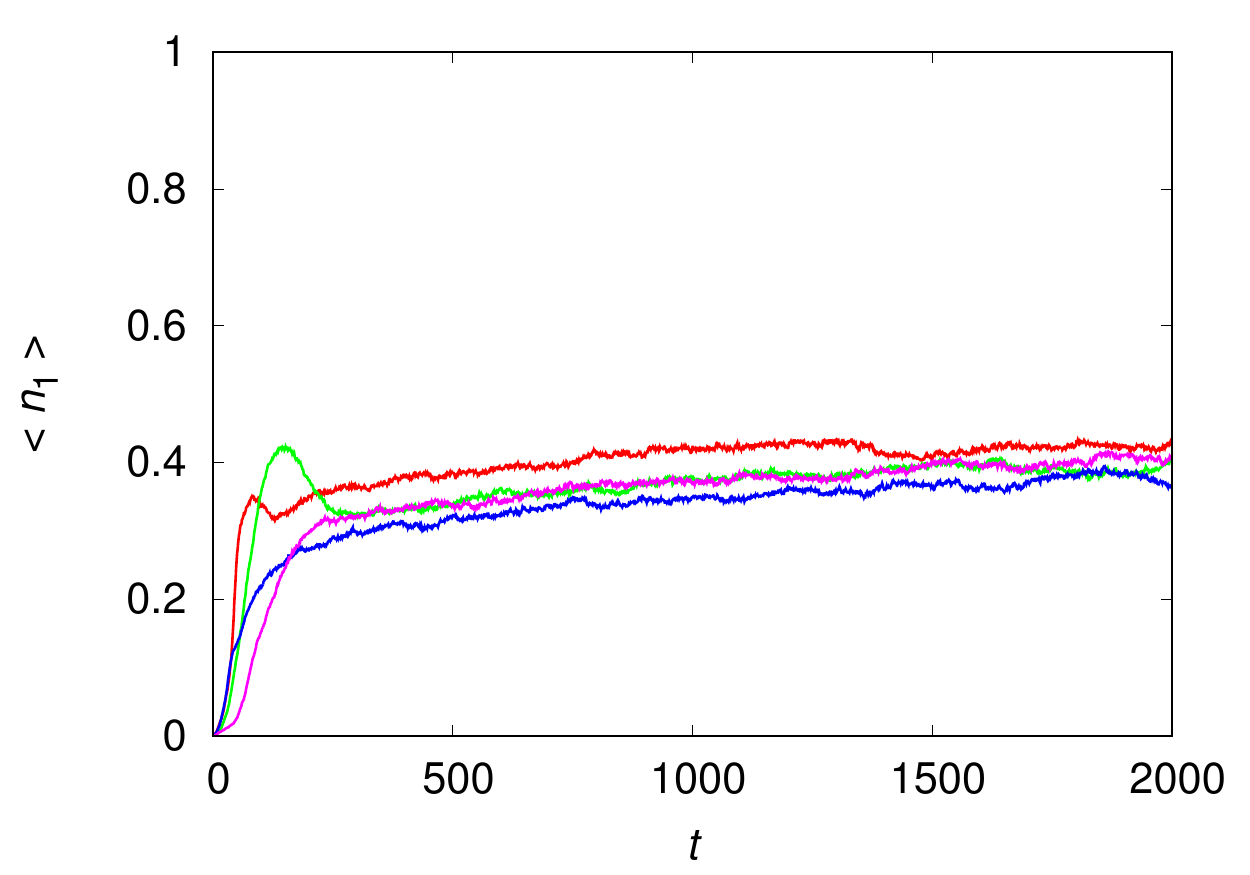}} 
  \subfigure{\includegraphics[width=0.4\textwidth]{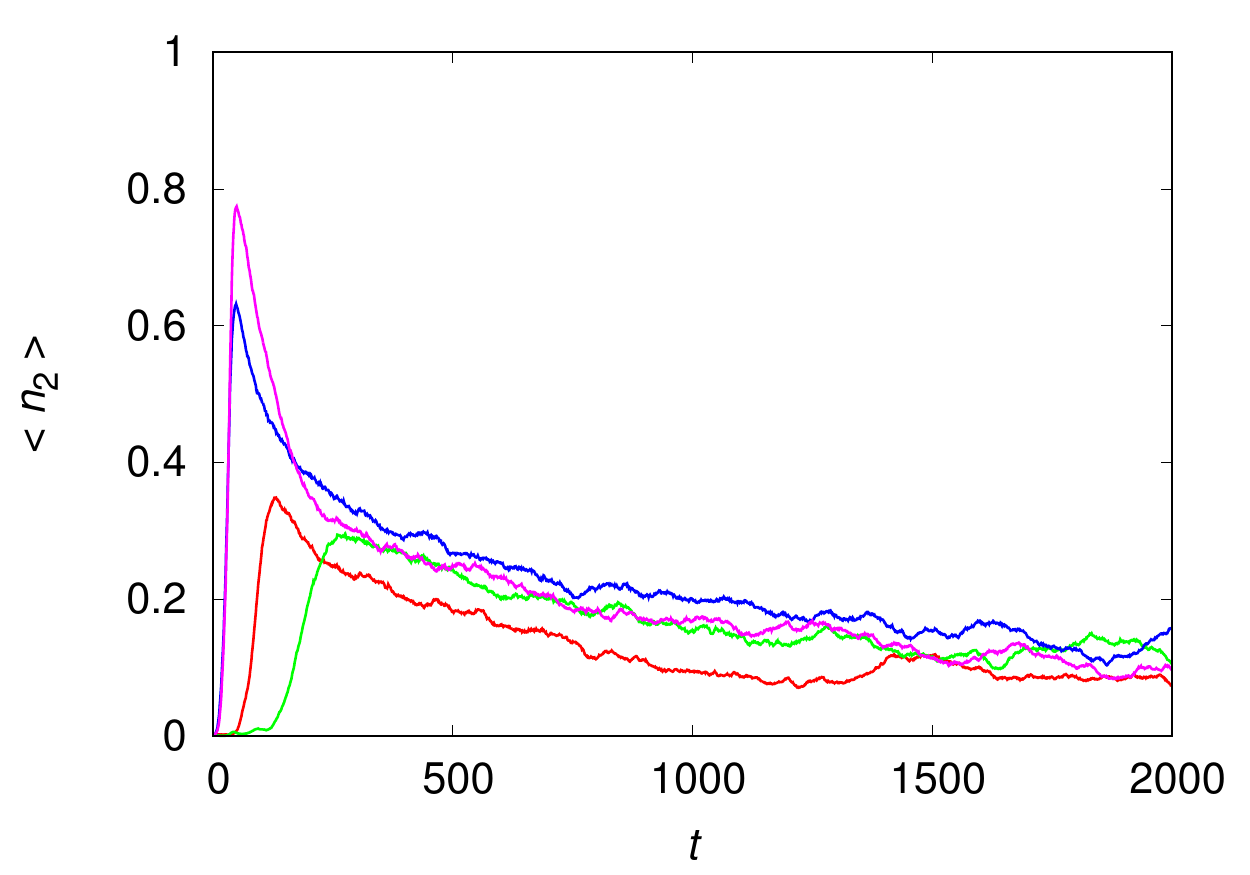}} 
   \subfigure{\includegraphics[width=0.4\textwidth]{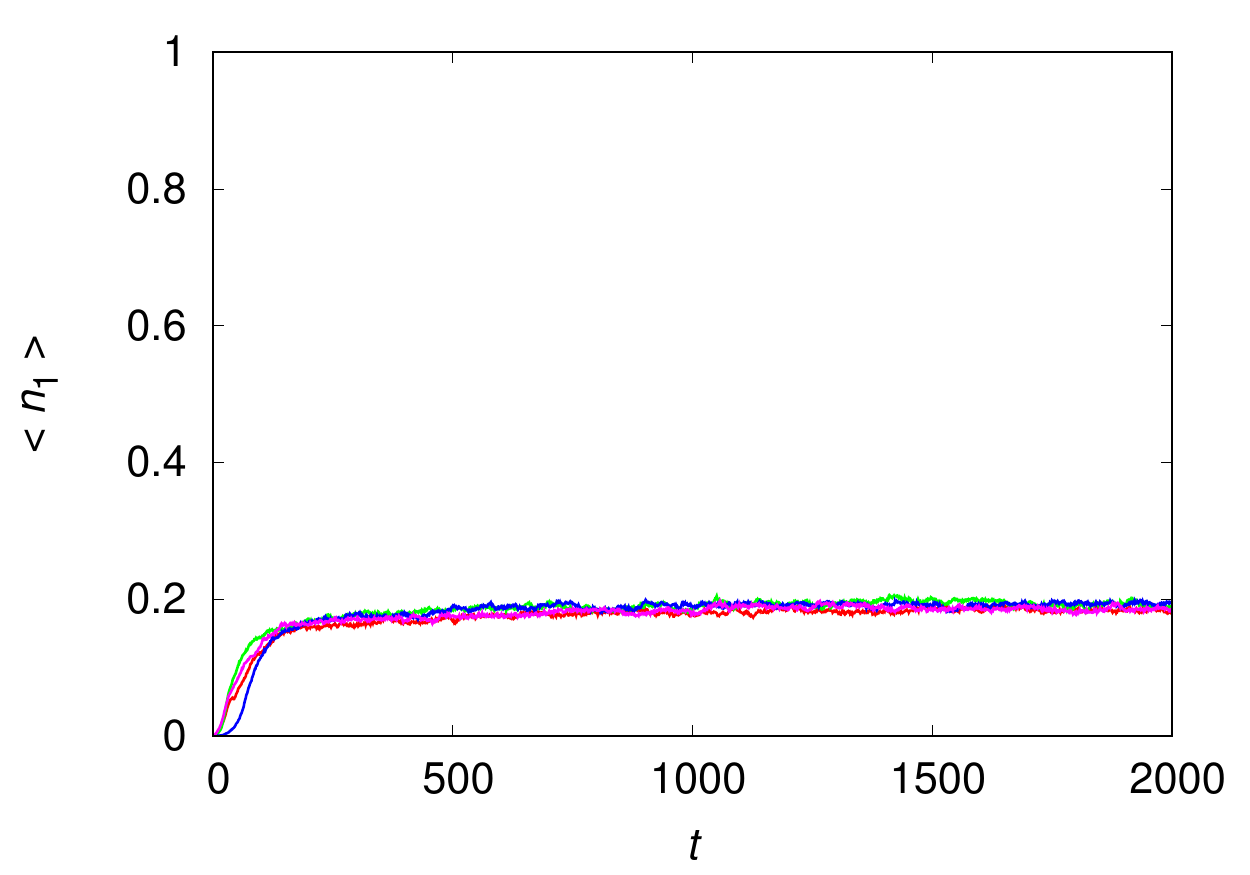}} 
    \subfigure{\includegraphics[width=0.4\textwidth]{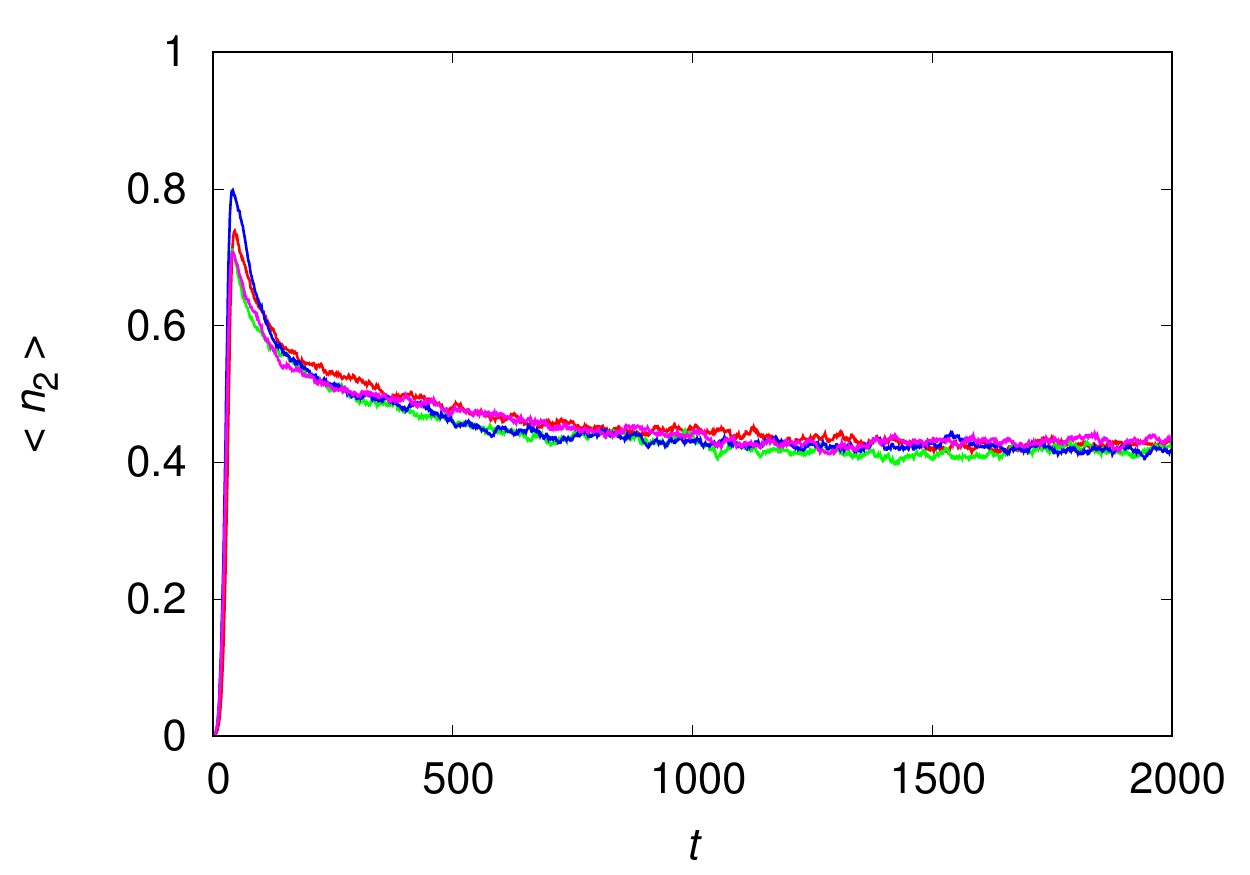}} 
\caption{Time evolution of sexual  species for four independent runs that led to coexistence. The rows show the species abundances for (top to bottom)  $p_{mig} = 0.1, 0.2, 0.3$ and $0.4$. The first column shows $\langle n_1 \rangle$  and the second column shows  $\langle n_2 \rangle$.   The parameters are $L= 20$, $K_{max}=100$, $c=0$,  $\sigma_e^2 =2$ and $\rho = 0$.
}  
\label{fig:S17}  
\end{figure}

\clearpage

\section{Variability of equilibrium variables among runs}\label{sec:S8}

Since in the main text we have focused only on the values of the metapopulation equilibrium  variables   averaged over independent runs, here we offer scatter plots of those variables to  assess   their variability among runs. We recall that, following the convention used in the Supplementary Material, the single bracket notation stands for an average over the last 100 generations of a run (as well as over patches, in the case of the relative abundances) and so we use single brackets  when considering properties of single runs.  In particular, in the scatter plots of figures \ref{fig:S18}, \ref{fig:S19} and \ref{fig:S20}, each symbol  represents the measures $\langle \Pi \rangle$, $\langle n_1 \rangle$ and $\langle n_2 \rangle$ for a single run that  led to coexistence. The total number of symbols (i.e., runs that led to coexistence)  in each scatter plot  is 1000. 

The great variability of $\langle \Pi \rangle$ for $\sigma_e^2 =1$ (and, to a minor extent,  for $\sigma_e^2 =0.5$) is due to the crossover between the regime of accidental coexistence, where coexistence happens in the neighborhood of a few patches characterized by extreme environment values (see figure \ref{fig:S9}) and the  regime of non-accidental coexistence, where coexistence happens within most patches of the grid (see figure \ref{fig:S10}). Figure \ref{fig:4} of the main text shows the rapid variation of $\langle \Pi \rangle$ in the region of the transition between these two regimes. However, in  the regime of non-accidental coexistence, the existence of which was the main thrust of our paper, the variability  among runs    is small for all measures considered. We note that the uncertainty on the estimate of the mean value of a variable, say $\langle \langle \Pi \rangle \rangle$, is proportional to the inverse square root of the number of independent runs, hence the claim that the error bars are smaller than the size of the symbols used to represent the data.

\begin{figure}[H]
 \subfigure{\includegraphics[width=0.48\textwidth]{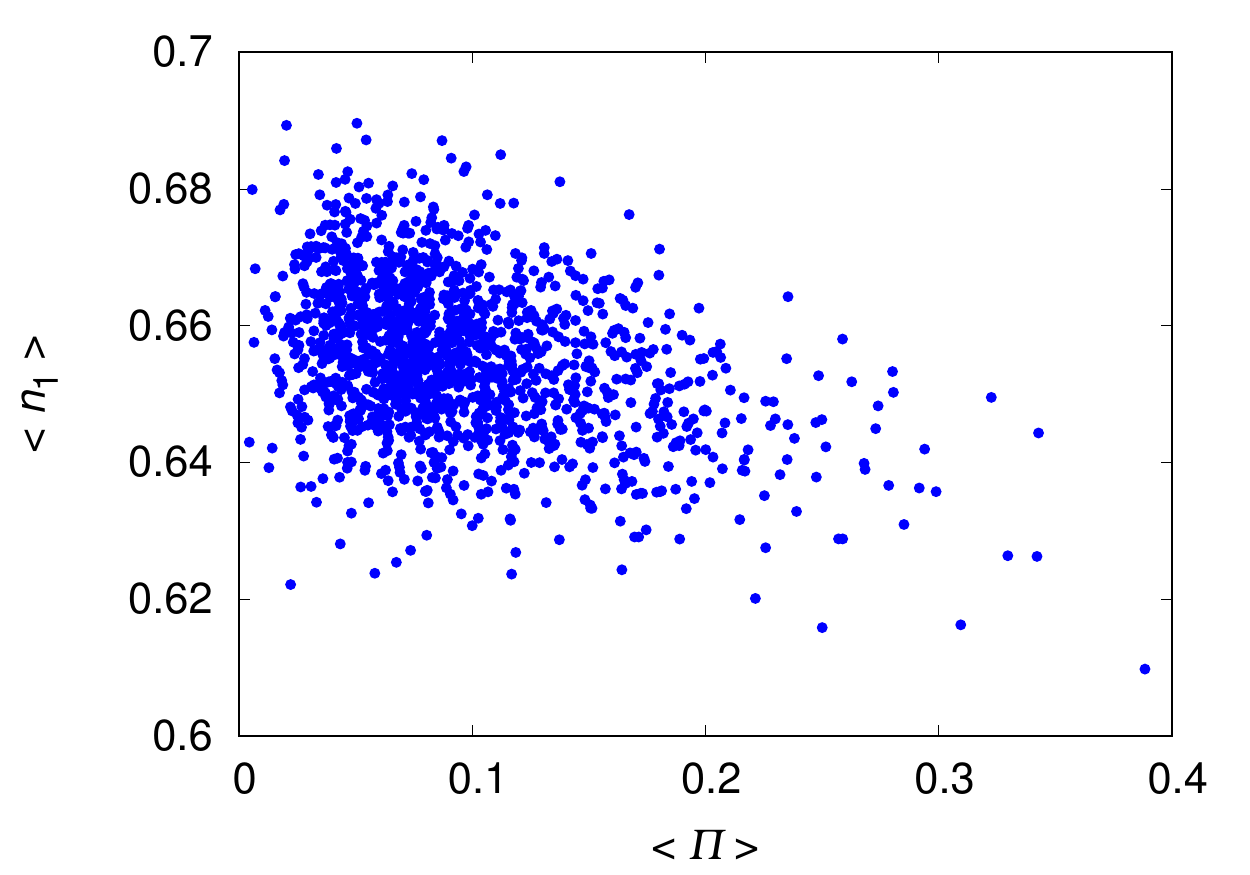}} 
 \subfigure{\includegraphics[width=0.48\textwidth]{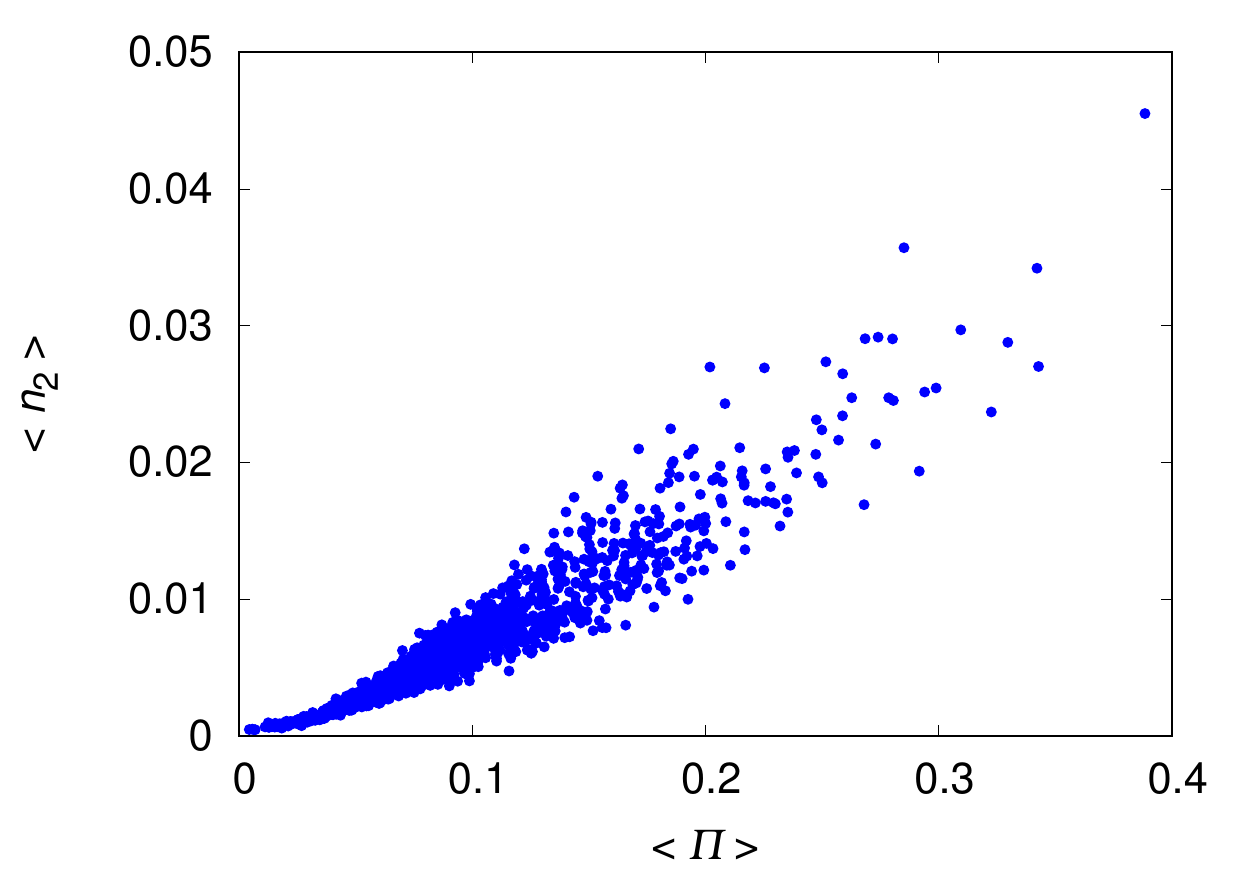}} 
\caption{Scatter plots of equilibrium properties of the metapopulation for $\sigma_e^2 =0.5$.  \textbf{Left Panel:} Fraction of patches where the two species coexist and relative abundance of species 1. \textbf{Right Panel:} Fraction of patches where the two species coexist and relative abundance of species 2. The parameters are $L= 20$, $K_{max}=100$, $p_{mig} = 0.3$, $c=0$ and $\rho = 0$.}
\label{fig:S18}  
\end{figure}

\clearpage

\begin{figure}[t]
 \subfigure{\includegraphics[width=0.48\textwidth]{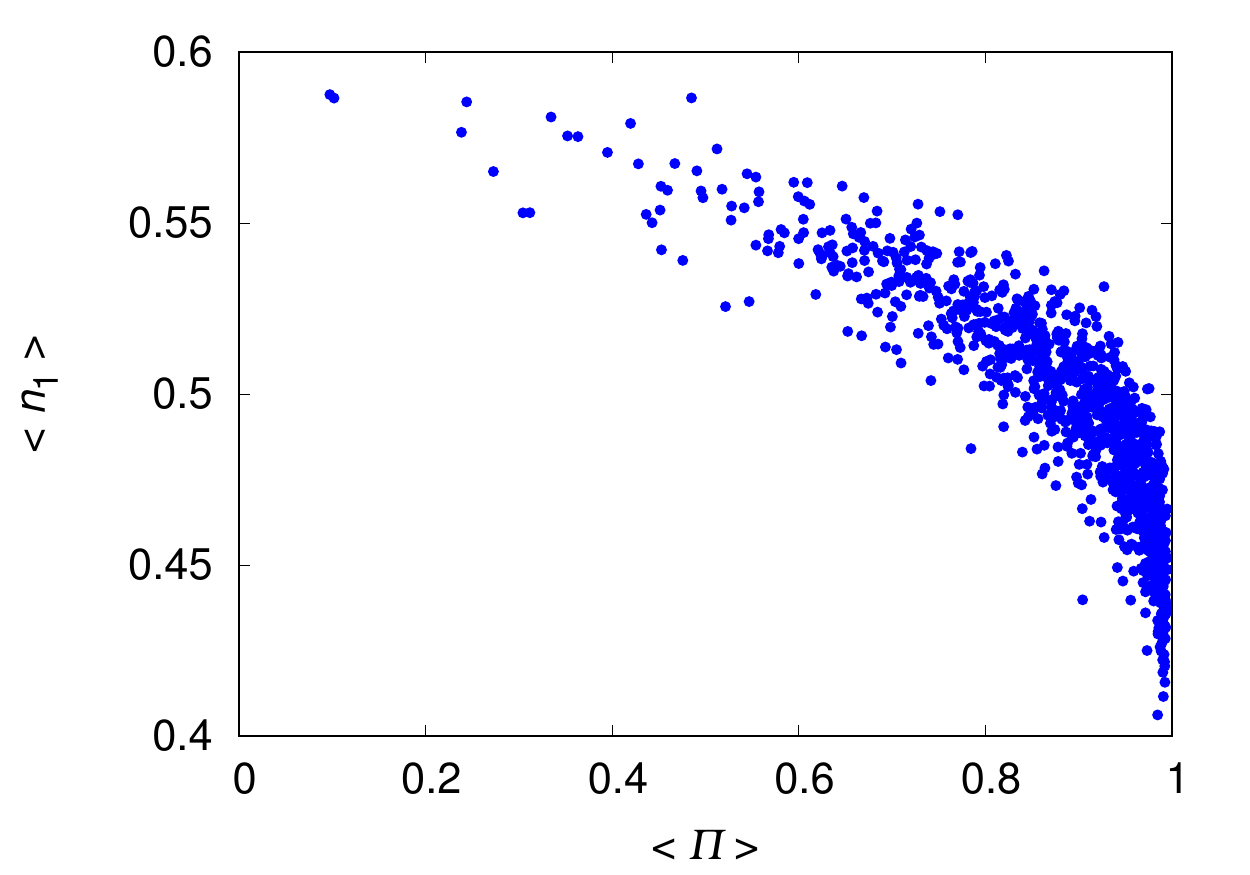}} 
 \subfigure{\includegraphics[width=0.48\textwidth]{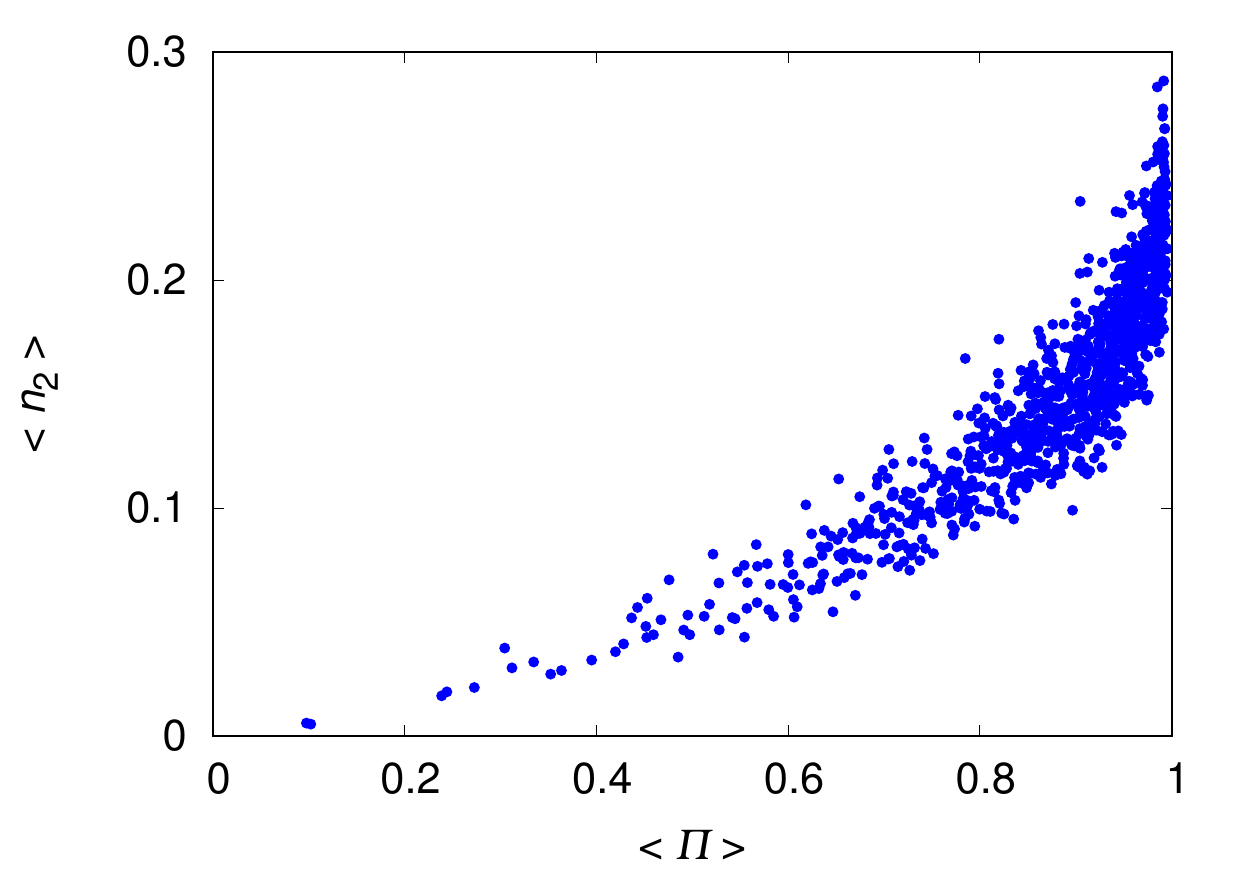}} 
\caption{Scatter plots of equilibrium properties of the metapopulation for $\sigma_e^2 =1$.  \textbf{Left Panel:} Fraction of patches where the two species coexist and relative abundance of species 1. \textbf{Right Panel:} Fraction of patches where the two species coexist and relative abundance of species 2. The parameters are $L= 20$, $K_{max}=100$, $p_{mig} = 0.3$, $c=0$ and $\rho = 0$.}
\label{fig:S19}  
\end{figure}

\begin{figure}[t]
 \subfigure{\includegraphics[width=0.48\textwidth]{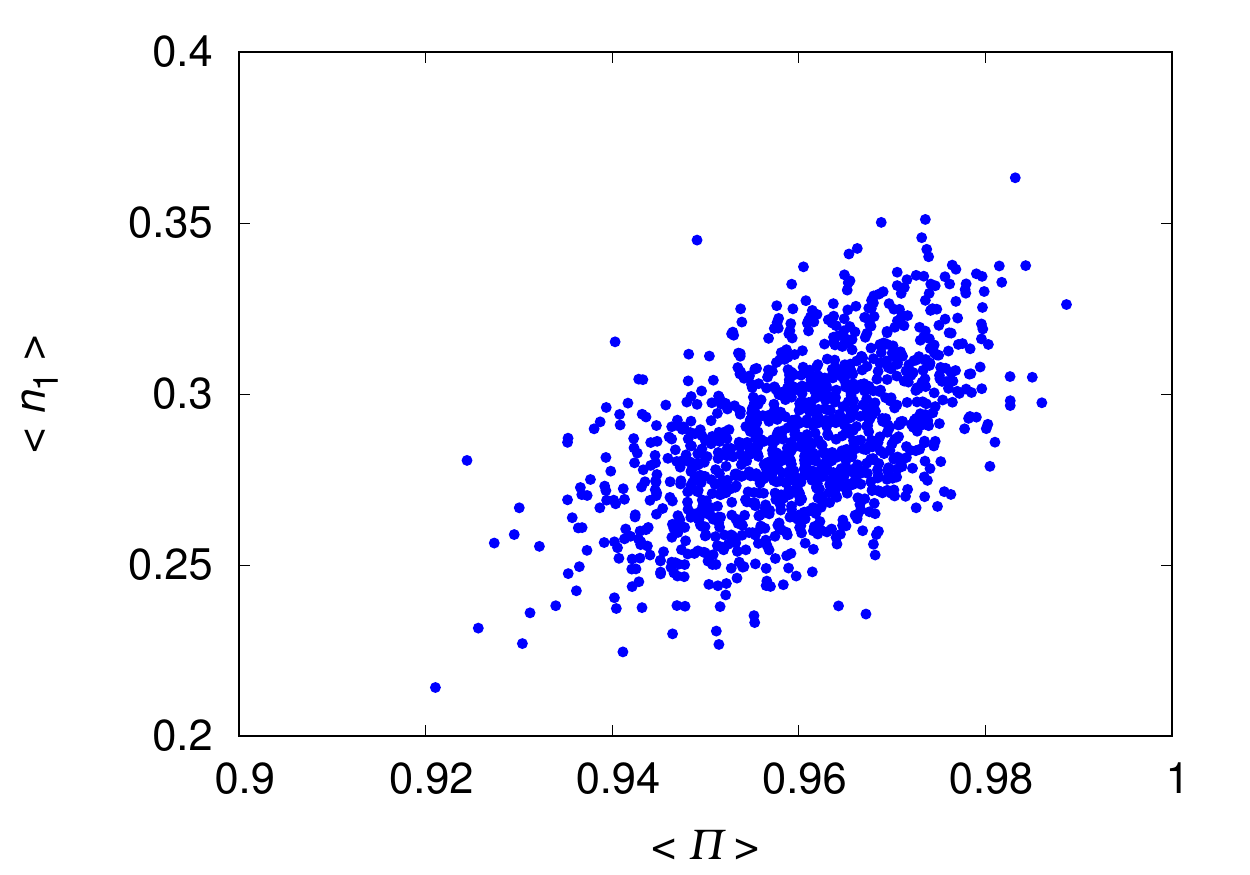}} 
 \subfigure{\includegraphics[width=0.48\textwidth]{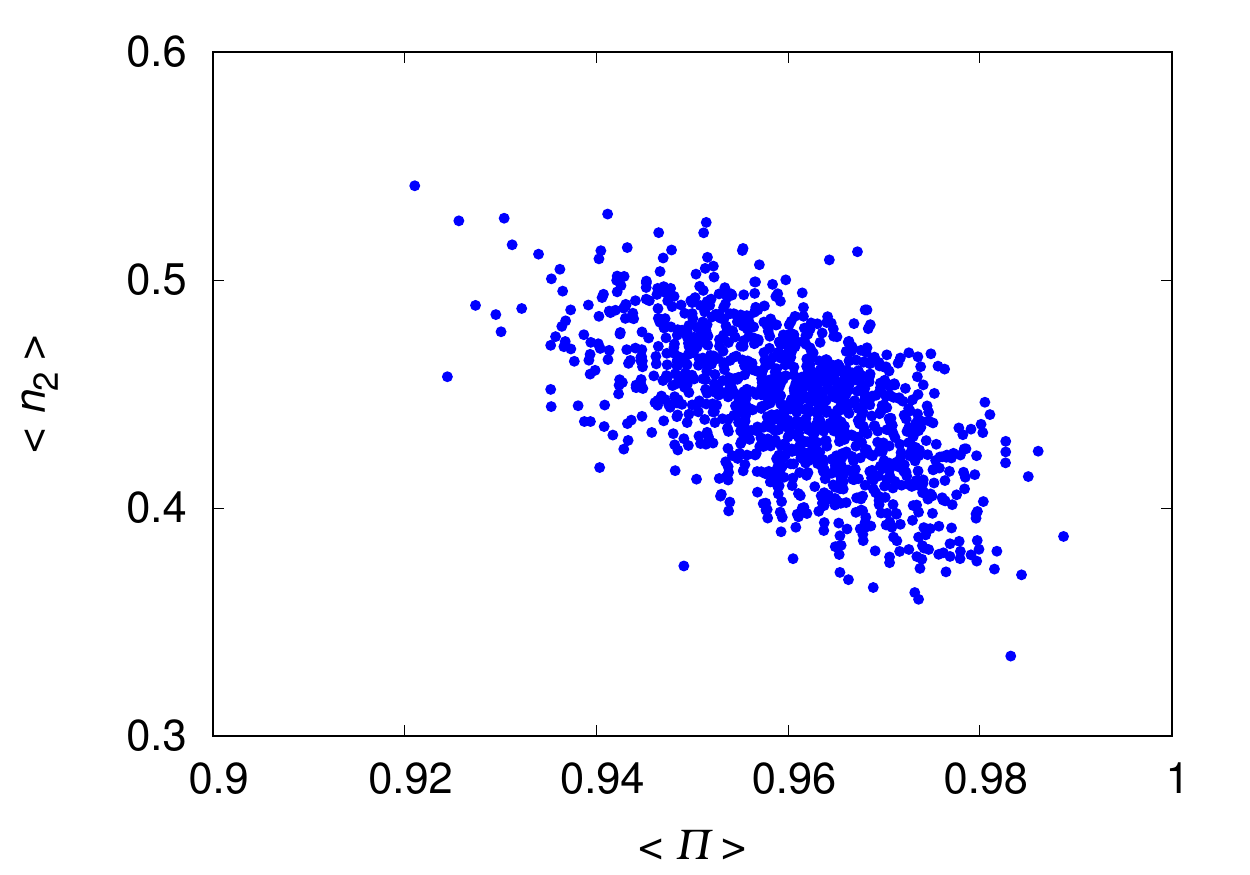}} 
\caption{Scatter plots of equilibrium properties of the metapopulation for $\sigma_e^2 =2$.  \textbf{Left Panel:} Fraction of patches where the two species coexist and relative abundance of species 1. \textbf{Right Panel:}  Fraction of patches where the two species coexist and relative abundance of species 2. The parameters are $L= 20$, $K_{max}=100$, $p_{mig} = 0.3$, $c=0$ and $\rho = 0$.}
\label{fig:S20}  
\end{figure}

\end{document}